\documentclass[aps,rmp,twocolumn,superscriptaddress
]{revtex4}

\usepackage{amssymb,amsmath,epsfig,wasysym,float}
\usepackage[sort&compress]{natbib}

\def\limacon{{lima\c{c}on}}{}

\newcommand{\tcoll}{\langle t_{\text{coll}} \rangle}
\newcommand{\ec}{\langle e^{-\kappa t_{\text{coll}}} \rangle_c}
\newcommand{\et}{e^{-\kappa t_{\text{coll}}({\bf x})}}

\begin{document}

\title{Leaking Chaotic Systems}

\author{Eduardo G. Altmann} 
\affiliation{Max Planck Institute for the Physics of Complex Systems, 01187 Dresden,
  Germany}
\author{Jefferson S. E. Portela}
\affiliation{Max Planck Institute for the Physics of Complex Systems, 01187 Dresden,
  Germany}
\affiliation{Fraunhofer Institute for Industrial Mathematics ITWM, 67663 Kaiserslautern, Germany}
\author{Tam\'as T\'el}
\affiliation{Institute for Theoretical Physics - HAS Research Group, E\"otv\"os University, P\'azm\'any P. s. 1/A, Budapest, H--1117, Hungary}

\begin{abstract}
There are numerous physical situations in which a hole or leak is introduced in an otherwise closed chaotic
system. The leak can have a natural origin, it can mimic measurement
devices, and it can also be used to reveal dynamical properties of the closed
system. A unified treatment of leaking systems is provided and applications to different physical
problems, both in the classical and quantum pictures, are reviewed. 
The treatment is based on the transient chaos theory of open systems, which is essential because real leaks have finite
  size and therefore estimations based on the closed system differ essentially from observations.
The field of applications reviewed is very broad, ranging from planetary astronomy and hydrodynamical flows, to plasma physics and
quantum fidelity.
The theory is expanded and adapted to the case of partial leaks (partial
absorption and/or transmission) with applications to room acoustics and optical microcavities in mind.
Simulations in the \limacon{} family of billiards illustrate the main text.
Regarding billiard dynamics, it is emphasized that a correct discrete-time representation
can be given only in terms of the so-called true-time maps, while traditional Poincar\'e
maps lead to erroneous results. Perron-Frobenius-type operators are generalized so that they describe true-time maps with partial leaks.\\
{\bf Published as: Rev. Mod. Phys. {\bf 85}, 869-918 (2013).}

\end{abstract}
\date{\today}

\maketitle
\tableofcontents


\section{Introduction} \label{sec:intro}\label{sec.1}
\subsection{Motivation}
\label{ssec.motiv}

Perhaps the most important distinction in the temporal evolution of a dynamical system is between persistent (asymptotic) and transient
  (finite-time) dynamics.
 Dynamical systems theory reflects this division and has developed specialized methods and tools to investigate persistent (e.g., strange attractors, asymptotic
  Lyapunov exponents) and transient (e.g., chaotic saddle, escape rates) chaotic dynamics~\cite{Ott-book,LaiTel-book,TelGruiz-book}. 
  These two approaches become connected when considering the effect of opening up a hole (or introducing a leak) in an otherwise
  closed chaotic system, converting by this persistent into transient chaos. 
Transient and persistent dynamics appear in both conservative and dissipative
  systems, and it is important to distinguish leakage (escape or removal of trajectories) from dissipation (contraction in the
  phase space). Introducing a leak  never generates an extra phase-space contraction and, e.g., a conservative system remains conservative
  after becoming leaky. 

More than a tool to investigate the relationship between different theories,
problems described by a closed chaotic system with a leak appear nowadays in a great variety of fields: 

\begin{itemize}

\item {\bf Room acoustics:} the decay of the sound energy 
characterized traditionally by the so-called reverberation time can be considered a
consequence of leaks: openings and absorbing surfaces on the room's boundary
 ~\cite{Bauer1990,Legrand1990,Legrand1991,Mortessagne1993}. Absorbing surfaces provide examples of partial leaks.

\item {\bf Chemical reactions}: unimolecular decay of excited chemical species have been modeled as an escape from a (chaotic) reactant
  region through a leak~\cite{Dumont:new,Ezra:new}.  

\item {\bf Hydrodynamical flows and environmental sciences:} the fact that certain regions of flows have 
special hydrodynamical features and might therefore change the properties of particles advected into these 
regions can be described by the so-called resetting mechanism~\cite{Pierrehumbert:1994}, which is
a kind of leak from the point of view of chaotic advection~\cite{Neufeld2000,Schneider2002,Schneider2003,Tuval2004,SFH:2005,Schneider2007}.

\item {\bf Planetary science and cosmology:} the (inelastic) collision of a small body with larger planetary objects 
leads to a drastic change in its dynamics compared to that in a point mass approximation of the larger bodies. In a first approximation the
problem can be treated as a loss due to leaks~\cite{Nagler:2004,Nagler:2005}. Similar ideas apply in cosmology~\cite{Motter2001}.

\item {\bf Optical microcavities:} Light rays in dielectric materials are partially transmitted and reflected (with the exception of regions where
  total internal reflection takes place).  Chaotic cavities can be constructed to provide a strong directionality of emission through such a
  partial leak, a requirement for laser~\cite{AltmannPRA,NS:1997,Lee:2004,Schwefel:2004,Ryu2006,Wiersig2008,DMSW:2009,Yan2009,Shinohara2009,Shinohara2010,Shinohara2011,Harayama2011}.

\item {\bf Plasma physics:} Particles in magnetic confinement devices are lost through collisions with sensors, antennas, or the chambers
  wall itself. These regions therefore play the role of a leak~\cite{Portela.review,Viana:2011,Portela2007,Evans:2002,Wingen:2007}.

\item {\bf Wave and quantum signatures of open systems:} Features related to that of a leaking classical dynamics appear in properties such as
  the (fractal) distribution of eigenstates~\cite{Kuhl2005,KNPS:2006,Casati:1999,Nonnenmacher:2008,Ermann2009,Pedrosa:2009,Novaes:2012}, the survival probability in
  simulations and experiments~\cite{Alt:1995,Alt1996,Fendrik:1997,Casati:Poincare,Friedman:2001,Kaplan:2001}, and in the fractal Weyl's law~\cite{LSZ:2003,Shep:2008,WM:2008,RPBF:2009,Kopp2010,Ermann:2010,Nonnenmacher:2011,Schomerus:2004}.

\end{itemize}

In dynamical-systems theory, the idea of leaking an otherwise closed chaotic systems was first proposed by Pianigiani 
and Yorke as early as 1979:
\begin{quote}
{\em Picture an energy conserving billiard table with smooth obstacles so that all
  trajectories are unstable with respect to the initial data. Now suppose a small hole is
  cut in the table so that the ball can fall through. We would like to investigate the
  statistical behavior of such phenomena}~\cite{PY:1979}.
\end{quote}
Their main motivation was precisely to investigate transient chaos, as opposed
to persistent chaos. The leakage procedure was therefore a tool to create
transiently chaotic systems. Interestingly, the development of the theory of transient chaos
happened not to follow this line over decades.

The importance of this mathematical approach becomes apparent when one realizes the
multitude of situations in which the leak region has a well-defined physical
interpretation. This aspect was first emphasized by Smilansky and 
coworkers, who pointed out that 
any measurement (both classical and quantum) leads unavoidably to a leakage of the system. They wrote in 1992:
\begin{quote}
{\em A discrete spectrum is a property of a closed system. However, the process of
  measuring the spectrum of a bounded system consists of coupling the system to an
  external continuum. Thus, for the purpose of measurement, the closed system is turned
  into a scattering system
}~\cite{Doron1992}.
\end{quote}
Physical realizations of the leak can thus be either the effect of measurement devices or intrinsic properties of the system, 
such as, e.g., absorbing boundaries.

Apart from physical leaks, there are also different {\em theoretical} motivations for considering leaking systems:

\begin{itemize}

\item Leakage is a tool to understand the dynamics of closed systems, providing thus a sort of {\em chaotic
spectroscopy}~\cite{Doron1992,Doron1992b}. 
More generally, systems with leaks help 
monitoring  or {\em peeping at} 
chaos~\cite{Bunimovich2007}  [see also \cite{Nagler2007}].
In this context, as in~\cite{PY:1979}, billiards with leaks
were the first systems investigated because they allow a natural connection between the
classical and quantum pictures~\cite{Bauer1990,Alt1996,Alt:1995}. 

\item Leaking systems have been explored in the context of synchronization of chaotic
oscillators~\cite{Jacobs1998}, and of the control of chaos~\cite{Paar1997, Paar2000,Buljan2001}. 

\item Leakage reveals the foliations inside the closed system~\cite{Schneider2002,Aguirre2003,Sanjuan2003,Aguirre:2008} that lead, e.g., to
  fractal exit boundaries~\cite{Bleher1988,Ree2002,Portela2007}. 

\item The distribution of
Poincar\'e recurrences, which is commonly used to quantify properties of closed
Hamiltonian dynamics~\cite{Chirikov1984,Zaslavsky:2002}, is equivalent to the survival
probability in the same system with a leak~\cite{Altmann2008}. 

\item Several quantifications of
wave or quantum chaos, such as Loschmidt echo~\cite{Gorin:2006,Jacquod:2009} or fidelity decay~\cite{Peres1984}, can be realized physically
in configurations that are analogous to introducing a localized leak in a closed system~\cite{GR:2007,Goussev:2008,Ares:2009,Kober:2011,Hohmann:2008}.

\end{itemize}

The common feature in all applications and theoretical procedures listed above is that one has some freedom when choosing the opening,
i.e. the leak in a well-defined closed chaotic system~\cite{Schneider2002}. 
This should be contrasted to genuinely open systems in which
the openness is intrinsic, and only slight parametric
changes are physically realistic, which typically do not allow to
go to the closed-system limit. 
Although both classes of systems are dynamically open, of our aims is to emphasize the benefits of considering leaking systems, which
are more precisely defined by two key elements: 
\begin{itemize}
\item [(i)] the existence of a well-defined closed system which can be used as a comparison;

\item [(ii)] the possibility of controlling (some) properties of the leak such as position, size, shape, or reflectivity.

\end{itemize}
Property (i) guarantees that one can compare transient and asymptotic dynamics and can be considered as a particular case of (ii) if
the possibility of reducing the leak size to zero is assured. 
Leaking systems can be both dissipative and conservative (Hamiltonian). Within this latter
category, we consider the problem of chaotic scattering [as typically defined, e.g., \cite{Gas-book}] to be beyond the scope of this
review because it lacks properties (i) and (ii) above\footnote{In some scattering cases it is possible to ``close'' the  inside of the
  scattering region (e.g., in the three disk problem, when the disks touch). However, in these cases the closing procedure is either
  arbitrary or unnatural from the point of view of scattering (e.g., the incoming trajectories are unable to enter the chaotic region).}

Our main approach in this Review Article is based on transient chaos theory, which is applied to the case of leaky systems and connected to different recent 
applications. Our aim is to be understandable by nonspecialists interested in learning what the implications of dynamical-systems theories
are to specific applications. At the same time, we emphasize how specific applications pose
new questions to the theory. Thus, we devote special attention to developing a theory consistent with the following two aspects
required by different applications:
\begin{itemize}

\item Leaks are not necessarily full holes, they might be ``semipermeable'', i.e, 
the energy content of trajectories entering a leak is partially transmitted and partially reflected. In such cases the leak is called a {\it
  partial} leak.

\item 
Discrete-time maps of open flows might lead to a loss of information over the temporal properties, 
and therefore it is essential to use the generalized concept of {\em true-time maps}~\cite{Kaufmann2001},
which will be defined in Sec.~\ref{ssec.billiard1}.

\end{itemize} 
We note here that even though our focus and numerical illustrations are on billiards (Hamiltonian systems), the
theoretical framework, and many of the specific results can be naturally extended to systems with dissipation.

In the remainder of this section we motivate the general problem through a historical example and a simple simulation.
In Sec.~\ref{sec.theory} we confront the simplest theory, based on the properties of the
closed system, with the appropriate transient chaos theory for open systems. A generalization of this theory to partial leaks is also given. 
Section~\ref{sec.operators} is devoted to a Perron-Frobenius-type operator formalism that is able to describe any kind of leaking dynamics.
The
main implications of transient chaos theory are explored in Sec.~\ref{sec.fullchaos}, including the case of multiple leaks and emission.
In Sec.~\ref{sec.weak} we discuss how to describe the generic
situation of weakly chaotic Hamiltonian systems (mixed phase space). Finally, in
Sec.~\ref{sec.applications} we use our results to give a detailed view on some of the problems we started this Section with.
Our conclusions
appear in Sec.~\ref{sec.conclusions}. In the Appendices~(Sec.~\ref{sec.appendices}) we
discuss some important but technical aspects of open billiards (like e.g., different types of measures
and algorithms). 

\subsection{Classical leaking: Kinetic theory and Sabine's law} \label{ssec.sabine}

Historically, perhaps the first problem involving systems with leaks was one related to the kinetic theory of gases.
Consider a container filled with ideal gas. 
How is the container emptied after a {\em small}
leak $I$ is introduced on its boundary? 

The answer can be obtained from an elementary application of the kinetic theory.
Here we follow basically the treatment of \cite{Bauer1990} and ~\cite{Joyce:1975}.
Let $I$ be a disk of area $\Delta A $ on the surface of the container
and $f({\bf v}, t)$ be the phase-space density of the particles, for which
\begin{equation}
\int f({\bf v},t) d^3 {\bf v} = \frac{N(t)}{V},
\label{eq.f}
\end{equation}  
where $N(t)$ is the number of particles in the container of volume $V$ at time $t$.   
The number of particles with velocity ${\bf v}$ leaving the system over 
a short time interval $dt$ is then
$dN=dt \Delta A {\bf v n} f({\bf v},t) d^3 {\bf v}$,  where ${\bf n}$ 
is the normal vector of the surface at the leak $I$.
The total number is then obtained by carrying out an integration over all velocities.
Thus, the time derivative of the number $N(t)$ of particles inside the container is 
\begin{equation}\label{eq.dndt1}
\frac{dN(t)}{dt}= - \Delta A \int {\bf v n} f({\bf v},t) d^3 {\bf v},
\end{equation} 
where the minus sign indicates that particles are escaping.

{\em Molecular chaos}, a basic ingredient of  kinetic theory, implies that an {\em equilibrium} phase-space density exists. In our
problem it is {\em homogeneous} (location independent) and   {\em isotropic}: all velocity directions are equally probable. 
In the limit of small~$\Delta A$ we can expect that there is a quasi-equilibrium distribution $f({\bf v},t)$ in the
  open system which sets in on a time scale shorter than the average lifetime. This quasi-equilibrium distribution shares the properties of that of closed systems. In this case, isotropy guarantees that the phase-space density depends only on the modulus $v$ of the velocity, and it is, therefore, 
convenient to use spherical coordinates for the integration. With $\theta$ being the
angle between velocity and the normal vector, ${\bf v.n}=v \cos{\theta}$, Eq.~(\ref{eq.dndt1})
reads as 
\begin{equation}
\frac{dN(t)}{dt}= - \Delta A \int_0^{\infty} \hspace{-0.1cm} v f(v,t) v^2 {d v}   \int_0^{\pi/2}
\hspace{-0.1cm} \cos{\theta} \sin{\theta} 
 d\theta  \int_0^{2\pi} \hspace{-0.1cm} d\phi.
\label{dN}
\end{equation} 

The spherical symmetry of the phase space density applied to (\ref{eq.f}) leads to
\begin{equation}
\int_0^{\infty} f(v,t) v^2 {d v} 4 \pi=\frac{N(t)}{V},
\end{equation} 
and implies that $w(v)=4 \pi f(v,t) v^2 V/N(t)$ is the probability density 
for the velocity modulus $v$ in the gas.  

Substituting this into (\ref{dN}), the first integral is found to be proportional to the average 
$\langle v \rangle$ of the velocity modulus. By carrying out all integrals, we   
find
\begin{equation}\label{eq.dndt2}
\frac{dN(t)}{dt}= - \frac{\Delta A \langle v  \rangle}{4 V} N(t).
\end{equation} 
As long as $\langle v \rangle$ is independent of time\footnote{In a thermodynamical system
the decay of particles eventually leads to a reduction in the pressure and temperature
inside the container, and thus to a reduction of~$v$. Here we are
interested 
in systems with constant~$v$. The exponential decay is then valid for any $t>0$.}
the decay of the particle number is thus exponential,
of the form of $\exp{(-\kappa t)}$, with an {\em escape rate} 
\begin{equation}\label{eq.kappav}
\kappa=\frac{\Delta A \langle v  \rangle}{4 V}.
\end{equation} 

For simplicity we focus here on an ensemble of identical particles with the {\em same}
velocity~$v$ colliding elastically, in which case~$\langle v \rangle \mapsto v$ in
Eq.~(\ref{eq.kappav})\footnote{Note that the dynamics of elastic collisions of identical particles is
equivalent to the dynamics of independent particles, as can be seen by exchanging
particle labels at collision.}. 
The reciprocal of the escape rate, which turns out to be the average lifetime, can then be written as
\begin{equation}
\langle \tau \rangle=\frac{1}{\kappa}=\frac{4V}{\Delta A v}.
\label{tau}
\end{equation} 
This is the time needed for the decay of the survivors by a factor of $e$.
Since the result is linear in $\Delta A$, and the velocity distribution is not only isotropic but also 
{homogeneous}, i.e. independent
of the position along the wall, the expression remains valid for {\em small} leaks $I$ of {\em any} shape, and $\Delta A$
is then the total leaking area. Since $\Delta A$ is small, $\langle \tau \rangle$ is large, and hence the assumption of a quasi-equilibrium distribution
becomes justified a posteriori.

An interesting, historically independent development is Sabine's law, a central object of 
architectural acoustics. This law says that the residual sound intensity in a room decays
exponentially with time
\cite{Joyce:1975,Mortessagne1993}. The duration to decay below the audible intensity is called the 
reverberation time, $T_r$, and was found experimentally by W. C. Sabine in 1898 to be
\begin{equation}
T_r=6 \ln{(10)}\; \frac{4V}{\Delta A c}.
\label{T}
\end{equation} 
Here $c$ is the sound velocity, and $\Delta A$ is the area
of the union of all openings of the room (or of all energy absorbing surfaces after proper normalization).
With $c=340$ m$/$s, the numerical value of $T_r$ in SI units is $T=0.16 V/\Delta A$.
Sabine's experiments also showed that the reverberation time for a pleasant
sound perception is on the order of a few seconds for a good auditorium, 
and he designed concert halls
(like, e.g., the Boston Music Hall) according to this principle.   

A comparison of (\ref{tau}) and (\ref{T}) reveals that Sabine's law is nothing but
an application of the exponential decay of the particle number evaluated with $v=c$ as the
particle velocity. 
What is escaping in this problem is however not particles, but the energy of the
sound waves. In the geometrical limit of room acoustics, one can consider
the decay of energy as the problem of particles which travel along sound rays    
and lose part of their energy upon hitting the leak or the absorbing surface.
The most remarkable property of Eq.~(\ref{T}) is 
its universality: the reverberation time is {\em independent} of the 
location of the sound source and of the shape of the room, provided the absorption is
weak and sound disperses uniformly around the room e.g., due to roughness or irregular geometry of the walls~\cite{Mortessagne1993}.   
 The pre-factor $6 \ln{(10)}$ in Eq.~(\ref{T}) 
results from the fact that in the acoustic context the decay below the audible 
intensity implies $60$ dB, i.e., a decay factor of $10^6$, instead of 
a factor of $e$ in (\ref{T}).
Sabine's law~(\ref{T}), dated back to 1898, appears thus to be the 
first application of leaking chaotic dynamical systems in the history of science!

We now take a closer look at the assumptions in the derivations above from the
perspective of the 
dynamics. In terms of the modern theory of dynamical systems, the
isotropy and homogeneity of the velocity distribution are a consequence of the following
two hypotheses:
\begin{itemize}
\item [{\bf H1:}] the leak size is {\em small}, so that the phase-space distribution does not change
  due to the openness; and
\item [{\bf H2:}] the particle dynamics inside the room is {\em chaotic}, more
  technically, the dynamics is {\em ergodic and strongly mixing} (implying exponential decay of correlations in time).
\end{itemize}

Under these assumptions, the exponential decay is valid also in other dimensions. 
For instance,  the escape rate in two-dimensional billiards is then found to be
\begin{equation}\label{eq.kappa2d}
\kappa=\frac{\Delta A v}{\pi V},
\end{equation} 
where $\Delta A$ is the length of the leak along the perimeter, and $V$
is the two-dimensional volume, the area, of the billiard table.
The replacement of factor $4$ by $\pi$ is due to the geometrical change from spherical
to planar polar coordinates.

It should be noted that in both cases the survival probability $P(t)$ up to time $t$
is 
\begin{equation}\label{eq.exp}
P(t)=e^{-\kappa t},
\end{equation}  
as obtained from Eq.~(\ref{eq.dndt2}), with initial condition~$P(0)=1$.
The probability $p(\tau)$ to leave around the escape time $\tau=t$ is the negative derivative
of~$P(t)$ and thus
\begin{equation}
p(\tau)=\kappa e^{-\kappa \tau},
\end{equation}  
and $P(t)=\int_t^{\infty} p(\tau) d\tau$. 
Since the exponential decay holds from the very beginning, the average
lifetime
\begin{equation}\label{eq.te}
\langle \tau \rangle =\int_0^{\infty} t' p(t') dt'= \int_0^{\infty}P(t') dt'
\end{equation}
is found to be $\langle \tau \rangle =1/\kappa$, which was used in Eq.~(\ref{tau}). The symbol~$\langle \ldots \rangle$
can be interpreted as an ensemble average.

Finally, it is instructive to write both expressions~(\ref{eq.kappav}) and~(\ref{eq.kappa2d}) of the escape rate as
\begin{equation}\label{eq.mut}
\kappa=\frac{\mu(I)}{\langle t_{\text{coll}} \rangle},
\end{equation}
where $\mu(I)={\Delta A}/{A}$ is the relative size of the leak
compared to the full wall surface, and can therefore be
considered as the {\em measure} of the leak (taken with respect to the
Lebesgue measure). The denominator has the dimension of time and is given by
\begin{equation}\label{eq.tau2d3d}
\langle t_{\text{coll}} \rangle=\frac{4V}{A v}, \;\;\;\; \mbox{or} \;\;\;\; \langle t_{\text{coll}} \rangle = \frac{\pi V}{A v},
\end{equation}
in the three- and two-dimensional case, respectively.
These~$\langle t_{\text{coll}} \rangle$'s turn out to be the precise expressions of the 
{\em average 
collision time}  between collisions
with the wall (or, after a multiplication by $v$, the mean-free-path),
well known for three- and two-dimensional {\em closed} rooms or billiards. As emphasized by~\cite{Joyce:1975,Mortessagne1993}, these results were obtained already in the late XIX century
by Czuber 
and Clausius. 
It is the average collision time that sets the characteristic time with which the average lifetime should be compared: for small leaks
  $\langle \tau \rangle \gg \tcoll$, i.e. the time scales strongly separate.

By definition, $\langle t_{\text{coll}} \rangle$ can be expressed as the average  over
the local collision times $t_{\text{coll}}({\bf x})$ for the phase-space coordinates
${\bf x}$ along the wall as
\begin{equation}\label{eq.tcoll}
\langle t_{\text{coll}} \rangle=\int  t_{\text{coll}}({\bf x}) d{\mu},
\end{equation}
where $\mu$ is the uniform phase space (Lebesgue) measure characteristic of 
conservative systems.
All equations found are taken with respect to the distributions characteristic of the 
{\em unperturbed} system. This is consistent with the small leak assumption ({\bf
  H1}
above) so that the escape rates obtained can be considered as
a leading order result in a perturbation expansion where averages 
can yet be taken with respect to distributions characterizing the unperturbed (closed) system.

In the modern applications mentioned in Sec.~\ref{ssec.motiv}, however, conditions
{\bf H1} (small leaks) and {\bf H2} (strong chaos) are typically not met. Here we
discuss in detail what happens in such cases. 
For instance, in any practical application the leak size is not, or cannot be made, infinitesimally small so that {\bf H1} is
violated and perturbation expansions break down. 

\begin{figure*}[!ht]
\includegraphics[width=1.6\columnwidth]{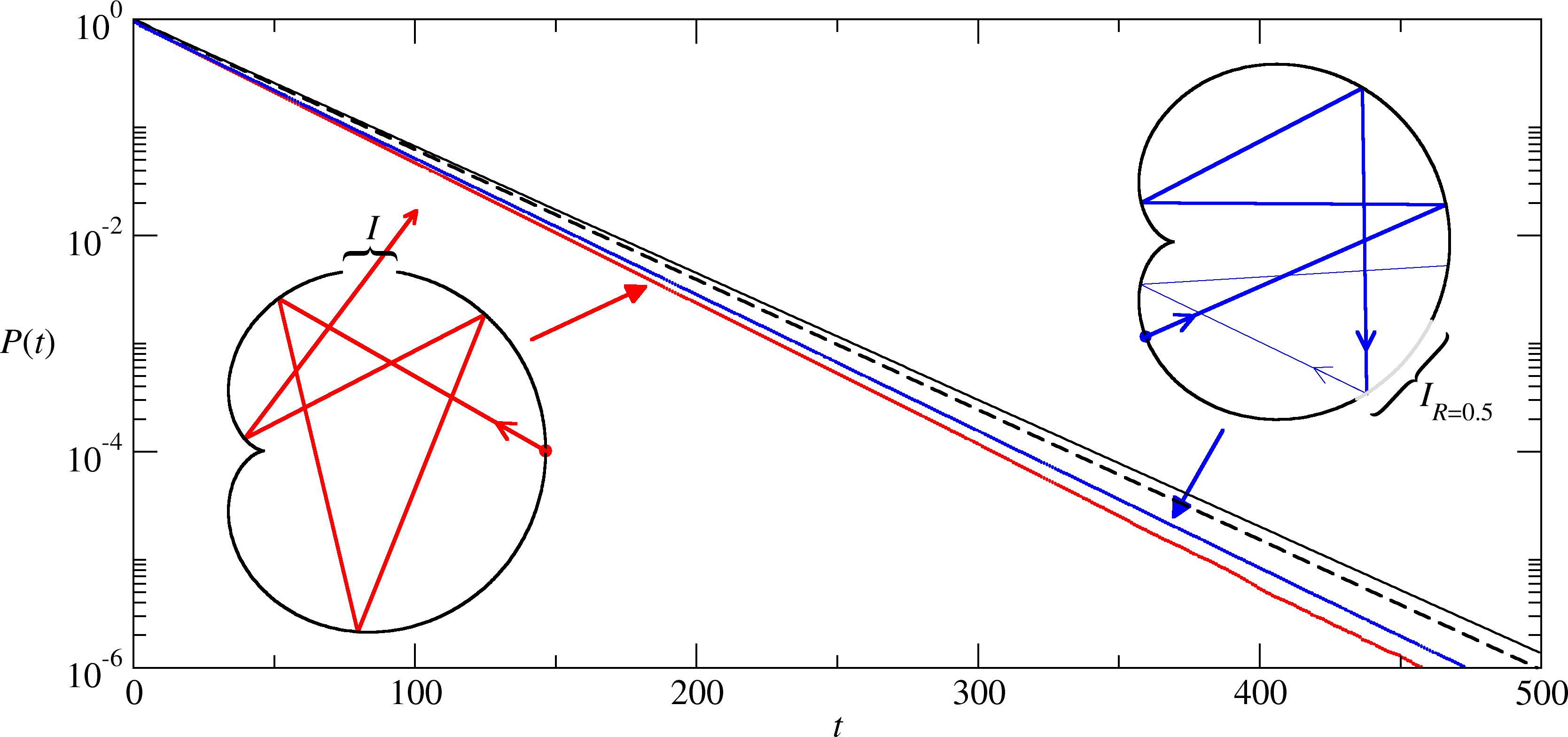}
\caption{(Color online) Escape of particles in a billiard with a finite leak.   The survival probability~$P(t)$ for the strongly chaotic cardioid billiard
  [Eq.~(\ref{eq.limacon}) with~$\varepsilon=1$] is shown for two different configurations of the leak~$I$ (same effective size $2 \Delta s$ but different
  leak positions $s_l$, and reflectivity $R$). The first (bottom, red) line corresponds to a full leak [see Eq.~(\ref{eq.leak1})] centered
  at the top of the billiard, $s_l=0.5$, with size~$2 \Delta s = 0.1 $ ($5\%$ of the perimeter) as shown in the left inset. The second  (next
  to the bottom, blue) line corresponds to a partial
  leak with~$R=0.5$   [see Eq.~(\ref{eq.leak2})] centered at $s_l=-0.25$ with a size~$2 \Delta s = 0.2$ as shown in the
  right inset.  The observed
  escape rates are~$\kappa=0.03002 \pm 0.00007$ 
(full leak) and $\kappa=0.02904 \pm 0.00003$ (partial leak), clearly  different from the predictions~$\kappa=0.0270$ (upper full line) based on 
Sabine's  law~(\ref{eq.mut})-(\ref{eq.tau2d3d}), and the naive estimate $\kappa^*=0.0277$, Eq.~(\ref{eq.kappastar}) (dashed line) with~$\mu(I)=0.05$. Initial conditions were uniformly distributed in the phase   space ($s,p=\sin \theta$, see Fig.~\ref{fig.closed}).}
\label{fig.robnik1}
\end{figure*}

We shall see that an exponential decay of the survival probability typically remains valid for  
finite leak sizes, at least after some initial period. 
{The estimation of the escape rate can be greatly improved} by considering a similar expression as in Eq.~(\ref{eq.mut}),  the measure of the leak divided by   
the average collision time, however, both taken with respect to a {\em different} measure:
\begin{equation}\label{eq.changemu}
\mu(I) \rightarrow \mu_c(I), \;\;\;\;\;\langle t_{\text{coll}} \rangle \;\rightarrow
\;\langle t_{\text{coll}} \rangle_c=\int  t_{\text{coll}}({\bf x}) d{\mu}_c.
\end{equation} 
The new relevant measure~$\mu_c$ differs from the original Lebesgue measure~$\mu$ since many particles
have left the system by the time of observation, and what counts is the set of 
long-lived particles. 
With finite leaks, the decay differs substantially from the naive estimate obtained by
using the original Lebesgue measure, as illustrated for our billiard example in Fig.~\ref{fig.robnik1}.
Even if precise definitions and further details appear only later, the conceptual difference between $\mu$ and $\mu_c$ is clear
  [compare Figs.~\ref{fig.closed}b and~\ref{fig.saddle2} for an illustration of the dramatic changes in the phase space of the billiard].

The theory of open dynamical systems 
tells us that this new c-measure is the so-called {\em conditionally invariant measure}
introduced by~\cite{PY:1979}, which is didactically introduced and investigated in 
Secs.~\ref{sec.theory}-\ref{sec.fullchaos}. The violation of hypothesis {\bf H2} of strong
chaos leads to even more radical changes, e.g. to a deviation from the exponential decay for
long times. This case will be investigated in Sec.~\ref{sec.weak}.

\subsection{Billiard dynamics and true-time maps}\label{ssec.billiard1}

In dynamical-systems theory,
the kinetic problem with fixed velocities and Sabine's picture of room acoustics
are described as billiard
systems, as noticed already by~\cite{Joyce:1975}. Billiards are defined as bounded volumes or areas inside which
particles move in a straight line with constant velocity~$v$ between collisions at the
boundary, where they experience specular, elastic reflection (i.e., the angle of incidence is equal to the angle  $\theta$ of reflection and
the absolute value of the velocity $v$ is conserved)~\cite{Chernov-book}. A recent sample of the research on billiards can be
found in~\cite{billiards.issue}.

For numerical and visualization convenience, we illustrate
our results in two-dimensional billiards. 
In this case the dynamics can be described in a
two-dimensional phase space, achieved by replacing the continuous-time dynamics by a corresponding
discrete-time system~${\bf f}$ that maps the position $s$ along the boundary and 
angle~$\theta$ of the~$n$-th collision into those of the $(n+1)$-th collision at the
boundary. By convention, the map $f$ connects the momenta right {\em after} the collisions. This procedure
corresponds to a Poincar\'e surface of section. The dimension of the full (four-dimensional) phase space is reduced by two (using momentum
conservation and the condition of collision). The shape of the billiard's  
boundary uniquely defines the dynamics of the 
particles, and system-specific properties depend sensitively on this
shape.  
It is
convenient to write the phase space of the map in terms of  Birkhoff coordinates 
${\bf x}=(s,p\equiv\sin \theta)$
in which case 
\begin{equation}
{\bf f}: ({\bf x}_n)\mapsto({\bf x}_{n+1})
\label{mapf}
\end{equation}
is area preserving~\cite{Berry1981,Chernov-book}.

A faithful representation of the temporal dynamics of billiards
requires augmenting (\ref{mapf})
by keeping 
track of the information about the time of each trajectory: 
\begin{equation}\label{eq.tn}
t_{n+1}=t_n+t_{\mathrm{coll}}({\bf x}_{n+1}),
\end{equation}
where $t_n$ denotes the time of the $n$-th collision at the boundary of the billiard, 
and $t_{\mathrm{coll}}$ denotes the time between two subsequent collisions. 
In what follows we associate $t_{\mathrm{coll}}$ with the Birkhoff coordinates of the {\em later} collision (${\bf x}_{n+1}$)
in order to be able to speak about the collision times within the leak when systems   
with leaks are considered (see Eq.(\ref{eq.leak1})). 

Equations (\ref{mapf})-(\ref{eq.tn}) are called the {\it true-time} map 
as coined by~\cite{Kaufmann2001}, which is also frequently used in the billiard context [see, e.g.,~\cite{Bunimovich2007}].  
More generally, true-time maps provide a link between discrete-time maps and continuous-time flows in the same spirit as described by
the mathematical concepts of {\em suspended} flows, {\em special} flows, or {\em flows under a function}~\cite{Gas-book,Katok:book}. They
have also been used in the context of transport models~\cite{Klages:2004,Matyas:2011}.

A true-time map is equivalent to the continuous-time representation, but leads to faster and more reliable results than a direct integration of the billiard flow. 
The different collision times can be taken into account also in the Perron-Frobenius representation of the dynamics, as shown in Sec.~\ref{ssec.operators}.
 
In contrast, 
the often used {\em Poincar\'e map}, represented by  
(\ref{mapf}) alone, provides a distorted image of time.
It implies associating with each pair of collision  the same time interval and thus loses contact with the temporal dynamics of the
continuous-time physical system (e.g., it can overestimate the importance of events with short
collision times). The Poincar\'e map generates a measure \emph{different} from 
that of the true-time map, and thus leads to erroneous results. 
When talking about maps in the billiard context, we, therefore, always mean 
true-time maps. (Poincar\'e maps of billiards will be mentioned again in Table~\ref{tab.ic.num} and Appendix~\ref{ssec.diff},
to illustrate the difference to true-time maps.)

\subsection{Example in a chaotic billiard}\label{ssec.billiard2}

\begin{figure}[!ht]
\centering
\includegraphics[width=0.8\columnwidth]{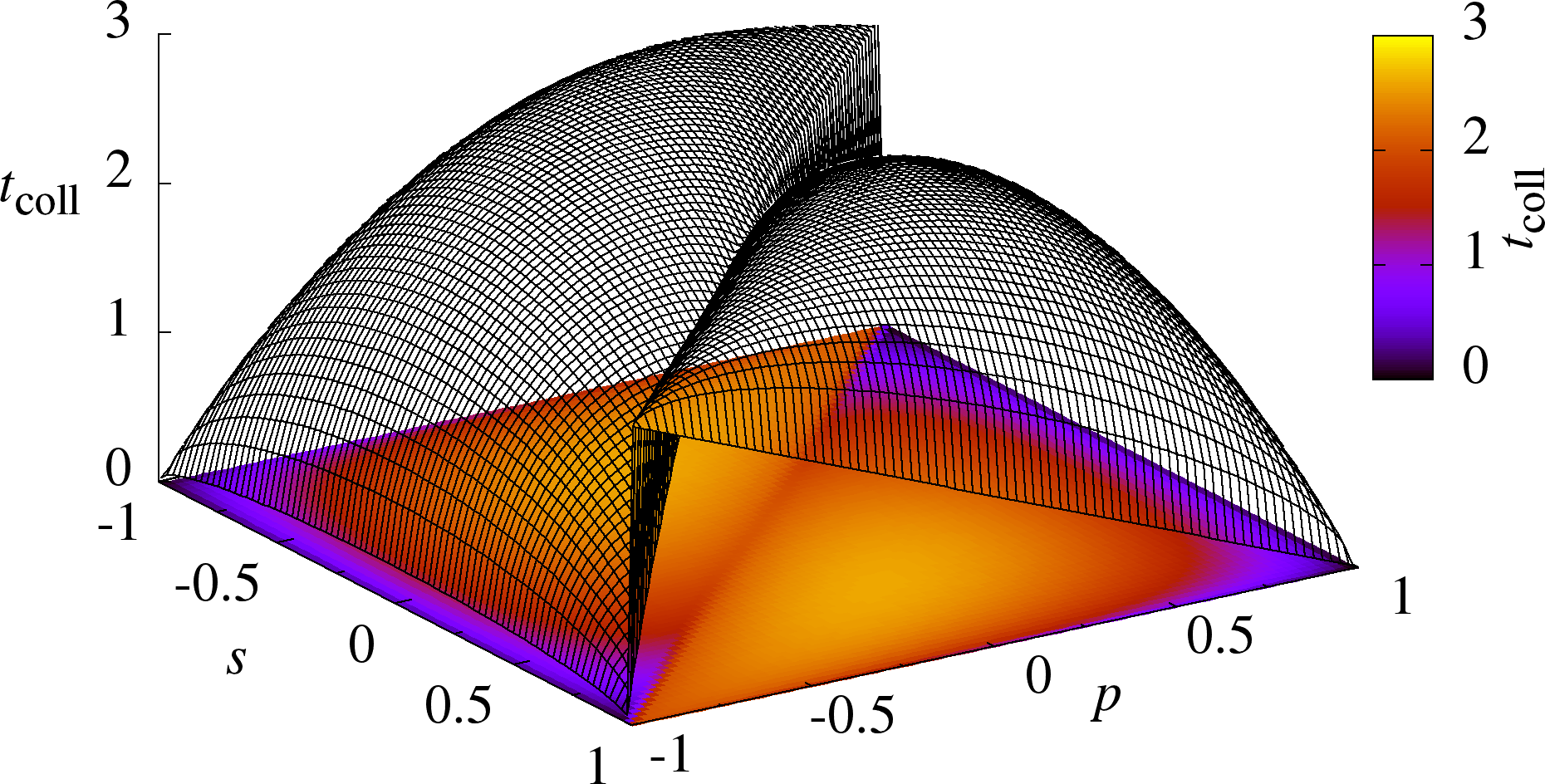}
\caption{(Color online) Collision time $t_{\text{coll}}({\bf x})$ as a function of the phase-space coordinates ${\bf x}=(s,p)$ in the cardioid
  billiard, Eq.~(\ref{eq.limacon}) with $\varepsilon=1$. Consistent with the convention in Eq.~(\ref{eq.tn}), $t_{\text{coll}}({\bf x})$ is
  defined as  the distance (or time, since $v\equiv1$) between ${\bf x}$ and the {\em previous} collision $f^{-1}({\bf x})$. The
  discontinuity close to the diagonal reflects the billiard's cusp at $s=\pm1$, see Fig.~\ref{fig.closed}a.}
\label{fig.tcoll3D}
\end{figure}

The main properties of two-dimensional billiards can be illustrated by
the family of \limacon{} billiards introduced by~\cite{Robnik1983} whose borders are defined in polar
coordinates~(${r},\phi$) by \limacon-like curves
\begin{equation}\label{eq.limacon}
\mathfrak{r}(\phi) = S(1 + \varepsilon \cos \phi),
\end{equation}
where~$S$ scales the size and~$\varepsilon$ controls the shape of the billiard. The
ratio~$S/v$ defines the unit in which time~$t$ is measured, which is the only effect
of~$S$ and $v$ on the dynamics. We therefore set $S=v=1$ in what follows,
which implies that the perimeter length is $A=8$, the billiard's area is 
$V=3\pi/2$, and the mean collision time (\ref{eq.tau2d3d}) is thus $\langle t_{\text{coll}} \rangle=3\pi^2/16$.
For convenience, throughout we use the convention that 
the perimeter coordinate $s$ is parameterized between $-1$ and $+1$ (see e.g.
Figs. \ref{fig.tcoll3D} and \ref{fig.closed}).  

For~$\varepsilon=0$ we recover the circular billiard, exhibiting 
regular dynamics. For~$\varepsilon=1$, Eq.~(\ref{eq.limacon}) defines the {\em cardioid} billiard, which is
ergodic and strongly mixing~\cite{Robnik1983,Wojtkowski1986}, satisfying the hypothesis of
strong chaos~{\bf H2} (Sec.~\ref{ssec.sabine}). For~$0<\varepsilon<1$, the billiard typically shows the coexistence of 
chaotic and regular components in the phase space~\cite{Dullin2001}, and exhibits weak chaos.
The collision time~$t_{\text{coll}}({\bf x})$ needed for the true-time map (\ref{mapf})-(\ref{eq.tn}) can be determined numerically and is
shown in Fig.~\ref{fig.tcoll3D} 
for the cardioid case $(\varepsilon=1)$. 
 
We now introduce a leak in such a closed billiard and test the limitations of
  Sabine's prediction.  For concreteness, consider removing $5\%$ of the top part of the perimeter of the strongly chaotic 
  cardioid billiard, as  shown in the left inset of Fig.~\ref{fig.robnik1}.  Numerical simulations of the survival probability of
  trajectories in this system yield an escape rate~$\kappa=0.030$, which differs substantially from the escape rate  $\kappa=0.8/3\pi^2=0.027$
  obtained from Sabine's original estimate 
  (\ref{eq.mut}) by using $\mu(I)=0.05$ and $\tcoll=3\pi^2/16$ . In fact, Sabine's estimates holds for infinitesimally small leaks only, and a naive extension for finite leaks will be presented in Sec.~\ref{ssec.closed}
and leads to Eq.~(\ref{eq.kappastar}), which is a generalization of Sabine's
prediction. 
This yields $\kappa^*=0.0277$ which is still about $10\%$ below the observed one.  
Although this difference appears to be small, it shows up
in the exponent of an exponential time dependence.
After $500$ time units, the number of observed survivors 
is a factor of $e^{0.003\times 500}=e^{1.5} \approx 4.5$ 
times smaller than the one based on the closed-system estimate.
In Fig.~\ref{fig.robnik1} it corresponds to the difference between the dashed (generalized Sabine's formula) and the bottom solid (direct
simulations) lines.

This very basic observation is just 
the simplest temporal manifestation of a series of discrepancies that will be discussed in
and that are all originated in the difference between the dynamics
of 
the closed and of the leaky systems (see Figs.~\ref{fig.closed} and~\ref{fig.transient} for the
illustration of the change in the 
phase space). All these 
illustrate the need for a deeper theoretical understanding of systems with leaks, beyond
the results obtained under simplifying assumptions such as those used to obtain Sabine's
law in Sec.~\ref{ssec.sabine}.

Before exposing the theory  in Sec.~\ref{sec.theory}, we define in full generality the 
problem of introducing a leak in an otherwise closed system. We 
emphasize that our motivation for using two-dimensional billiards is visual convenience and
direct connection to applications.
The idea of introducing leaks in dynamical systems
applies to a much broader class of systems where the results of this paper can also be
applied, such as e.g., non-billiard type Hamiltonian systems, dissipative systems,
and also in higher dimensions.

\subsection{Definition of the leak}\label{ssec.leak}

Consider a closed system described by a map ${\bf f}_{\text{closed}}({\bf x})$. Here
we are mainly interested in maps ${\bf f}_{\text{closed}}$ that admit chaotic motion, but the
introduction of a leak is independent of this requirement. Choose the leak~$I$ as a subset of the phase space
$\Omega$. In its simplest version, a particle is regarded as having
escaped the system after entering the region $I$. The dynamics can thus be
described by the following map: 
\begin{equation} \label{eq.leak1}
{\bf x}_{n+1}={\bf f}({\bf x}_n) =
 \left \{ \begin{array}{ll} {\bf f}_{\text{closed}}({\bf x}_n) & \mbox{ if } {\bf x}_n \notin I \\
                                       \mbox{ escape } & \mbox{ if } {\bf x}_n \in I .\\
                                       \end{array} \right.
\end{equation}
Since escape is considered to occur one step {\em after} entering~$I$, map ${\bf f}$ is defined in~$I$. 

In the example shown in the left inset of Fig.~\ref{fig.robnik1}, the leak~$I$ is centered at the boundary point~$s_l=0.5$ with width~$2\Delta s=0.1$. 
 In general, a
leak~$I$ can be centered at any phase-space position~${\bf x}_l = (s_l,p_l) \in \Omega$. 
The leak mentioned above corresponds thus to~$I=[s_l-\Delta s, s_l+\Delta s]\times [-1,1]$, representing
a rectangular strip parallel to the $p$-axis. 
A prominent physical example of
leaks represented by strips parallel to the $s$-axis
is that of dielectric cavities. In this case light rays
coming from a medium with higher refractive index ($n_{\text{in}}> n_{\text{out}}$) are totally reflected if they collide with
$|p|>p_{\text{critical}}=n_{\text{out}}/n_{\text{in}}$,  where
$p_{\text{critical}}=\sin(\theta_{\text{critical}})$ is the critical momentum ($\theta_{\text{critical}}$
is the critical angle). The leak is then $|p|<p_{\text{critical}}$, $s$ arbitrary.
 
A general leak~$I$ can have
arbitrary shape (e.g., circular, square, oval, etc.) and can also be
composed of disjoint regions:~$I=\cup I_i$. In this last case, a
natural question is
that of the nature of the set of initial conditions which lead to each $I_i$, i.e., of
the properties of the escape basins $B_i$. This problem will be discussed in Sec.~\ref{ssec.basins}. 
For presentational convenience we focus on leaks at the billiard's boundary, in which case
we can still faithfully represent the phase space with Birkhoff coordinates. (For leaks inside the billiard, a 
representation in the full phase space is needed.)

There are also physically relevant types of leaks that go beyond the definition in
Eq.~(\ref{eq.leak1}). For instance, in room acoustics, or in the above mentioned 
dielectric cavities, it is very natural to consider
objects with {\em partial} reflection and partial absorption (or transmission). In this case 
we associate each 
particle with an intensity~$J$ that monotonically decays due to collisions at the leak. The dynamics of particles is given by the closed map~${\bf
  x}_{n+1}=f_{\text{closed}}({\bf x}_n)$, but the intensity of each particle will change as
\begin{equation} \label{eq.leak2}
J_{n+1}=\left \{ \begin{array}{ll} J_n & \mbox{ if } {\bf x}_n \notin I \\
                                              R({\bf x}_n)  J_n & \mbox{ if } {\bf x}_n \in I,\\
                                              \end{array}\right.
\end{equation}
where the {\it reflection coefficient}~$0\leq R<1$ might also depend on the phase-space position~{\bf
  x} within the leak. The full leak defined in Eq.~(\ref{eq.leak1}) is recovered by taking $R \equiv 0$ ($J_n \equiv J_0$) in
Eq.~(\ref{eq.leak2}).  
Altogether, a leak~$I$ is defined by its size, position, shape, and
reflectivity. 
In Sec.~\ref{sec.fullchaos} we show that all these different characteristics of the leak affect the observable quantities of interest.

\section{Theory for finite leaks}\label{sec.theory}\label{sec.2}

\subsection{Theory based on closed-system properties}\label{ssec.closed}

\begin{figure}[!b]
\includegraphics[width=1\columnwidth]{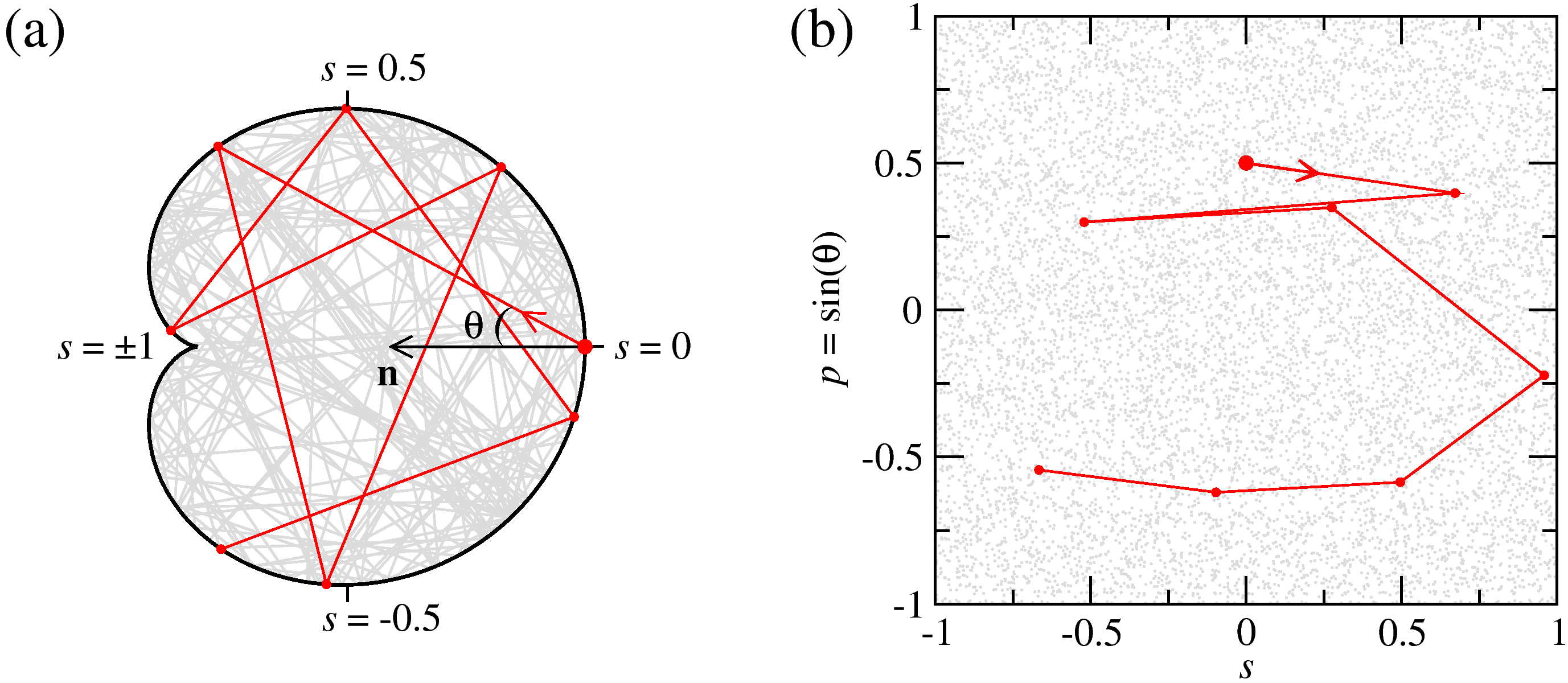}
\caption{(Color online) Dynamics in the closed cardioid billiard [Eq.~(\ref{eq.limacon}) with
  $\varepsilon=1$]. (a) Configuration space with parameterization of the perimeter~$s\in[-1,+1]$
  and collision angle~$\theta$. (b) Phase space depicted in Birkhoff
  coordinates~${\bf x}=(s,p=\sin(\theta))$ obtained at the collisions with the boundary. Two
  trajectories are shown in (a) and (b), one long (gray lines / black dots) and one short 
  (symbols with lines). 
}
\label{fig:closed}\label{fig.closed}
\end{figure}

The spirit behind Sabine's theory described in Sec.~\ref{ssec.sabine} is to calculate the
observable quantities of the open system based on the
properties of the closed system. While the results of this theory are exact only for
infinitesimally small leaks, it is natural to extend them 
to systems with finite leaks.  As already shown above,
the dynamics of two-dimensional  billiards can 
be conveniently represented by the true-time map
(\ref{mapf})-(\ref{eq.tn}).
Since for \limacon{} billiards both $s$ and~$p$ change in~$[-1,1]$, ${\bf f}$ preserves the measure~$d\mu=\frac{1}{4}\cos(\theta)d\theta ds$.
Figure \ref{fig:closed} illustrates how this map is applied in the case of the cardioid
billiard. Upon the $n$-th collision with the wall
the length $s_n$ along the perimeter is determined (measured from the point lying farthest
from the cusp) along with $p_n=\sin{\theta_n}$. Any trajectory in the configuration space 
(like the red and gray curves in 
Fig. \ref{fig:closed}a) is thus mapped on a sequence of points in discrete time in Fig. \ref{fig:closed}b, and the time is monitored via Eq.~(\ref{eq.tn}) in the
knowledge of $t_{\text{coll}}({\bf x})$ 
shown in Fig.~\ref{fig.tcoll3D}.

This billiard is strongly chaotic and the measure $\mu$
is the Lebesgue measure. 
This
means that the predictions of the theory based on the closed system are extremely
simple: trajectories are assumed  to follow the natural invariant density of the closed
system, $\rho_\mu({\bf x})=1/4$, i.e., they are uniformly distributed in ${\bf x}=(s,p)$.

We apply this theory to estimate the escape rate of a chaotic system with a finite
leak. The average collision time $\tcoll$ for the \emph{closed} system [Eq.~(\ref{eq.tcoll})] is independent of the leak
size, and Eq.~(\ref{eq.tau2d3d}) remains valid. The
escape rate resulting from this estimation will be denoted~$\kappa^*$. It again depends only on the size (measure) of the leak~$\mu(I)$,
but this time we do not assume~$\mu(I)$ to be small.
For instance, a leak $I=[s_l-\Delta s,s_l+\Delta s]\times [p_l-\Delta p,p_l+\Delta p]$ has size $2\Delta s$ along the $s$ axis, height 
 $2 \Delta p$ in $p$, 
area $4\Delta s \Delta p$ and a measure $\mu(I)=\Delta s \Delta p$. This is the measure of trajectories escaping on the
time scale $\tcoll$ of one collision. The survival probability after $n=t/\tcoll$
collisions can be estimated as
\begin{equation}\label{eq.Ptstar}
P(t)=\left(1-\mu(I)\right)^{t/\tcoll} = e^{-\kappa^* t},
\end{equation}
which yields a {\em naive estimate} for the escape rate
\begin{equation}\label{eq.kappastar}
\kappa^*=\frac{-\ln(1-\mu(I))}{\tcoll}.
\end{equation}
This can be considered a generalization of Sabine's law because it is a natural extension of
Eq.~(\ref{eq.mut}) to finite~$\mu(I)$. 
 Formula (\ref{eq.kappastar}) is usually attributed to Eyring's work in 1930 \cite{Mortessagne1993}, but
  see \cite{Joyce:1975} for a detailed historical account. 
The naive prediction
in Fig.~\ref{fig.robnik1} was determined with 
(\ref{eq.kappastar}) and still considerably differs from the measured decay.
It is important to note that while Sabine's theory is exact for infinitesimally small
leaks, Eq.~(\ref{eq.kappastar}) is just an approximation of the finite-size case.
Although it leads to an improved understanding of the problem of room acoustics,
it neglects the fact that the presence of a large leak essentially changes 
the dynamics because only a small portion of the closed system's orbits 
has sufficiently long lifetime to give a considerable contribution
to both the escape rate and the average collision time. 
A precise understanding of the dynamics in systems with finite leaks, including an explanation of
the behavior observed in Fig.~\ref{fig.robnik1}, can be given only if one abandons the
approach based on closed systems and adopts a description in terms of the theory
of transient chaos~\cite{LaiTel-book}.

\subsection{Theory based on transient chaos}\label{ssec.transient}

\begin{figure}[!t]
\includegraphics[width=1\columnwidth]{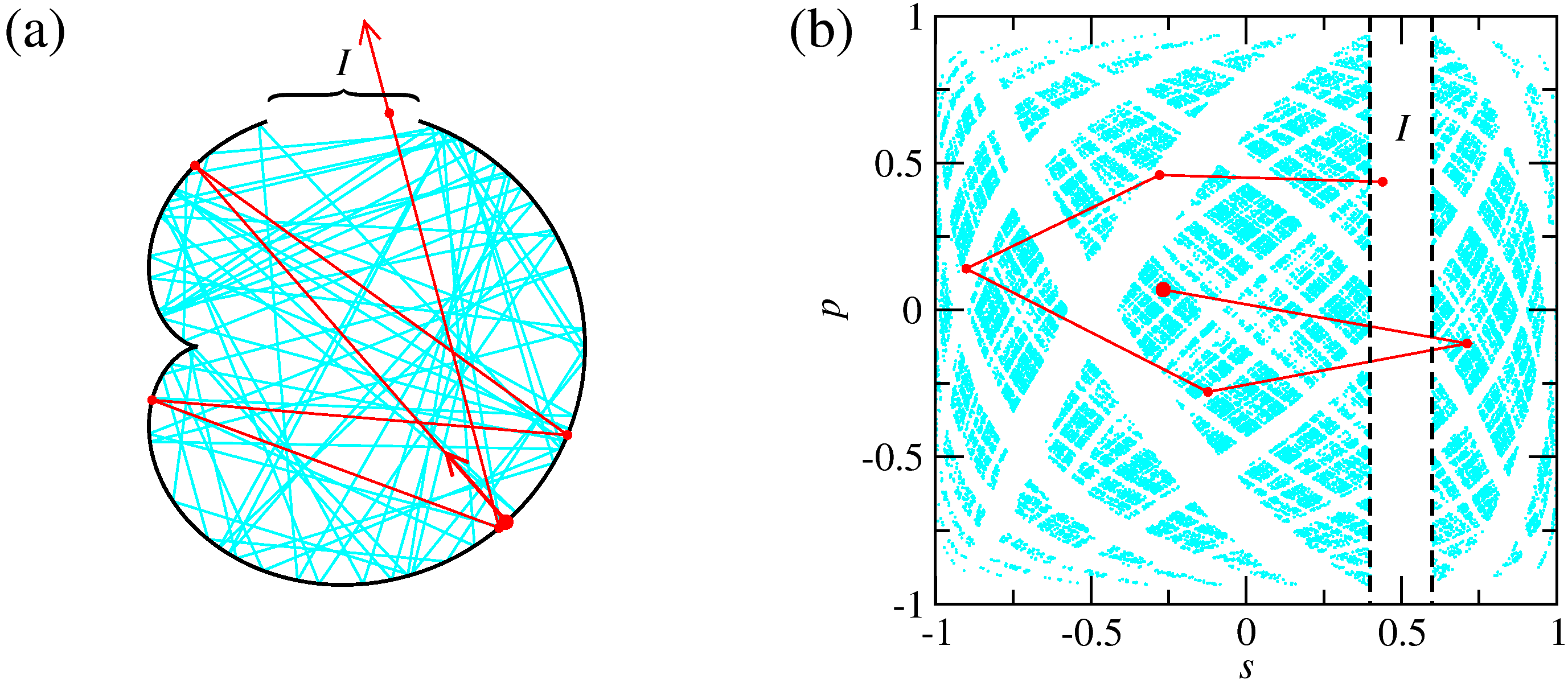}
\caption{ (Color online) Dynamics in a leaky cardioid billiard [Eq.~(\ref{eq.limacon}) with
  $\varepsilon=1$]. (a) Configuration space with the leak~$I$ centered around $s_l=0.5$
  with~$\Delta s= 0.1$ (in the momentum space $p_l=0$ and $\Delta     p=1$). One short lived (symbols with lines) and
  one long-lived (lines) orbit are shown. (b) Phase space of the true-time map with the chaotic saddle
  (dots) and the short lived trajectory (line). 
}
\label{fig:open}\label{fig.transient}
\end{figure}

The basic idea of transient chaos theory is to look at the invariant set of orbits that
never leave the system for both~$t\rightarrow\pm\infty$.
A key statement of the theory is that there is a nonattracting chaotic set in the
phase space that is responsible for the transiently chaotic dynamics~\cite{Ott-book,TelGruiz-book,Gas-book,LaiTel-book}. 
This set is a {\em chaotic saddle} and is of course drastically different from the chaotic set
of the closed system.   
To illustrate this difference we present in Fig. \ref{fig:open}
a leaky billiard, its chaotic saddle, and a short lived trajectory. 
It is apparent that the long-lived orbits are rather exceptional and the saddle
is very sparse: it is a measure zero object (with respect to Lebesgue measure), a set that exhibits
double fractal character. The difference between the closed system's Sabine-type
theories and the ones based on transient chaos can pictorially best be 
expressed by comparing Figs. \ref{fig:closed}b and \ref{fig:open}b.
It becomes evident that transient chaos is supported by a strongly selected 
and extremely ordered {\em subset} of the closed system's trajectories. Hence the measures ($\mu$ and $\mu_c$) with
which averages should be taken are fundamentally different in the two cases.

The saddle is responsible for the exponential decay of the survival probability
\begin{equation}
P(t) \sim e^{-\kappa t},
\label{kappa}
\end{equation}
where $\sim$ indicates an asymptotic equality in $t$. The escape rate~$\kappa$ is a property of the saddle and is independent of the initial distribution
of the trajectories used to represent an ensemble.

The invariant set of transient chaos is called a saddle because it 
possesses a stable and an unstable manifold. 
The stable (unstable) manifold
is composed of all trajectories that approach the chaotic saddle for $t\rightarrow\infty$
in the direct (inverted) dynamics. These manifolds of attracting and repelling character are 
extremely important to understand the properties of the open system. This is the
  reason why the term chaotic saddle is more appropriate than the often used term repeller (see, e.g., \cite{Gas-book}), which
  suggests (erroneously) that only
  unstable directions exist.

\begin{figure}[!t]
\includegraphics[width=\columnwidth]{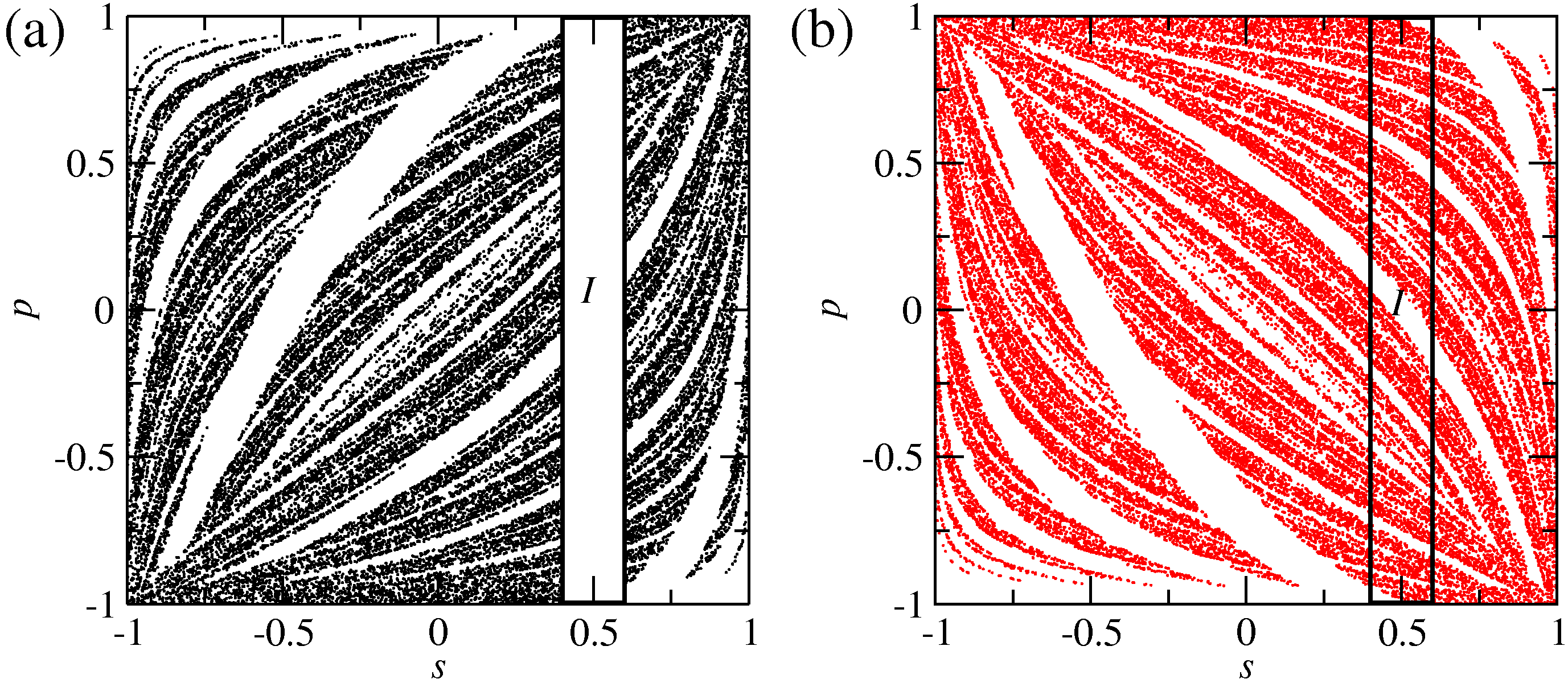}
\caption{(color online) (a) Stable and (b) unstable manifolds of the cardioid billiard shown in Fig.~\ref{fig.transient}, obtained by the
  sprinkler method ($N_0=10^8,t^*=120,\Delta t^*=40$, see  Appendix~\ref{ssec.appendixnew}). }
\label{fig:saddle}\label{fig.saddle}
\end{figure}

Here we present the so-called {\em sprinkler method}, that  
can be used to calculate not only a chaotic saddle but also its stable and unstable 
manifolds~\cite{TelGruiz-book,LaiTel-book}. One starts with $N_0 \gg 1$ trajectories distributed
uniformly over the phase space. 
One then chooses a time $t^*\gg 1/\kappa$ and follows the time
evolution of each initial point up to $t^*$. Only trajectories
that do not escape are kept, whose number is approximately
$N_0 e^{(-\kappa t^*)}$. If $\kappa t^*$ is sufficiently large (but not too large 
such that only a few points remain inside), trajectories with this long  
lifetime come close to the saddle in the course of dynamical evolution, 
implying that their initial points must be in the immediate vicinity of the
stable manifold of the saddle (or of the saddle itself), and their end points
be close to the unstable manifold of the saddle. The latter is so because 
most points still inside after time $t^*$ are about to
leave.
The {points from the middle} of these trajectories ($t \approx t^*/2$)
are then in the vicinity of the saddle. 
 In the spirit of true-time maps,
we used 
a generalization of this method (see Appendix~\ref{ssec.appendixnew})
to generate the chaotic saddle of Fig.~\ref{fig:open}b and the corresponding
manifolds plotted in Fig.~\ref{fig:saddle}a,b.

From the construction above it is clear that the particles being in the process of escape are distributed along the unstable
manifold. When compensating the loss due to escape by pumping in new particles according
to an appropriate way, which corresponds in practice to multiplying the density by $\exp{(\kappa t)}$, 
we obtain  
an invariant density as the one shown in Fig.~\ref{fig.saddle2}. This stationary distribution is known to be
the {\em conditionally invariant measure}. 

{ Traditionally,
a measure~$\mu_c$ is said to be conditionally invariant (c-measure for short) if for any subset $E$ of the region of interest $\Omega$
~\cite{PY:1979,Demers:2006}  
\begin{equation}\label{eq.conditional}
\frac{\mu_c({\bf f}^{-1}(E))}{\mu_c({\bf f}^{-1}(\Omega))}=\mu_c(E).
\end{equation}
This means that the c-measure is not directly invariant under the map
${\mu_c(f^{-1}(E))} \neq \mu_c(E)$, but it is preserved under the incorporation of the
compensation factor $\mu_c(f^{-1}(\Omega))<1$ (the c-measure of the set remaining in~$\Omega$ in one iteration). 

Although many c-measures might exist~\cite{Collet:2000,Demers:2006}, here
we are interested in the {\em natural} c-measure which is concentrated 
along the saddle's unstable manifold, 
and has been demonstrated to be relevant in 
several transient-chaos-related phenomena~\cite{LaiTel-book}.
Fig. \ref{fig.saddle2} shows the c-measure over the full phase space of the cardioid and indicates 
that the distribution is rather irregular. 
This should be compared 
with the smooth Lebesgue measure characterizing the closed 
system. 

\begin{figure}[!t]
\includegraphics[width=0.65\columnwidth]{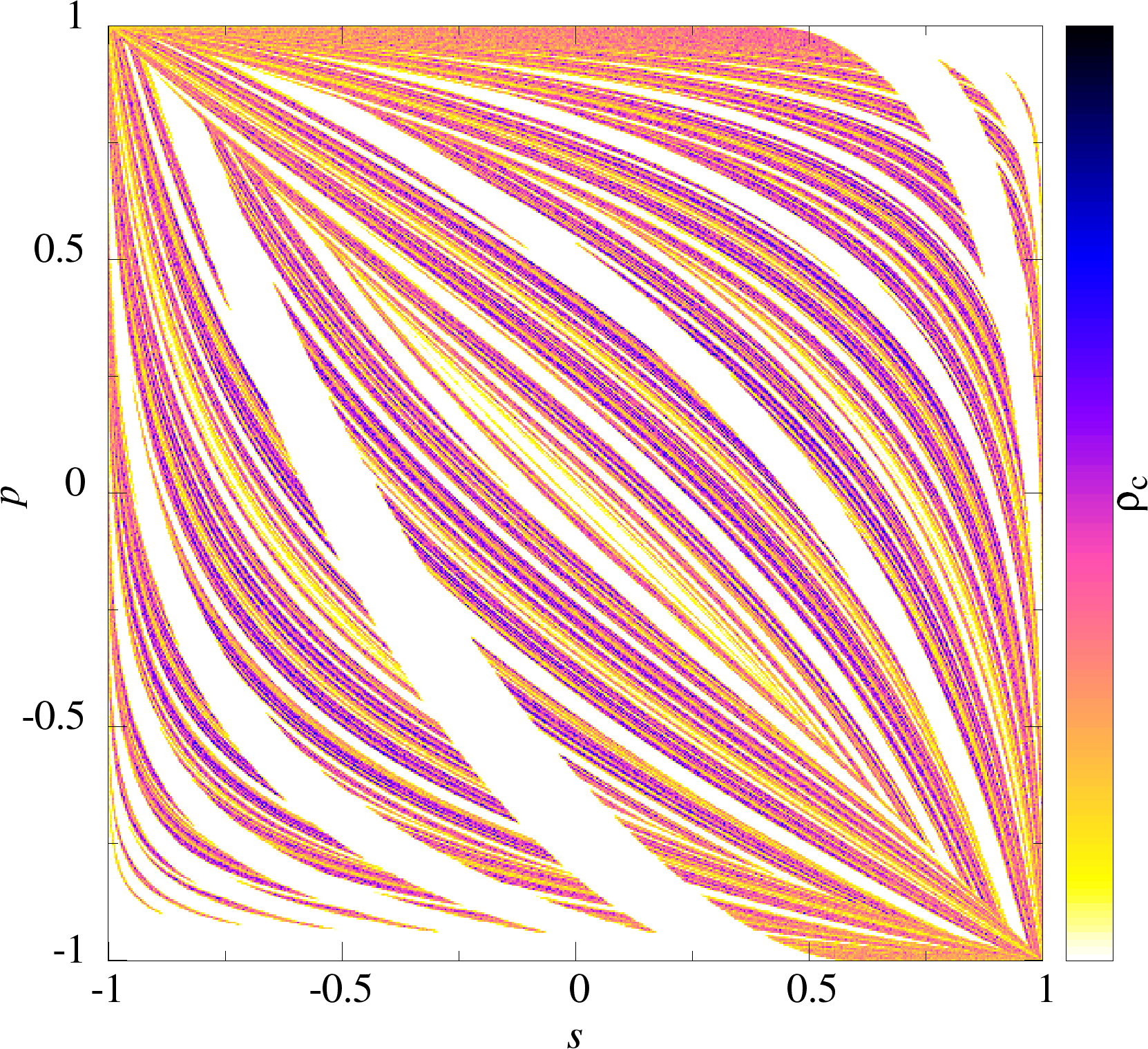}
\caption{(Color online) Density of trajectories on the unstable manifold shown in Fig.~\ref{fig.saddle}b. This distribution corresponds to  the c-measure,
  the measure according to which averages are to be taken in the transient chaos context. Note that the c-measure is defined within the
  leak.
 ($N_0=10^8,t^*=80,\Delta t^*=80$, see  Appendix~\ref{ssec.appendixnew}) } 
\label{fig:saddle2}\label{fig.saddle2}
\end{figure}
 


\noindent {\bf Dimensions of the invariant sets:} Both the 
chaotic saddle and its manifolds
are fractal sets, as can clearly be seen from Figs. \ref{fig:open}b and 
Fig. \ref{fig:saddle}a,b. Commonly, there are (at least) two different 
dimensions used to quantify the fractality of these sets, the box-counting
dimensions, $D_0$, and the information dimensions $D_1$.
The former characterizes the mere geometrical pattern, the latter
also the distribution of particles on the pattern \cite{Ott-book}.

The chaotic saddle of a two-dimensional map has a 
clear direct product structure: it
can locally be decomposed into two Cantor-set-like components, one along each manifold. The dimensions along the unstable (stable)  manifold is called the partial
dimension in the unstable (stable) direction and is marked by an upper index $1 (2)$.
None of the partial dimensions can be larger than one. 
The dimension $D_0$ and $D_1$ of the chaotic saddle is the sum of the two corresponding partial
dimensions~\cite{TelGruiz-book,LaiTel-book}:
\begin{equation}
D_0=D_0^{(1)}+D_0^{(2)}, \;\;\;\;\; D_1=D_1^{(1)}+D_1^{(2)}.
\label{sumD}
\end{equation}
The saddle might also contain very rarely visited, and thus atypical,  regions.
Consequently, the
information dimension $D_1$ 
cannot be larger than the box-counting dimension $D_0$,
which naturally holds for the partial dimensions, too: $D_1^{(j)} \le D_0^{(j)}$, $j=1,2$.
The value of the box-counting dimension is found often to be close to that of
the information dimension and it is then sufficient to use only one of them.  

The manifolds  are also of direct product structure, but 
one component of them is a line segment, an object of partial dimension $1$ (see Fig. \ref{fig:saddle}a,b).
Since the saddle can be considered as the
intersection of its own manifolds, the dimension $D_{0,1}^{(u(s))}$ of the unstable (stable)
manifold is $1$ plus the partial dimension in the stable (unstable) direction
\begin{equation}
D_{0,1}^{(u)}=1+D_{0,1}^{(2)}, \;\; \; D_{0,1}^{(s)}=1+D_{0,1}^{(1)}.
\label{sumDu}
\end{equation}    
It should be noted that the information dimensions $D_1^{(u)}$ of the unstable
manifold is nothing but the information dimension of the c-measure that sits on this
manifold. 

A simplifying feature occurs in Hamiltonian systems:
due to time reversal symmetry the partial dimensions coincide\footnote{
{This symmetry explains why Fig.~\ref{fig:saddle}a,b  are mirror images of each other with respect to the $s$ axis
(after the leak is removed from the unstable manifold). The closed cardioid has an additional symmetry $(s,p) \mapsto (-s,-p)$. }}.
Therefore in Hamiltonian cases, we have
\begin{equation}
D_{0,1}=2 D_{0,1}^{(1)}, \;\;\;\;  D_{0,1}^{(u)}=D_{0,1}^{(s)}=1+D_{0,1}^{(1)}=1+D_{0,1}/2, 
\label{DHam}
\end{equation}
i.e., all relevant dimensions can be 
expressed by the partial dimensions~$ D_{0,1}^{(1)}$ in the unstable direction.

It is an important result of transient chaos theory that this
partial information dimension can be expressed in a simple way by the escape rate and the
average continuous-time Lyapunov exponent  $\bar{\lambda}$ on the chaotic saddle:
\begin{equation} 
D_1^{(1)} = 1 - \frac{\kappa}{\bar{\lambda}}.
\label{KG1}
\end{equation}
This relation, the {Kantz-Grassberger relation}~\cite{KG:1985} 
 states that the dimension 
observed along the unstable direction deviates from $1$ the more,
the larger the ratio of the escape rate (a characteristics of the global 
instability of the saddle) to the average Lyapunov exponent (a characteristics
of the local instability on the saddle) is\footnote{Relation (\ref{KG1}) is valid in dissipative cases as well, but
the partial dimension along the stable manifold
is then  
$D_1^{(2)}=D_1^{(1)}{\bar{\lambda}}/{\vert \bar{\lambda}' \vert}$, 
where $\bar{\lambda}'$ is the average negative Lyapunov exponent 
on the saddle.}.


\noindent {\bf Implications for systems with leaks:} All results described so far are valid for open systems in general, they are not
particular to systems with leaks. In the latter case, however, the chaotic saddle and its manifolds depend
sensitively on the leak~$I$. 
This implies that the unstable manifold's dimension, and the c-measure in general, might strongly
depend on the size, location, and shape of the leak. 

We assume throughout this work that (a) $I$ is not too large so that trajectories do
not trivially escape after a short time 
($P(t)\neq0$, for any~$t$); and (b)  ergodicity and the 
chaotic properties of the closed system lead to {\it one and only one} chaotic saddle after the
leak is introduced. We also exclude the possibility of trajectories re-entering the billiard after hitting a leak. This assumption is naturally
satisfied in convex billiard, but should be enforced in the \limacon{} billiard for leaks around the cusp\footnote{Physically we can imagine that an
absorbing material is placed on the border of the closed billiard so that trajectories crossing the leak are immediately absorbed.}.

We can take advantage of the fact that a leaky system  is obtained from a strongly chaotic closed
system. For instance, it is possible to explicitly the chaotic saddle by extracting from the original
phase space all the ${\bf f}_{\text{closed}}$ images and preimages of the set used as leak $I$ in 
${\bf f}$. 
To see this, in Fig.~\ref{fig.iterleak} leak $I$ of Figs.~\ref{fig.transient} and \ref{fig.saddle} is shown together
with its forward and backward images. Compared to Fig.~\ref{fig.saddle},  we see
that the white regions in the plot of the stable (unstable) manifold correspond to the backward (forward) iterates of the leak.

Due to area preservation and
ergodicity of the closed system, the result of removing infinitely many images can only be a set of measure zero. 
The chaotic saddle in a leaking system
is the set of points that remain in the {\em complement} of the leak~$I$ and all its images for
both forward and backward iterations. {Indeed, the complement of the union of the sets in panels (a) and (b) of Fig.~\ref{fig.iterleak}
already provides a good approximation to Fig.~\ref{fig.saddle}(a) and (b), respectively, and the complement of both panels is a good
approximation to Fig.~\ref{fig.transient}(b).} The chaotic saddle is a fractal subset of the original chaotic set
(the full phase space in our strongly chaotic example). Furthermore, all invariant sets of
the leaky system (periodic orbits, manifolds of the saddle, etc.)  are  {\em subsets} of
those in the corresponding closed system.

\begin{figure}[!t]
\includegraphics[width=\columnwidth]{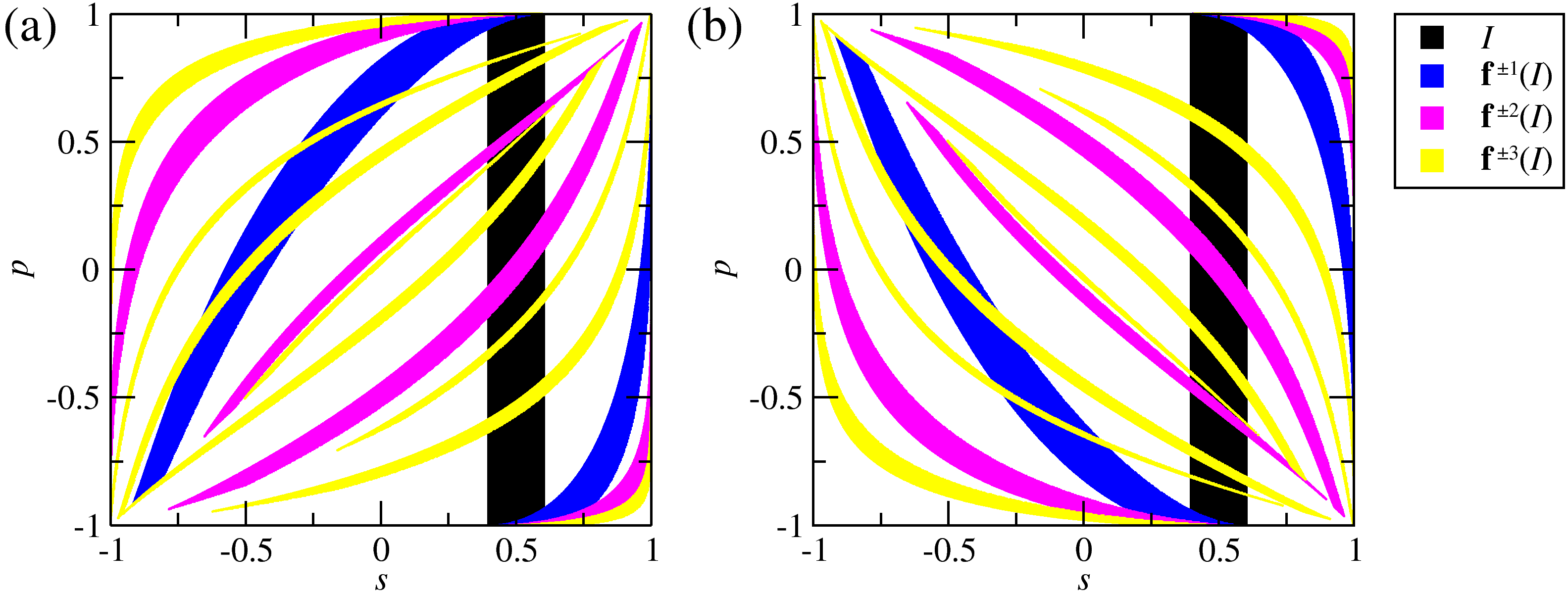}
\caption{(Color online) Phase space $(s,p)$ of the cardioid billiard showing the (closed system) iterates of the black vertical stripe
  that corresponds to $I$ in Figs.~\ref{fig.transient} and \ref{fig.saddle}). (a) Backward iterates. (b) Forward iterates.
  In both cases the first (dark gray/blue), second  (gray/magenta), and third (light gray/yellow) iterates are shown.}
\label{fig.iterleak}
\end{figure}


\noindent {\bf An improved escape rate formula:} We now notice that since escape in Eq.~(\ref{eq.leak1}) is considered to
occur one step {\em after} entering~$I$, the map  
${\bf f}$ with leak is defined in~$I$ and thus the unstable manifold of the chaotic saddle
enters~$I$ (see Fig.~\ref{fig.saddle}b). {It is also possible to compute measures of the leak, as indicated in Eq.~(\ref{eq.changemu}). 
The region of
 interest~$\Omega$ 
in Eq.~(\ref{eq.conditional}) is identified with the closed system's phase space, $\mu_c(\Omega)=1$, and the compensation factor in this
equation is obtained as
\begin{equation}\label{eq.conditional2}
\mu_c({\bf f}^{-1}(\Omega))=\mu_c(\Omega \backslash I)=1-\mu_c(I).
\end{equation}
}
We now recall that the c-measure is invariant and is distributed along the
unstable manifold. 
The escape rate can be estimated by the same simple calculations that lead to
Eq.~(\ref{eq.kappastar}) by replacing the natural measure~$\mu(I)$ by the c-measure $\mu_c(I)$ of the
leak. Carrying out also the averaging of the collision times 
with respect to the c-measure [see Eq.~(\ref{eq.changemu})], we find that
\begin{equation}\label{eq.kappa}
\kappa(I) \simeq -\frac{\ln(1-\mu_c(I))}{\tcoll_c},
\end{equation}
where subscript $c$ stands for the c-measure of the true-time map and $\simeq$ indicates approximate equality.
The validity and limitations of the improved formula~(\ref{eq.kappa}) will be carefully discussed in Sec.~\ref{ssec.escaperate},
but by now it is instructive to discuss the implications of this equation. It clearly shows that for finite leaks the escape rate cannot be
obtained from the properties of the closed system and the c-measure should be used, a measure which deviates essentially from that of the
closed system. The difference between $\mu(I)$ (the area of $I$) and $\mu_c(I)$ (the proportion of dots within $I$) is visually clear from Fig.~\ref{fig.saddle}b.

We now discuss the case of open maps, in contrast to true-time maps. Their escape
rate we denote by $\gamma$ (to sharply distinguish from the continuous-time or true-time map escape rate $\kappa$) implying that the discrete-time survival probability 
$P(n)$ decays as $e^{-\gamma n}$.    
The escape rate given by
\begin{equation}
e^{-\gamma(I)}=1-\mu_c(I) \;\;\; \rightarrow 
\gamma(I)=-\ln(1-\mu_c(I))
\label{gamma}
\end{equation}
has been known since~\cite{PY:1979} and can also be obtained directly from Eq.~(\ref{eq.conditional2}) for leaky maps~\cite{Paar2000,Altmann2008}.
It expresses the fact that the c-measure $\mu_c$ of the leak is the proportion of 
particles escaping via the leak within an iteration step. Since starting from the 
c-measure the decay is exponential from the very beginning, the proportion of the surviving particles after one time unit is $\exp{(-\gamma)}$, of those who escape is 
$1-\exp{(-\gamma)}$, and thus $\mu_c(I)=1-\exp{(-\gamma)}$ which is equivalent to (\ref{gamma}).

Equation~(\ref{eq.kappa}) leads to Eq.~(\ref{gamma}) when $t_{\text{coll}}({\bf x})\equiv1$. It shows also that when using the true time of
the system together with a surface of section at the billiard's boundary, it is essential to  
take into account that the average collision time differs from $\tcoll$ given in Eq.~(\ref{eq.tau2d3d}). As already
anticipated in Eq.~(\ref{eq.changemu}), with finite leaks the correct average
collision time is given by
%
$\tcoll_c=\int t_{\text{coll}}({\bf x}) \rho_c({\bf x}) d{\bf x},$
where $\rho_c$ is the density of the c-measure characterizing the 
system in the presence of leak $I$.
We note that different corrections for $\kappa$ due to the collision times were 
suggested by~\cite{Mortessagne1993,Ryu2006}. While~\cite{Mortessagne1993} uses a Gaussian approximation for the distribution of the collision times of long-lived trajectories, \cite{Ryu2006} took into account only the collision times inside the leak. None of them is equivalent 
to (\ref{eq.kappa}) or to the exact expressions in Sec.~\ref{ssec.escaperate}.  

Another general statement we make about systems with leaks is that when the size of the
leak goes to zero, the properties of the open system tend to those of the closed system\cite{deMoura1999,Aguirre2003},
i.e., the theory of Sec.~\ref{ssec.closed} becomes correct. In
particular $\mu_c(I)\rightarrow\mu(I)\rightarrow0$, which implies
that~$\kappa$ in Eq.~(\ref{eq.kappa}) tends to~$\kappa^*$ in Eq.~(\ref{eq.kappastar}) and both tend to Sabine's
prediction~$\kappa(I)=\mu(I)/\tcoll$, Eq.~(\ref{eq.mut}). In terms of dimensions, $D_{0,1}\rightarrow 2$. A nontrivial
  closed-system approximation~$D^{*(1)}_1$ of the information dimension can be obtained from the  Kantz-Grassberger relation~(\ref{KG1}) with
  $\kappa(I)=\kappa^*$, $\bar{\lambda}(I)=\bar{\lambda}$ (Lyapunov exponent of the closed system)~\cite{Neufeld2000}. 



\noindent {\bf Periodic orbits in maps with leaks:} We also review the dynamics in general maps with leaks. 
Periodic orbits analysis~\cite{Cvitanovic-book} is a powerful method to investigate chaotic systems and also
illustrates the spirit of leaking systems. Generally, a dense set of unstable 
periodic orbits is embedded into the chaotic saddle and this set can be used to obtain 
an expression for the escape rate
of the saddle. As pointed out by~\cite{Altmann2009}, in systems with leaks
this can be done either using the periodic orbits of the open system (that never hit the
leak) or using only the periodic orbits of the closed system that hit the leak. To illustrate
the simple arguments that lead to this, let us consider the case of computing the escape
rate~$\gamma$ of a generic discrete mapping~${\bf f}$. First we split the
set~$\Gamma_n^{(\text{all})}$ of all periodic orbits of length $n$ (i.e., all orbits that have an integer
period equal to $n,n/2,n/3,...$) of the closed system into two sets:
$\Gamma_n^{(\text{inside})}$, the orbits that have at least one point inside~$I$, and the
complementary set~$\Gamma_n^{(\text{outside})}$, i.e., all orbits for which all points are outside the leak $I$. In the limit of large~$n$ the following relation holds for
hyperbolic systems~\cite{Ott-book,Dorfman-book} 
\begin{equation}\label{eq.upo}
e^{-n\gamma}=\sum_i \frac{1}{ |\Lambda (\Gamma^{(\text{outside})}_{i,n})|},
\end{equation}
where the sum is over all points~$i$ of periodic trajectories 
in~$\Gamma_n^{(\text{outside})}$, and~$\Lambda$ is
the largest eigenvalue of the n-fold iterated map ${\bf f}^n$ on the orbit. 

Next, we notice that in the closed system~$\gamma=0$ (no escape). Therefore,
\begin{equation}\label{eq.upo2}
1=\sum_i \frac{1}{|\Lambda (\Gamma^{(\text{all})}_{i,n})|},
\end{equation}
where the sum is over all points~$i$ in $\Gamma^{(all)}_n$. Subtracting~(\ref{eq.upo})
from~(\ref{eq.upo2}) we obtain
\begin{equation}\label{eq.upo3}
1-e^{-n\gamma}=\sum_i \frac{1}{|\Lambda (\Gamma^{(\text{inside})}_{i,n})|},
\end{equation}
where the sum now is over all points~$i$ in $\Gamma_n^{(\text{inside})}$, i.e., all points
that belong to periodic orbits that have at least one point in~$I$. Altogether this means
that even if we are allowed to probe the system only through the leak, the identification
of the periodic orbits entering~$I$ 
suffices for the computation of the main properties of the chaotic
saddle that exists inside the system. This can be applied also to more efficient methods based on expansions of the zeta function~\cite{Cvitanovic:zeta}.

A simple relation can be obtained in uniformly expanding (piecewise linear) maps in which leaks are selected as elements of a Markov partition.
In this case it is possible to prove that for two different leaks $I_1$ and $I_2$ the relation $\gamma(I_1)>\gamma(I_2)$ 
holds if and only if the shortest periodic orbit in $I_1$ is shorter than the one in $I_2$~\cite{Bunimovich:2012}.
This follows also from Eq.~(\ref{eq.upo3}) with constant $\Lambda$ (as in piecewise linear maps with constant slope).


\subsection{Initial conditions and average escape times} \label{subsec:ld_leak_Poincare}\label{ssec.poincare}

Typical observable quantities in transient chaos theory, such as~$\kappa$ and the
fractal dimensions, are independent of the choice of the density
of initial conditions~$\rho_0({\bf x})$ because they are directly related to the 
properties of the invariant chaotic saddle. More precisely, the
underlying assumption is that 
$\rho_0({\bf x})$ overlaps the stable manifold of this saddle. The stable manifold of the chaotic saddle typically 
spreads through the phase space in a filamentary pattern (e.g., as in Fig.~\ref{fig.saddle}a) and therefore smooth 
$\rho_0({\bf x})$'s will typically fulfill this requirement.  Even in this typical case, there are important quantities that {\em do depend} on
$\rho_0({\bf 
    x})$ such as any quantity averaged over a large number $N$ of trajectories. The dependence on initial conditions and the universal
asymptotic decay of the survival probability $P(t) \sim
  e^{-\kappa t}$ can be reconciled by noticing that for short times, $t<t_s$, the escape of trajectories is non-universal, and $P(t) \neq
  e^{-\kappa t}$. Even if $t_s$ is short, a large fraction of the trajectories may escape for $t<t_s$.

Here we discuss in more detail the simplest yet representative case of the average  lifetime~\cite{Altmann2009}
\begin{equation}\label{eq.meant}
\langle \tau \rangle_{\rho_0} \equiv \lim_{N\rightarrow \infty} \frac{1}{N} \sum_{i=1}^{N} \tau_i= \int_0^\infty \tau p(\tau) d\tau = \int_0^\infty P(t') dt',
\end{equation}
obtained with different initial densities $\rho_0({\bf x})$, and hence with different survival probabilities $P(t)$,
where 
$\tau_i$ is the lifetime of trajectory $i$. 
Here
we used $p=-dP/dt$  [see Eq.~(\ref{eq.te})], $P(0)=1$, and $P(t)\rightarrow 0$ faster than~$1/t$. 

For maps, the averaged discrete lifetime is
\begin{equation}
\langle \nu \rangle_{\rho_0} = \sum_{n'=0}^\infty n' p(n')
\end{equation}
where $p(n)$ is the probability to escape in the $n$th step.
Notice that for true-time maps ((\ref{mapf}), (\ref{eq.tn})), in general, $\langle \tau \rangle_{\rho_0} \neq \langle \nu  \rangle_{\rho_0}
\tcoll_c$, with $\tcoll_c$ given by Eq.~(\ref{eq.changemu}).  Instead, 
$$\begin{array}{rl}
\langle \tau \rangle_{\rho_0}
&= \lim_{N\to\infty}\dfrac{1}{N} \sum_{i=1}^{N} \sum_{j=1}^{\nu_i} t_{\text{coll}}({\bf x}^{(i,j)})\\
&= \lim_{N\to\infty}\dfrac{1}{N} \sum_{i=1}^{N} \nu_i\bar{t}_{\text{coll}}^{(i)}\\
&= \big\langle \nu \bar{t}_{\text{coll}} \big\rangle_{\rho_0},
\end{array}$$
where $\nu_i$ is the total number of collisions along the $i$-th trajectory,
${\bf x}^{(i,j)}$ is the position of the $j$-th collision ($j=1\ldots
\nu_i$) of trajectory $i$ that has initial condition ${\bf x}^{(i,0)}$, $\bar{t}_{\text{coll}}^{(i)}\equiv\frac{1}{\nu_{i}} \sum_{j=1}^{\nu_i} t_{\text{coll}}({\bf x}^{(i,j)})$ is the mean collision time of
trajectory $i$, and the average $\langle \ldots \rangle$ is taken over $i=1\ldots N$ trajectories. The reason for this difference is that for short
times $\bar{t}_{\text{coll}}^{(i)}$ differs significantly from $\tcoll_c$.

We would like to see if $\langle \tau \rangle_{\rho_0}$ and $\langle \nu \rangle_{\rho_0}$ can be expressed as a function of other easily measurable quantities. We also try to find a relation between $\langle \tau \rangle_{\rho_0}$ and $\langle \nu \rangle_{\rho_0}$ for the following particular 
initial densities $\rho_0({\bf x})$:

\subsubsection{Conditionally invariant density: $\rho_c$} We take initial conditions according to the c-measure $\rho_0({\bf
    x})=\rho_c({\bf x})$. As explained in Sec.~\ref{ssec.transient}, $\rho_c({\bf x})$ describes the escaping process and is achieved by
  rescaling the surviving trajectories from an arbitrary smooth initial density. Therefore, for $\rho_0({\bf x})=\rho_c({\bf x})$ we find $p(t)=\kappa e^{-\kappa t}$
  for all $t>0$ and from Eqs.~(\ref{eq.meant}) and~(\ref{eq.kappa}) the simple relation:
\begin{equation}\label{eq.tau-rhoc}
\langle \tau \rangle_c =  \dfrac{1}{\kappa} \simeq -\frac{\tcoll_c}{\ln(1-\mu_c(I))}
\end{equation}
follows.

For maps with escape rate $\gamma$, the normalization of $p(n)$ implies
$\sum_{n=1}^\infty p(n)=1$, and thus 
$p(n)=(e^\gamma-1) e^{-\gamma n}$ (since $e^{\gamma}-1 \approx \gamma$ for $\gamma\ll1$).
This leads to a different result~\cite{Altmann2009} 
\begin{equation}\label{eq.nu-rhoc}
\langle \nu \rangle_c = \dfrac{1}{1-e^{-\gamma}} = \dfrac{1}{\mu_c(I)}.
\end{equation}
In the last equality we used Eq.~(\ref{gamma}).
It is important to note that for maps obtained from flows, the c-densities~$\rho_c$ of the map and flow (or true-time map) are usually different due
to the nontrivial collision time distribution.

\begin{figure}[!b]
\includegraphics[width=1\columnwidth]{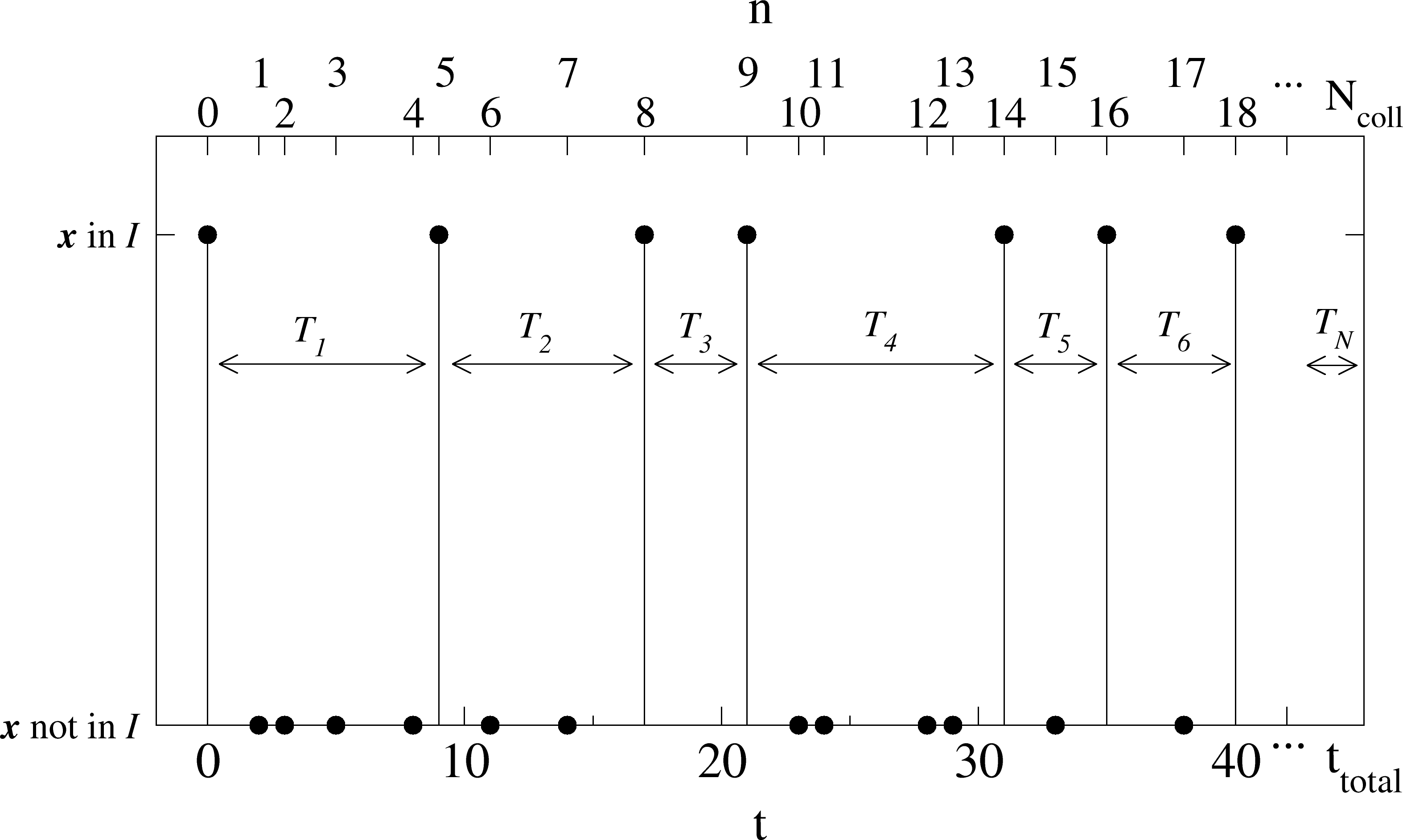}
\caption{Schematic illustration of Poincar\'e recurrences in a closed billiard. 
The symbols indicate the times~$t$ of the $n$-th collision with the boundary of a single trajectory. The recurrence times $T_i$'s are defined as the times between successive collisions in~$I$. In the total time
   $t_{\text{total}}$ there are $N_{\text{coll}}$ collisions, out of which $N$ collisions are inside the region~$I$.}
\label{fig.poincare}
\end{figure}

\begin{table*}
\begin{center}
\begin{tabular}{|l|c c c|c|}
\hline
{}
&{}
&{Large leaks}
&{}
&{Limit of small leaks}
\\
{}
&{}
&{Finite $\mu(I)\neq\mu_c(I)$}
&{}
&{$\mu_c(I)=\mu(I) \rightarrow 0$}
\\\hline
{$\rho_0({\bf x})$}
&{ c-measure: $\rho_c$}
&{ recurrence: $\rho_r$}
&{ natural, smooth: $\rho_{\mu,s}$}
&{$\rho_{r,c,\mu,s}$}
\\\hline
{\bf Mean time}
&{}
&{}
&{}
&{}
\\
{continuous, t}
&{$\langle \tau \rangle_c = \dfrac{1}{\kappa} \simeq \dfrac{-\tcoll_c}{\ln(1-\mu_c(I))}$}
&{$\langle \tau \rangle_r = \dfrac{\tcoll}{\mu(I)}=\langle \nu \rangle_r \tcoll\neq \dfrac{1}{\kappa}$}
&{$\langle \tau \rangle_{\mu,s} \approx \dfrac{1}{\kappa}$}
&{$\langle \tau \rangle = \dfrac{\tcoll}{\mu(I)}$}
\\
{discrete, n}
&{$\langle \nu \rangle_c=\dfrac{1}{1-e^{-\gamma}}=\dfrac{1}{\mu_c(I)}$}
&{$\langle \nu \rangle_r=\dfrac{1}{\mu(I)}$} 
&{$\langle \nu \rangle_{\mu,s} \approx\dfrac{1}{\mu_c(I)}$}
&{$\langle \nu \rangle =\dfrac{1}{\mu(I)}=\dfrac{1}{\gamma}$}
\\\hline
{\bf Escape rate}
&{}
&{}
&{}
&{}
\\
{continuous, t}
&\multicolumn{3}{|c|}{$\kappa \simeq -\dfrac{\ln(1-\mu_c(I))}{\tcoll_c} 
 \neq \dfrac{\ln(1-\mu(I))}{\tcoll} = \kappa^*$}
&{$\kappa=\dfrac{\mu(I)}{\tcoll}=\dfrac{1}{\langle \tau \rangle}$}
\\
{discrete, n}
&\multicolumn{3}{|c|}{$\gamma=-\ln{(1-\mu_c(I))}\neq -\ln(1-\mu(I)) \equiv \gamma^*$}
&{$\gamma=\mu(I)=\dfrac{1}{\langle \nu \rangle}$}
\\\hline
\end{tabular}
\caption{ Dependence of the average lifetime $\langle \tau \rangle$ and $\langle \nu \rangle$ for flows and maps, respectively,
on the initial distribution 
$\rho_0({\bf x})$, and expressions for the escape
  rate~$\kappa, \gamma$. The natural invariant measure of the leak (closed system)~$\mu(I)$ and the conditionally invariant
  measure~$\mu_c(I)$ of the leak~$I$ coincide only in the limit of small leaks. From~\cite{Altmann2009}}\label{tab.ic} 
\end{center}
\end{table*}

\subsubsection{Recurrence density:~$\rho_r$}
As pointed  by~\cite{Altmann2008,Altmann2009}, there is an initial density~$\rho_0({\bf x})=\rho_r({\bf x})$ connected to the
 problem of {\it Poincar\'e recurrences} that leads to simple results for~$\langle \tau \rangle_r$ and $\langle \nu \rangle_r$. 
The Poincar\'e recurrence theorem asserts that in a closed
dynamical system with an invariant measure~$\mu$, almost any trajectory (with respect to $\mu$) chosen inside a
region~$I$ with~$\mu(I)>0$ will return to $I$ an infinite number of times. The times~$T_i$'s between two consecutive returns are called
Poincar\'e recurrence times, a central concept in dynamical-systems 
theory~\cite{Haydn:2005}. The (cumulative) distribution of recurrence times~$P_r(T)$ is also used to quantify chaotic properties of specific
systems~\cite{Chirikov1984}. Figure~\ref{fig.poincare} illustrates the Poincar\'e recurrence set-up in billiard systems. Using the notation
introduced in Fig.~\ref{fig.poincare}, the average recurrence time $\bar{T}$ of a single long trajectory is calculated as 
\begin{equation}\label{eq.kac}
\bar{T} = \frac{1}{N} \sum_{i=1}^N T_i = \frac{t_{\text{total}}}{N}=\frac{N_{\text{coll}}\tcoll}{N} = \frac{\tcoll}{\mu(I)},
\end{equation}
where $T_i$ is the $i$th recurrence time along the trajectory, $\tcoll$ is the average collision time of a typical trajectory starting
inside the leak that, due to ergodicity, coincides with the closed system~$\tcoll$ given by
Eq.~(\ref{eq.tcoll}), $N_{\text{coll}}$ is the total number of collisions up to time $t_{\text{total}}$, $N$ is the number of such collisions inside $I$, and
$\mu(I)=N/N_{\text{coll}}$ is the fraction of collisions on~$I$.  Equation~(\ref{eq.kac}) is valid for large $t_{\text{total}},N,N_{\text{coll}}$.

For maps, analogously, we obtain that the average discrete recurrence time $\bar{\mathcal{N}}$ is given by
\begin{equation}\label{eq.kac2}
\bar{\mathcal{N}} = \frac{1}{\mu(I)}.
\end{equation}
The inverse relation between measure and recurrence time shown in Eqs.~(\ref{eq.kac}) and~(\ref{eq.kac2}) is known as Kac's lemma~\cite{kac-book,Zaslavsky:2002}. For higher moments of the return time distribution see \cite{Cristadoro:2012}.

The connection to systems with leaks is achieved by {\it identifying} the recurrence region and the leak~$I$. One can find
an appropriate initial density~$\rho_0({\bf x})=\rho_r({\bf x})$ for the open case for which the survival probability in the presence of
leak $I$ coincides with the recurrence time distribution to $I$ in the closed system
\begin{equation}\label{PP}
P_r(T)=P(t)
\end{equation} 
for any $t=T$. 
This can be done by using the positions ${\bf x} \in I$
of the $N$ recurrent points  as initial conditions (see Fig.~\ref{fig.poincare}). Because of the ergodicity of the closed chaotic system, in the
limit of long times, the points of this single 
trajectory are distributed according to the natural density~$\rho_\mu({\bf x})$ of the closed system, justifying the notation $\mu(I)$ 
in the equation above.
In the case of the billiard systems discussed here, $\rho_r({\bf x})$
corresponds to initial conditions in the leak, uniformly distributed in ${\bf x}=(s,p)$, but with velocities pointing inward, i.e., not escaping through the leak.
Physically, this corresponds to shooting trajectories into the billiard  through the leak. 

If time is counted discretely, 
$\rho_r({\bf x})$ corresponds to the next iteration of the uniform distribution
(natural measure of closed system $\rho_\mu({\bf x})$) of initial conditions in ${\bf x} \in I$. In the general case of an invertible
map ${\bf f}$ this is obtained by applying the 
Perron-Frobenius~\cite{Dorfman-book,Gas-book} operator as~\cite{Altmann2009}:
\begin{equation}\label{eq.rhor}
\rho_r({\bf x}) = \frac{\rho_\mu({\bf f}^{-1}({\bf x})\cap I)}{J({\bf f}^{-1}({\bf x})\cap I)\mu(I)} \;\; \text{ for } \;\; {\bf x}\in f(I),
\end{equation}
where~${\bf f}^{-1}(x)\cap I$ denotes the points that come from $I$, $J$ is the Jacobian of the
map, and the factor $\mu(I)$ ensures normalization. Note that the single iteration introduced in the definition of~$\rho_r$ is {\em compensated} at
the end because the escape is considered also one time step after entering~$I$, see Eq.~(\ref{eq.leak1}). With $\rho_0({\bf x})=\rho_r({\bf x})$, $P(t)=P_r(T=t)$ for all
$t\ge0$, showing that the problem of 
Poincar\'e recurrence can be interpreted as a specific problem of systems with leaks. In particular, $P_r(T) \sim e^{-\kappa T}$ with the escape rate
$\kappa$ of the system opened up in $I$. The average lifetime is given by Eq.~(\ref{eq.kac}) as
$$\langle \tau \rangle_r=\bar T = \frac{\tcoll}{\mu(I)} 
\neq \frac{\tcoll_c}{\mu_c(I)} \neq -\frac{\tcoll_c}{\ln(1-\mu_c(I))}.$$
For maps, Eq.~(\ref{eq.kac2}) implies that
$$ \langle \nu \rangle_r = \bar{\mathcal{N}} = \frac{1}{\mu(I)} \neq \frac{1}{\mu_c(I)}.$$  
Note that 
both relations above reveal that
for~$\rho_r$ there is no difference in the dependence on the measure for maps and flows,
in contrast to  Eqs.~(\ref{eq.tau-rhoc}) and~(\ref{eq.nu-rhoc}) obtained with $\rho_c$
as the initial density.

\subsubsection{Closed-system density: $\rho_\mu$, or any smooth  $\rho_s$} The most popular initial condition for systems with leaks is by far $\rho_\mu$. This
corresponds to introducing  the leak {\em after} the trajectories have reached equilibrium inside the system. While this is sometimes implicitly
assumed as natural, the results of this section show that~$\rho_\mu$ is not the only possibility, and certainly not the simplest one. For
$\rho_\mu$ and any other smooth initial distribution~$\rho_0=\rho_s$, the surviving trajectories relax towards the c-density 
$\rho_c({\bf x})$, the natural density of the open system. 
The numerical investigations of~\cite{Altmann2009} indicate that results with $\rho_\mu$ or other smooth densities~$\rho_s$ are usually similar to those obtained by
$\rho_c$, and more different from those obtained by $\rho_r$. This can be understood by noting that the hyperbolicity of the chaotic saddle
leads to a fast convergence  $\rho_0({\bf x}) \rightarrow \rho_c({\bf x})$. While a convergence also holds for $\rho_r$, it is made much slower
by $\rho_r$'s lack of any 
overlap with the chaotic saddle, with orbits taking, therefore, a 
longer time to approach the saddle~\cite{Altmann2009}. 


\begin{table}
\begin{center}
\begin{tabular}{|l|c c c|}
\hline
{$\rho_0({\bf x})$}
&{ $\rho_c$}
&{ $\rho_r$}
&{ $\rho_{\mu,s}$}
\\\hline
{\bf Mean time}
&{}
&{}
&{}
\\
{continuous,t}
&{$\langle \tau \rangle_c =15.24$}
&{$\langle \tau \rangle_r =18.50$}
&{$\langle \tau \rangle_{\mu,s} = 14.23$}
\\
{discrete,n}
&{$\langle \nu \rangle_c=8.28$}
&{$\langle \nu \rangle_r=10.0$}
&{$\langle \nu \rangle_{\mu,s} =7.78$}
\\\hline
{\bf Escape rate}
&{}
&{}
&{}
\\
{continuous,t}
&\multicolumn{3}{|c|}{$\kappa=0.06559\neq\kappa^*=0.05693$}
\\
{discrete,n}
&\multicolumn{3}{|c|}{$\gamma_{P\text{map}}= 0.1286 \ne \gamma^*=0.1054$}
\\\hline
\end{tabular}
\caption{ Numerical results for the average lifetime in the cardioid billiard with a leak $s_l=0.5$, $\Delta s=0.1$ (as in
  Figs.~\ref{fig:open},~\ref{fig:saddle},~\ref{fig:saddle2},~\ref{fig.convergence},~\ref{fig:app}, and~\ref{fig:rhoMap}). Other data: $\tcoll=1.85055$, $\mu(I)=0.1$, $\mu_c(I)=0.1175$, $\tcoll_c=1.916$. In oder to illustrate the
  case of maps, instead of the true-time map, exclusively for this
   simulation we have used the Poincar\'e map of the billiard. 
   In order to minimize the effect of 
  sliding orbits (see Sec.~\ref{ssec.diff}) we used in all simulations 
  the following restrictions: a cut-off in the maximum collision time at $83$
  collisions ($t=158$ in Fig.~\ref{fig:app}),  
and $\rho_s$ is taken to be constant in $s\in[-1,1]$, $p\in[-0.9,0.9]$. $\rho_c$ was built by iterating $\rho_s$. 
For $\gamma_{P\text{map}}$ we used a more restrictive cut-off, $2\langle \tau
\rangle_{\mu,s}$, because it is more sensitive to the sliding orbits.
 The quantities $\mu_c(I)$ and $\tcoll_c$ were calculated using Eq.~(\ref{avgBQ})}\label{tab.ic.num} 
\end{center} 
\end{table}


The results of this section are summarized in Tab.~\ref{tab.ic}. 
All formulae here 
were confirmed within a $3\%$ error in the cardioid billiard with a finite leak (Tab.~\ref{tab.ic.num}).
In the limit of small leak,
the simple case~$\langle \tau \rangle = 1/\kappa$ discussed in Eq.~(\ref{eq.te}), and Sabine's law in the form of (\ref{eq.mut}) are recovered, independent of the initial distribution. 

\subsection{Extension to partial leaks} \label{ssec.partial}

\begin{figure*}[!ht]
\includegraphics[width=1.9\columnwidth]{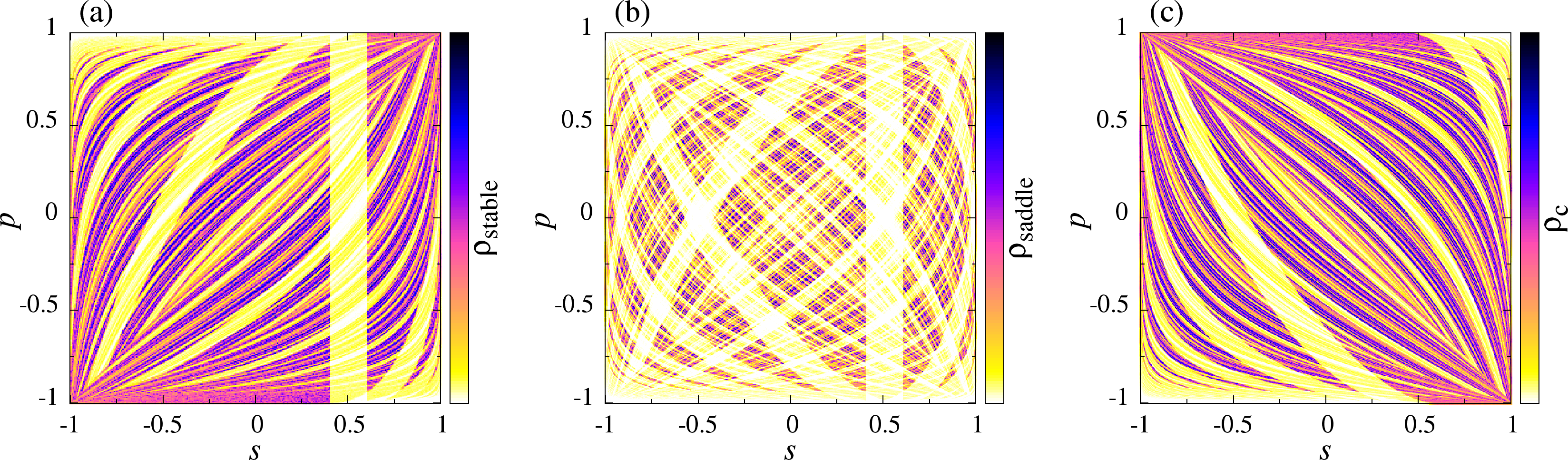}
\caption{(Color online) Invariant densities 
for the cardioid billiard considered in Figs.~\ref{fig.transient}, \ref{fig.saddle}, and
  \ref{fig.saddle2} with partial leak $R=0.1$ for ${\bf x} \in I$. The different panels show 
  the
  densities associated to the (a) stable manifold, (b) chaotic saddle, and (c) unstable manifold. The densities were obtained considering
  the intensity weighted trajectories at times $1/\tilde{\kappa}^* < 60<t<80$ and plotting at (a) their initial conditions, (b) their
  positions around the time $t/2$, and (c) their positions over the time interval $60<t<80$. See Appendix~\ref{ssec.appendixnew} for details.} 
\label{fig:partial}\label{fig.partial}
\end{figure*}

So far we have restricted our theory to the case of full leaks that completely transmit the trajectories falling on them.  At the end of
Sec.~\ref{ssec.leak} we emphasized the need to extend the theory to leaks that partially transmit and partially reflect
trajectories. 
This problem can be handled by associating to each trajectory~$i$ an intensity~$J^{(i)}_n$ that decreases
with time $J^{(i)}_{n+1}= R({\bf x}) J^{(i)}_n$ due to collisions at the leak regions with reflectivity~$R({\bf x})<1$, as described by (\ref{eq.leak2}).
 Here we show how the results
of Sec.~\ref{ssec.transient} can be extended to this case.

For simplicity, we consider that all $N$ trajectories of the initial ensemble start with intensity $J_0=1$. Non-homogeneous energy intensities  
can be achieved by manipulating the density of initial trajectories~$\rho_0({\bf x})$, as discussed in the previous section. For longer
times the energy density depends not only on the density of trajectories $\rho({\bf x},t)$ but also on their intensities, which can be thought as weights~$0\le J \le 1$
attributed to each trajectory.  The survival probability $P(t)$ of trajectories is physically reinterpreted as the fraction $\tilde{P}(t)$ of the total
energy still inside the system. 
In strongly chaotic systems, $\tilde{P}(t)$ is expected to decay asymptotically as
$$ P(t) \sim e^{-\tilde{\kappa} t},$$
with $\tilde{\kappa}$ as the {\it energy escape rate.}
In practice, $\tilde{P}(t)$ can be obtained as a sum over all $N$ trajectories
\begin{equation}\label{eq.ptilde}
\begin{array}{ll}
\tilde{P}(t) & \approx \frac{1}{N} \sum_{i=1}^N J^{(i)}_t \\ \\
 & =\int_\Omega J_t({\bf x}) \rho({\bf x},t) d{\bf x} \equiv \int_\Omega \tilde{\rho}({\bf x},t) d{\bf x},
\end{array}
\end{equation}
where $\rho({\bf x},t)$ is the density and $J_t({\bf x})$ is the 
intensity of trajectories
in~${\bf x}$ at time~$t$. The intensity of each trajectory $i$ at time $t$ is given by $J^{(i)}_t=\Pi_{j=1}^{\nu_i(t)} R({\bf x}^{(i)}_j)$,
with $\nu_i(t)$ the number of collisions of trajectory $i$ until time $t$ and ${\bf x}^{(i)}_j$ the position of these collisions. We define the combined density 
\begin{equation}\label{eq.rhotilde}
\tilde{\rho}\equiv \rho J,
\end{equation}
which is,  by convention, not normalized for $t>0$.
The appearance of $\tilde{\rho}$ in Eq.~(\ref{eq.ptilde})  reflects the more general fact that the division between trajectories and intensities has no observable consequence, despite
their natural interpretation in physical terms and their natural implementation in ray simulations. It is only their combination,
Eq.~(\ref{eq.rhotilde}), that leads to a physically relevant 
density for the case of partial leaks, both for the c-measure and for estimations based on $\mu$. A mathematical description coherent with
this interpretation considers operators acting on densities, and is given in the next section. In the remaining of this section we consider
how partial leaks modify the closed system (Sec.~\ref{ssec.closed}) and transient chaos (Sec.~\ref{ssec.transient}) theories discussed above.

We start with a straightforward  extension of the naive estimate of the escape rate.
When estimating the transmission happening through the leak one has
to account for the partial transmission because for a trajectory in~${\bf x}\in I$ only a fraction $1-R({\bf x})$ of its intensity is lost.
Therefore, the measure of the leak ($\mu(I)=\int_I d\mu$) used in 
Eq.~(\ref{eq.kappastar}) to compute the naive estimate~$\kappa=\kappa^*$
has to be replaced by 
\begin{equation}\label{eq.mupartial}
\mathfrak{m}(I)=\int_I (1-R({\bf x})) d\mu =  \int_I (1-R({s, \theta})) \frac{1}{4}\cos \theta d\theta ds,
\end{equation}
where $R({\bf x})$ is the (position dependent) reflection coefficient. 
The argument of Sec.~\ref{ssec.closed} leads to the naive estimate
\begin{equation}
\tilde{\kappa}^* = \frac{-\ln(1-\mathfrak{m}(I))}{\tcoll},
\label{eq.naiveR}
\end{equation}

see also~\cite{Joyce:1975,Ryu2006}.
This shows that the closed-system
estimation for a leak with $R=0.5$ is the same as the one for a leak of half the size, which was tested in Fig.~\ref{fig.robnik1} (and
proved to be inaccurate).

The essential extension of the transient chaos theory is to consider a (modified) c-density  $\tilde{\rho}_c$ which
should be normalized. As~$\rho_c$, $\tilde{\rho}_c$ can also be achieved
by rescaling. Here the {\em intensity} should be multiplied by a  
factor~$e^{\tilde{\kappa} t}$ to compensate the global decay of $\tilde{P}(t)$. This rescaling implies that the c-measure
$\tilde{\mu}_c(E)$ of a region~$E\in\Omega$ in a partially open system is the fraction of the intensity at
time~$t\rightarrow\infty$ that is in~$E$. 
A more proper estimate of the escape rate~$\tilde{\kappa}$ from this measure has to include, as in Eq.~(\ref{eq.mupartial}),  
the partial transmission through the leak as~\cite{AltmannPRA}
\begin{equation}\label{eq.mucpartial}
\mathfrak{m}_{c}(I)=\int_I (1-R({\bf x})) d\tilde{\mu}_c = \int_I (1-R({\bf x})) 
\tilde{\rho}_c({\bf x}) d{\bf x}.
\end{equation}
The analogue of the improved escape rate formula~(\ref{eq.kappa}) for a system with partial leaks is thus
\begin{equation}\label{eq.kappaR}
\tilde{\kappa} \simeq -\frac{\ln(1-\mathfrak{m}_c(I))}{\tcoll_{\tilde{c}}},
\end{equation} 
where $\tilde{c}$ in $\tcoll_{\tilde{c}}$ indicates average taken with respect to $\tilde{\rho_c}$.

The complete extension of the transient chaos theory presented in Sec.~\ref{ssec.transient} is considerably more subtle and requires the
extension of the invariant sets discussed above (chaotic saddle and its invariant manifolds). We illustrate our general considerations
introducing uniform reflectivity $R=0.1$ in the leak of the cardioid billiard of Figs.~\ref{fig.transient}-\ref{fig.saddle}. The procedures used previously (see
Figs.~\ref{fig.transient}, \ref{fig:saddle}, and~\ref{fig.saddle2}) were employed to obtain Fig.~\ref{fig.partial}, but now the intensities
of the trajectories were used as weights attached to each trajectory, see Appendix.~\ref{sec.ap-algorithm} for details. 
The energy escape rate was found numerically to be $\tilde{\kappa}=0.058$, obviously less than  $\kappa=0.066$ for the full leak case
  of the same size.
The support of the regions with non-negligible densities shown in Fig.~\ref{fig.partial} shows the
stable manifold, chaotic saddle, and unstable manifold, respectively. In the case of partial leaks it is important to distinguish the
manifold of the trajectories from the 
manifold of the (trajectories weighted with) intensities~$J$. 
Whenever the reflection coefficient is non-vanishing,  $R({\bf x})\ne0$ for all ${\bf x} \in \Omega$, 
trajectories survive forever (and correspond thus to that of the closed problem).
The chaotic set of trajectories is then the full phase space, its manifolds are also space-filling, 
and are not very informative. The interesting patterns present in all panels of Fig.~\ref{fig.partial} motivate us to interpret the results
in terms of manifolds of the (trajectories weighted with) intensities. This generalizes the case of full leak ($R=0$ for ${\bf x} \in I$).  

Consider first the case of the stable manifold~$W^S$. While in the full leak case~$W_{\text{full}}^S$ is obviously outside the leak (e.g., in
Fig.~\ref{fig.saddle}a),  
in the partial leak case $\tilde{W}^S$ should certainly include trajectories which start in the leak and never return. These
{\em new} trajectories are nothing but the closed map  preimage of $W_{\text{full}}^S$ in $I$. Indeed, in
Fig.~\ref{fig.partial}(a) the filamentary structure extends inside~$I$  (with much lower intensity). These new trajectories modify the manifolds
in the whole phase space because their preimages can be outside the leak.  More
generally, it is natural to consider $W_{\text{full}}^S$ as the set of all points that carry
a nonvanishing intensity~$J$ for arbitrarily large~$t$~\cite{AltmannPRA}
\begin{equation}\label{eq.wu}
{\bf x} \in \tilde{W}^S \Leftrightarrow  \text{ for } t\rightarrow +\infty, \; J_t({\bf x}) \rightarrow J_{+\infty}({\bf x}) > 0,
\end{equation}
where $J_{\infty}({\bf x})$ is the asymptotic intensity.
The same reasoning applies to the unstable manifold~$\tilde{W}^U$, in which case the limit $t\rightarrow -\infty$ is taken in Eq.~(\ref{eq.wu}).
The presence of new orbits, compared to the case with full leak, is clearly seen when comparing Figs.~\ref{fig.saddle2}
and~\ref{fig.partial}~(c). Figure~\ref{fig.partial}(c) shows the c-density $\tilde{\rho}_c({\bf x})$ of the partial leak case.
The chaotic saddle (CS) is shown in Fig.~\ref{fig.partial}(b) and should be compared to Fig.~\ref{fig.transient}(b). Again, it can be
thought as $CS = \tilde{W}^S \cap \tilde{W}^U$, i.e., the points that remain with $J>0$ for $t\rightarrow \pm \infty$. 
The natural extensions of the concepts of chaotic saddle and its invariant manifolds 
are expected to hold in all systems with partial leaks (see also~\cite{WM:2008}). 

A complete description of distributions such as those shown in Fig.~\ref{fig.partial} cannot be achieved by simply including the new
trajectories  mentioned above. For instance, trajectories that collide infinitely many times with $R<1$ but still less frequently than other
trajectories might also contribute to the asymptotic properties. Indeed, it appears  natural that the support
of the distributions such as those shown in Fig.~\ref{fig.partial} will fill an area of the phase space and therefore have a trivial
fractal dimension $D_0=2$. Given the peaked structures of the densities, we can speculate that the information dimension~$D_1$ and the
  generalized dimension $D_q$ with $q>0$ might be below~$2$ 
[see ~\cite{TelGruiz-book} for simple examples of sets with trivial $D_0$ and nontrivial $D_1$].
It remains, however, to be verified whether relations~(\ref{sumDu})-(\ref{KG1}) are applicable to systems with partial leaks.

In the next section we introduce a formalism based on Perron-Frobenius operators, that directly calculates the
(normalized) density~$\tilde{\rho}_c$ and the artificial (non-observable) distinction between $J$ and $\rho_c$ becomes 
superfluous.

\section{Operator formalism}\label{sec.operators}\label{sec.3}
\label{sec.operator}

In the previous sections we used extended trajectories which depend not only on the phase space coordinate ${\bf x}$ 
of the map but also on the true time $t$ and on the intensity $J$. These two quantities are labels which are attached to each trajectory and 
change depending on (but do not affect) the sequence of collisions ${\bf x}_i$, $i=0 \ldots n$. 
In this section we develop a more elegant formalism based on operators acting on densities $\rho({\bf x},t)$ that naturally accounts for both of 
these aspects, allowing for a more rigorous treatment. In the spirit of systems with leaks, and similarly to
Sec.~\ref{sec.theory},  we start with the case of closed systems that will be used for comparison with the leaking case.

\begin{figure}[!ht]
\centering
\includegraphics[width=0.5\columnwidth]{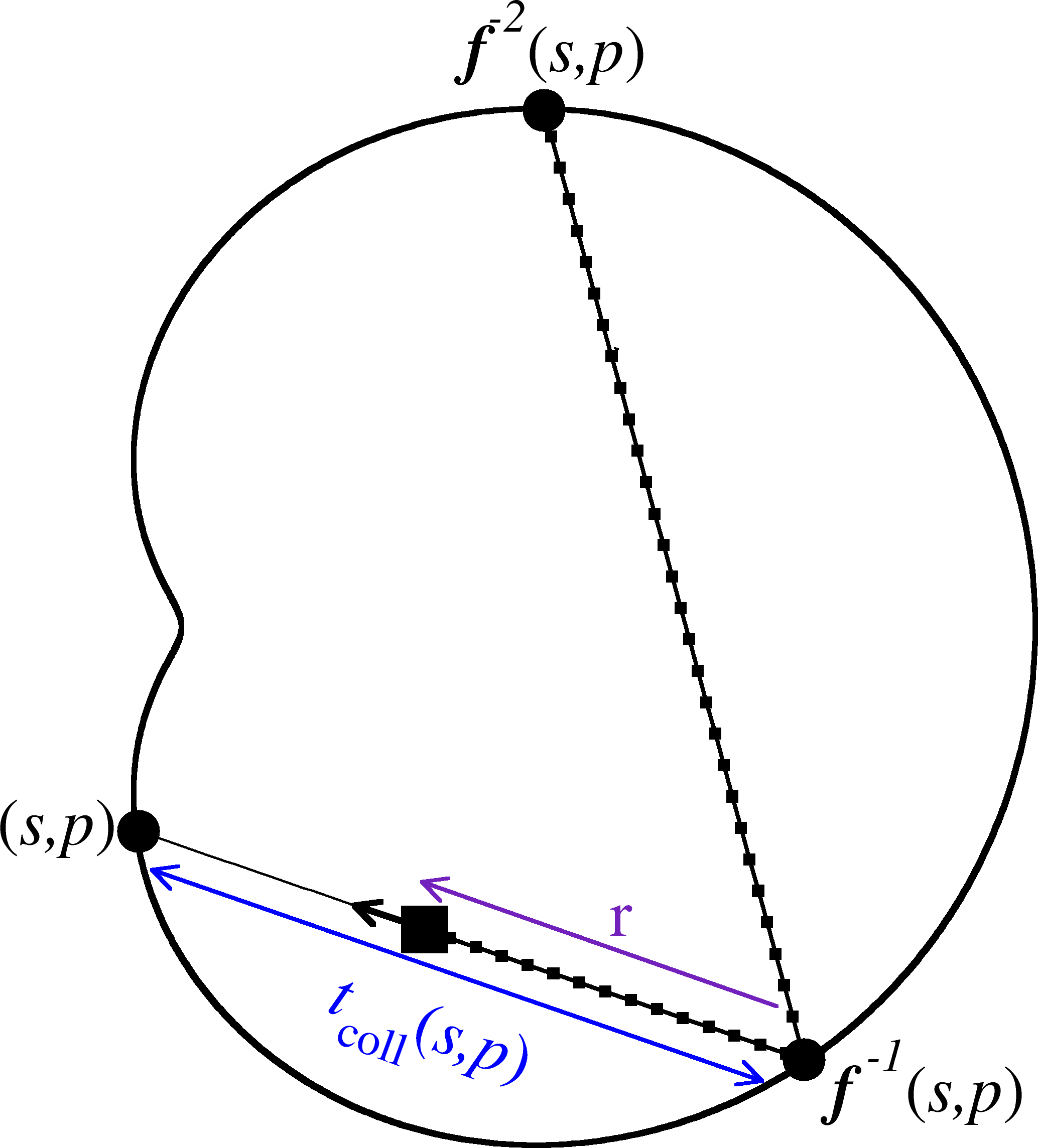}
\caption{(Color online) Schematic illustration of flow and map coordinates for billiards. 
We use the convention that a particle  in the flow ($\blacksquare$) 
has coordinates $(s,p,r)=({\bf x},r)$. Note that {\bf x} is the endpoint of a flight within the billiard. Due to the specular nature of the reflection, the
$p$ coordinate does not change during a collision, ${\bf x}\equiv(s,p)$ corresponds thus to the map coordinate denoted by $\bullet$ at the boundary of
the billiard defined right after collisions. Smaller
dots represent other particles in the flow that move along the same trajectory. 
} 
\label{fig.mapflow}
\end{figure}

Before discussing how densities evolve in time, it is essential to clarify the relationship between the different characteristic densities in
billiard systems. Here we search for exact relationships between the measure~$\mu_F$ of the flow and 
the measure~$\mu$ of the true-time map (for more details see Sec.~\ref{sec.ap-algorithm}).  
In closed Hamiltonian systems the natural measure $\mu$ is the Lebesgue measure.

A trajectory of the billiard flow can be represented by the Birkhoff coordinates ${\bf x}=(s,p)$ of its {\em next} collision with the boundary and the time
$r$ {\em after} the previous collision, as illustrated in Fig.~\ref{fig.mapflow} (see~\cite{Chernov-book} for a rigorous mathematical
formulation). We call $r$ the coordinate time and it also corresponds to a distance in the billiard, 
since the velocity of the particle has been chosen to be unity.
Of course, the coordinate time fulfills $0 \le r \le
t_{\mathrm{coll}}({\bf x})$ where the collision time $t_{\mathrm{coll}}({\bf x})$ is defined as 
the time~$t$ between the collision in ${\bf x}$ and the {\em last} collision\footnote{These definitions are convenient because
they assure that $t_{\mathrm{coll}}({\bf x})$ and the c-measure
are defined inside the leak ${\bf x} \in I$.}. 

\subsection{Closed system}\label{ssec.opclosed}
    
In the time-continuous representation, let $\rho_F({\bf x},r)$ denote the   
probability density for finding a flow trajectory at $({\bf x},r)$. 
The corresponding invariant measure $d \mu_F$ of the flow
can then be written as
\begin{equation}
d \mu_F = \rho_F ({\bf x},r) d{\bf x} dr.
\label{muF}
\end{equation} 

Consider now the true-time map in standard Birkhoff coordinates ${\bf x}=(s,p)$.
Let $\rho_{\mu} ({\bf x})$ denote the probability density of the invariant measure in this map, which is a constant as seen in Fig.~\ref{fig.closed}.
The corresponding measure $d \mu$ is
\begin{equation}
d \mu  = \rho_\mu ({\bf x}) d{\bf x}.
\label{mu}
\end{equation}
We want to connect $\mu$ to  $\mu_F$. Since 
the dynamics between collisions is a uniform motion of unit velocity, 
$\rho_F$ is independent of~$r$ and, therefore, the
flow density in variable ${\bf x}$ is proportional to that of the map
\begin{equation}\label{eq.A}
\rho_F ({\bf x},r) = A \rho_\mu({\bf x}),
\end{equation}
where $A$ is a constant. Its value follows from the normalization of 
the measure $\int d \mu_F=1$:
\begin{equation}
\int \rho_F ({\bf x},r) dr d{\bf x} = \int \; 
\int _0^{t_{\mathrm{coll}}({\bf x})} A \rho_\mu({\bf x})  dr d{\bf x} $$ $$
=A \int \; t_{\mathrm{coll}}({\bf x}) \rho_\mu({\bf x})  d{\bf x}  =1.
\end{equation}   
Since the integral in the last equality equals the average collision time in the closed system's true-time map, we have
\begin{equation}\label{eq.A2}
A =\frac{1}{\tcoll}.
\end{equation}      
From this relation, and Eqs.~(\ref{muF}), (\ref{mu}), and (\ref{eq.A}), it follows that the flow measure and the map measure are related as
\begin{equation}
d \mu_F = \frac{d r}{\tcoll} d\mu.
\label{mumuF}
\end{equation} 
This is a well known relation connecting flows and maps in closed billiards (see, e.g., \cite{Chernov-book}). Recalling that $\tcoll$ can be related to simple geometric properties of the billiard as in
Eq.~(\ref{eq.tau2d3d}), this formula shows that the mean collision time provides a simple and elegant connection between properties of the
flow and the map. For instance, Lyapunov exponents of the flow are equal to those of the true-time map divided by $\tcoll$.

From Eq.~(\ref{mumuF}) one can also estimate the escape rate by 
assuming that the existence of such a leak does not 
influence the validity of this relation. The rate of change of $\mu_F$ over 
coordinate time~$r$ is in this type of perturbation approach 
\begin{equation}\label{eq.fmuc}
\mathcal{F}=\frac{d\mu_F}{dr} = \frac{d\mu}{\tcoll},
\end{equation}
and this rate is independent of coordinate time $r$. 
The fraction of trajectories escaping through leak $I$ over a time unit can be computed as $\mathcal{F}(I)=\int_I d{\bf x} \mathcal{F}$. Consider a time interval of the length
of the average collision time $\tcoll$.
The decay of surviving trajectories is 
exponential, $P(t) = e^{-\kappa t}$, for any time
$t$ for trajectories distributed according
to the c-density, which is then approximately~$\rho_\mu({\bf x})$. 
We can thus estimate $\kappa$ using Eq.~(\ref{eq.fmuc}) and writing the proportion of particles that escape the billiard in a time~$t=\tcoll$
as $1-e^{-\kappa \tcoll}$:
\begin{equation}\label{eq.escaperateclosed}
\mathcal{F}(I) \tcoll = \mu(I) = 1- e^{-\kappa \tcoll}.  
\end{equation}
This leads to the naive estimate $\kappa^*=-\ln(1-\mu(I))/\tcoll$ stated in Eq.~(\ref{eq.kappastar}). 

Finally, the dynamics of densities in closed maps can be rewritten in terms of the Perron-Frobenius
operators~\cite{Gas-book,Dorfman-book,LaiTel-book}
\begin{equation}
\rho_{n+1}({\bf x}') = 
                \frac{\rho_{n}({\bf x})}{\mid \mathcal{J}({\bf x})\mid}_{\vert_{{\bf x} \in {\bf f}^{-1}({\bf x}')}},
\label{eq.FP}
\end{equation}
where $\mathcal{J}({\bf x})$ is the Jacobian at point ${\bf x}$. In Hamiltonian systems $\mathcal{J}({\bf x})=1$, and the constant Lebesgue
density $\rho_\mu({\bf x})=$constant is a stable fixed point of Eq.~(\ref{eq.FP}).
Since there is no escape, the largest eigenvalue of the Perron-Frobenius operator
is unity. 
In the next subsection we repeat the procedures presented above for systems with large leaks, i.e., we derive a
connection between the flow and map measures, an expression for the escape rate, and 
establish a Perron-Frobenius formalism. 
A direct connection between the Perron-Frobenius operator Eq.~(\ref{eq.escaperateclosed}) of the open and leaky systems has recently been investigated for cases with
Markov partitions by~\cite{Froyland:2010}.

\subsection{Flow and map measures in billiards with leaks}\label{sec.averagecollision2}

From the point of view of the escape process, the analogue of the natural measure 
is the c-measure. 
Even if the c-measure definition in Eq.~(\ref{eq.conditional}) applies for {\it maps} ({\bf x} coordinates), it is essential to
extend this concept to flows and {\it true-time maps} ({\bf x} and r coordinates) in order to take the real time of trajectories
into account.
Indeed, it is known that averages taken with respect to flows
(or true-time maps)   
and to traditional maps differ considerable in open systems~\cite{Kaufmann2001}.


The main difference with respect to the closed case is the dependence of the c-measure $\mu_{Fc}$ of the flow on the coordinate time
$r$. The survival probability decays in {\it time} $t$ as $e^{-\kappa t}$, independent of the {\bf x} coordinate. It is thus natural to 
associate the coordinate time $r$ with $t$ so that $\mu_{Fc}$ decays as $e^{-\kappa r}$. }Therefore, Eq.~(\ref{eq.A}) has to be replaced by 
\begin{equation}\label{eq.Ac}
\rho_{Fc} ({\bf x},r) = A_c \rho_c({\bf x}) e^{-\kappa r},
\end{equation}
where $\rho_{Fc}$ is the density of the flow's c-measure, and $\rho_c$
is the density of the true-time map c-measure, $\mu_c$. This latter is the measure
we  used in the main part of this paper.
The density $\rho_c({\bf x})$ is independent of $r$, as seen in Eq.~(\ref{eq.Ac}), and thus it can be thought to remain constant from
the previous collision ($r=0$) until the collision at {\bf x} ($r=t_{\text{coll}}({\bf x})$, see Fig.~\ref{fig.mapflow}). This constant
value is proportional to $\rho_{Fc}({\bf 
x},r=0)$. The proportionality factor $A_c$ follows 
from $\int d \mu_{Fc}=1$ as
\begin{equation}\label{eq.}
\int \rho_{Fc} ({\bf x},r) dr  d{\bf x} = \int \; \int _0^{t_{\mathrm{coll}}({\bf x})}  A_c \rho_c({\bf x}) e^{-\kappa r} dr d{\bf x} $$ $$
= A_c \int   \; \rho_c({\bf x}) \frac{1}{\kappa}(1-e^{-\kappa t_{\mathrm{coll}}({\bf x})}) d{\bf x}=1.
\end{equation}   
This yields
\begin{equation}\label{eq.Ac2}
A_c =\frac{\kappa}{1-\ec},
\end{equation}      
which tends to $1/\tcoll$ for $\kappa \rightarrow 0$ as in Eq.~(\ref{eq.A2}).
The map and flow c-measure $\mu_{Fc}$ and  $\mu_c$ 
are related as
\begin{equation}\label{mumuF2}
d \mu_{Fc} =  A_c e^{-\kappa r}d r d\mu_c.
\end{equation} 
This is the generalization to open systems of relation~(\ref{mumuF}).
In contrast to the case of closed billiard, the relation between flow and true-time map is not given by $\tcoll$ (or $\tcoll_c$) alone, but
includes an explicit dependence on $\kappa$ and the coordinate time~$r$. 
In Sec.~\ref{ssec.projected} we also connect $\mu_{Fc}$ and $\mu_c$ to the flow measure {\em projected} to the billiard's boundary.

\subsection{Exact escape rate formula}\label{ssec.escaperate}

We can now obtain an exact formula for the escape rate following the arguments used 
for closed systems at the end of
Sec.~\ref{ssec.opclosed}. From Eq.~(\ref{mumuF2}) one can again introduce the rate of change over coordinate time~$r$ as
$\mathcal{F}_c={d\mu_{Fc}}/{dr}$ which is, in contrast to Eq.~(\ref{eq.fmuc}), not independent of coordinate
time $r$. From Fig.~\ref{fig.mapflow} and Eq.~(\ref{eq.Ac}), the true-time density flowing out at ${\bf x}$ is given by
$\rho_{Fc}(f({\bf x}),r=0)=A_c \rho_c(f({\bf x}))$. Therefore
the fraction of trajectories that escape through leak $I$ in time $\tcoll_c$ is
$\int_0^{\tcoll_c} \mathcal{F}_c(f(I)) 
dr$, with $f(I)=$  ``escape'' as given in Eq.~(\ref{eq.leak1}). In
view 
of (\ref{eq.Ac2}) and (\ref{mumuF2}), 
Eq.~(\ref{eq.escaperateclosed}) turns into
\begin{equation}
\int_0^{\tcoll_c} \mathcal{F}_c(f(I)) dr = \mu_c(f(I)) \frac{1-e^{-\kappa \tcoll_c}}{1-\ec}= 1- e^{-\kappa \tcoll_c}. 
\end{equation}
Here in the last equality we used again the fact that for the c-measure the fraction of trajectories escaping up to time $t$ is $1-\exp{(-\kappa
t)}$. Since the escape happens immediately (i.e. $t_{\text{coll}}(\text{escape})=0$ and thus
$\mu_c(f(I))=\mu_c(I)$), we obtain an implicit relation for $\kappa$ as:
\begin{equation}\label{eq.escaperate}
\ec = 1-\mu_c(I).
\end{equation}
This formula, which is a new result and will be derived more formally in Sec.~\ref{sec.3}, establishes a relation between
the average of an expression containing $\kappa$ times the collision time and
the measure of the leak, both averages taken with respect to the c-measure of the
true-time map.  It is thus a generalization of the Pianigiani-Yorke formula 
(\ref{gamma}), valid for usual maps, which is recovered from Eq.~(\ref{eq.escaperate})
in the limit of $t_{\text{coll}}\equiv1$. It is an exact expression, the culmination of different approximations of $\kappa$ discussed
previously, as summarized in Tab.~\ref{tab.sum}.

\begin{table}
\begin{center}
\begin{tabular}{|c|c | c |}
\hline

{Escape rate formula}
&{$\kappa$ }
&{Equation}

\\\hline
{Sabine's estimate}
&{$\frac{\mu(I)}{\langle t_{\text{coll}} \rangle}$}
&{(\ref{eq.mut})}
\\
{Naive (Eyring) estimate ($\kappa^*$)}
&{$-\frac{\ln(1-\mu(I))}{\langle t_{\text{coll}} \rangle}$}
&{(\ref{eq.kappastar})}
\\
{Improved estimate ($\kappa_1$)}
&{$-\frac{\ln(1-\mu_c(I))}{\tcoll_c}$}
&{(\ref{eq.kappa})}
\\
{Exact expression}
&{$\ec = 1-\mu_c(I)$}
&{(\ref{eq.escaperate})}
\\\hline
\end{tabular}
\caption{Summary of the escape rate formulas for strongly-chaotic systems with a leak $I$.
The measures are $\mu(I)=\int_I \rho_{\mu}({\bf x}) d {\bf x}$ and $\mu_c(I)=\int_I \rho_c ({\bf x}) d {\bf x}$. The averages $\langle
\ldots \rangle$ and $\langle \ldots \rangle_c$ correspond to averages in $\mu$ and $\mu_c$, respectively. 
The corresponding formulas for systems with {\em partial} leaks are given in Eqs.~(\ref{eq.naiveR}),
 (\ref{eq.kappaR}), and (\ref{eq.kappacorrect}).}
\label{tab.sum}
\end{center}
\end{table}

It is worth applying the cumulant expansion to the left hand side of (\ref{eq.escaperate}):
\begin{equation}\label{eq.expansion}
\ln(\langle e^{-\kappa t_{\text{coll}}}\rangle_c) = \sum_{r=1}^{\infty} \frac{(-\kappa)^r}{r!} C_r(t_{\text{coll}})=\ln(1-\mu_c(I)),
\end{equation}
where $C_r(y)$ are the cumulants of $y$ ($C_1 = \langle y \rangle_c, C_2 = \sigma_c^2(y)$, etc.). 
Keeping only the first term of the expansion, we
find a first order approximation, $\kappa_1$,
of the escape rate 
as
$$\ln(1-\mu_c(I)) = -\kappa_1 \tcoll_c.$$
This is the improved escape rate formula~(\ref{eq.kappa}), obtained in Sec.~\ref{sec.theory} as a generalization (based
on the theory of transient chaos) from the naive estimate~(\ref{eq.kappastar}). 

We can estimate the deviation between $\kappa_1$ and $\kappa$ by including the second term of expansion~(\ref{eq.expansion}). For the second
order approximant $\kappa_2$, a quadratic equation is obtained: 
$$ \ln(1-\mu_c(I))= -\kappa_2 \tcoll_c +\frac{\kappa_2^2}{2} \sigma^2_{t_{\text{coll,c}}},$$
where $\sigma^2_{t_{\text{coll,c}}}$ is the second cumulant of the collision time distribution $t_{\text{coll}}({\bf x})$ taken with respect
to the c-measure $\mu_c$.
This yields the explicit form for $\kappa_2$:
\begin{equation}\label{eq.kappa2}
\begin{array}{ll}
 \kappa_2 &= \left[{\tcoll_c}-\sqrt{\tcoll^2_c+2 \sigma^2_{t_{\text{coll}},c}\ln(1-\mu_c(I))}\ \right]/{\sigma^2_{t_{\text{coll,c}}}} \\
       
       & \approx \kappa_1 \left( 1+\frac{\kappa_1}{2}\frac{\sigma^2_{t_{\text{coll,c}}}}{\tcoll_c}\right),
\end{array}
\end{equation}
where the approximation is valid for small variance~$\sigma_{t_{\text{coll,c}}}$ of collision times.
In a similar spirit, corrections 
due to the uneven distribution of $t_{\text{coll}}$ have also been obtained by~\cite{Mortessagne:1992}, see also~\cite{Joyce:1975}.

The approximation $\kappa_2 \approx \kappa_1$ is valid for 
\begin{equation}\label{eq.cond}
\frac{\sigma^2_{t_{\text{coll}},c}}{\tcoll^2_c} \ln(1-\mu_c(I)) \ll 1.
\end{equation}
Our numerical simulations (see Appendix~\ref{sec.ap-algorithm}) yield independent estimations for $\kappa$ and $\kappa_{1,2}$. In our
typical configuration (e.g., Fig.~\ref{fig.saddle}), the agreement between $\kappa_1$ and $\kappa_2$ was on the order of
$0.6\%$ and between $\kappa_2$ and $\kappa$ of $0.06\%$ (below the precision of the results reported in Tab.~\ref{tab.ic.num}). 
From Eq.~(\ref{eq.cond}) we see that
there are two effects that can make $\kappa_1$ to be a good approximation of $\kappa$: (a) small leak: $\mu_c(I)\ll 1$, 
and (b) $\sigma_{t_{\text{coll}},c}/\tcoll_c \ll 1$. The latter
corresponds to a small variance of the collision time distribution, and is the limit under which the true-time map reduces to a usual map
[and  Eq.~(\ref{gamma}) is recovered]. 
Note, however, that the naive estimate $\kappa^*$, Eq.~(\ref{eq.kappastar}), does not follow in any
order of the cumulant expansion.  
As previously noticed by~\cite{Joyce:1975,Joyce:1978}, $\kappa^*$ coincides with $\kappa$ only under the physically unrealistic conditions that: (i) the
distribution in the room is uniform $\rho_c=\rho_\mu$, e.g. non-specular collisions immediately randomize the trajectories or absorption is
uniformly distributed;
and (ii) $t_{\text{coll}}$ is constant. If only condition (i) is satisfied, Eqs. (\ref{eq.escaperate}) and~(\ref{eq.kappa2})
could be applied using $\mu$ instead of $\mu_c$. Joyce also pointed out that an analogue of Eq.~(\ref{eq.kappa2}) was also published by
Kuttruff in the 1970's.
The $\kappa^*$ estimate~(\ref{eq.kappastar})  is thus the result of a not fully consistent
argumentation: it intends to correct for the finite size of the leak, but fully forgets about changing the measure to the c-measure, a
correction of comparable magnitude.  The naive estimate~$\kappa^*$ has also been obtained by~\cite{Bunimovich2007} as the first term of an expansion in the
leak size.     

\noindent {\bf Partial leaks:} We are now in a position to heuristically generalize (\ref{eq.escaperate}) for systems with partial leaks~$R\neq0$ discussed in
Sec.~\ref{ssec.partial}. The key observation is that the 
left- and right-hand sides of Eq.~(\ref{eq.escaperate}) can be interpreted as {\em global}  and {\em local} quantities,
respectively. The {\em global} quantity in the partial leak case remains unchanged: in one {\em iteration} of the true-time map the c-density at position
${\bf x}$ decays as $ e^{-\tilde{\kappa}   t_{\text{coll}}({\bf x})} $, 
and the global estimation of the proportion $\mathcal{S}$ of the remaining energy after a time~$t_{\text{coll}}$ is obtained simply as the average of this factor over the full
phase space~$\Omega$ 
\begin{equation}\label{eq.global}
\mathcal{S}=\int_\Omega \tilde{\rho}_c({\bf x}) e^{-\tilde{\kappa} t_{\text{coll}}({\bf x})} d{\bf x} = \langle e^{-\tilde{\kappa} t_{\text{coll}}} \rangle_{\tilde{c}}.
\end{equation}
We denote the c-density of problems with partial leaks as $\tilde{\rho}_c$, and index $\tilde{c}$ refers to such c-densities.
The same quantity is obtained in a {\em local} approach by 
considering the proportion of trajectories {\em not} crossing the 
leak in one iteration of the true-time
map. In the full leak case this is  $ 1-\int_I \rho_c({\bf x}) d{\bf x}= 1-\mu_c(I)$ which appears on the right hand side of Eq.~(\ref{eq.escaperate}).
When the leak is partially reflecting, the portion  $\int_I R({\bf x})\tilde{\rho}_c({\bf x}) d{\bf x}$
of the leak's measure enhances the number of survivors. The full proportion $\mathcal{S}$  is 
\begin{equation}\label{eq.survivors}
\begin{array}{lll}
\mathcal{S} &= 1-\int_I \tilde{\rho}_c({\bf x}) d{\bf x}+\int_I R({\bf x})\tilde{\rho}_c({\bf x}) d{\bf x} \\
 &=1-\int_I (1-R({\bf x}))\tilde{\rho}_c({\bf x}) d{\bf x} \\
 & = \langle R({\bf x}) \rangle_{\tilde{c}},
\end{array}
\end{equation}
where we used that the last integral [which is exactly $\mathfrak{m}_{c}(I)$ of Eq.~(\ref{eq.mucpartial})] can be written as an integral
over the  full phase space $\int_{\Omega} (1-R({\bf x}))\tilde{\rho}_c({\bf x}) d{\bf x}$, since outside the leak
$R\equiv1$, and 
we used the normalization $\int_{\Omega} \tilde{\rho}_c({\bf x}) d{\bf x}=1$. This means that the
remaining energy portion is the full phase space average of the 
reflection coefficient~$\langle R \rangle_{\tilde{c}}$. Equating the global~(\ref{eq.global}) and local~(\ref{eq.survivors}) expressions  we obtain the
formula  
\begin{equation}\label{eq.kappacorrect}
\langle e^{-\tilde{\kappa} t_{\text{coll}}} \rangle_{\tilde{c}} = \langle R \rangle_{\tilde{c}}.
\end{equation}
This is a general exact relationship for the energy escape rate $\tilde{\kappa}$ that applies to 
all cases discussed in this paper and is one of our main new results. 
In the next section we provide a derivation of this expression based on Perron-Frobenius operators. 
Equation~(\ref{eq.kappacorrect}) goes over
into relation~(\ref{eq.escaperate}) for full leaks, i.e. for $R={0}$ inside the leaks.

It should finally be noted that the value of our escape rate relations (\ref{eq.kappacorrect}) and 	(\ref{eq.escaperate}), just like
that of the Pianigiani-Yorke formula~(\ref{gamma}), is conceptual. They do not provide an efficient way for determining the escape rate (that is very well
handled numerically), rather they illustrate how the escape rate of the flow
follows from properties of the true-time map.

\subsection{Operators for true-time maps with partial leaks}\label{ssec.operators}

Consider an invertible open map ${\bf f}$. Its escape rate $\gamma$ is known to appear as the 
largest eigenvalue of an operator, the Perron-Frobenius operator. This operator
is defined \cite{Tel:1987,LaiTel-book} by the iteration scheme of a density function $\rho$
\begin{equation}
{e^{-\gamma}} \rho_{n+1}({\bf x}') = 
                \frac{\rho_{n}({\bf x})}{\mid \mathcal{J}({\bf x})\mid}_{\vert_{{\bf x} \in {\bf f}^{-1}({\bf x}')}},
\label{FP2}
\end{equation}
where $\mathcal{J}({\bf x})$ is the Jacobian at point ${\bf x}$. 
By considering the right hand side to be the result of an operator acting on function $\rho$,
the left hand side shows that $e^{-\gamma}$ is an eigenvalue (in the space of positive 
$\rho$-s, the largest eigenvalue) of this operator.  
Equation (\ref{FP2}) expresses that the total probability in a small region
at
step $n$ is the same as in  the image of that region  under map ${\bf f}$, when 
taking into account a factor $e^\gamma$ for compensating the escape. The escape rate follows from the
requirement that the integral of $\rho_n$ over a fixed phase space region
containing the chaotic saddle remains finite in the limit $n \rightarrow \infty$.
The limit distribution $\rho_{\infty}$ is then the density $\rho_c$ of the
conditionally invariant measure concentrated on the unstable manifold of the chaotic saddle. This is the well-known picture for open maps \cite{Tel:1987,LaiTel-book} . 

As a generalization of this idea, in a true-time problem where the distribution
of collision times $t_{\mathrm{coll}}({\bf x'})$ is known,
the continuous-time
escape rate $\kappa$ is determined by the iteration scheme
\begin{equation}
e^{-\kappa t_{\mathrm{coll}}({\bf x'})} \rho_{n+1}({\bf x}') = 
                \frac{\rho_{n}({\bf x})}{\mid \mathcal{J}({\bf x})\mid}_{\vert_{{\bf x} \in {\bf f}^{-1}({\bf x}')}}.
\label{FP2true}
\end{equation}
In this equation the true-time property is incorporated on the left hand side in the multiplicative factor containing the escape rate, 
consistent with our convention for $t_{\text{coll}}$ in Eq.~(\ref{eq.tn}). 
This is a higher-dimensional extension of the true-time formalism used for one-dimensional maps by~\cite{Kaufmann2001} and, in a slightly
different context, by~\cite{Gaspard:1996,Gas-book}. 

The escape rate $\kappa$ again can be considered as an eigenvalue, and its value follows from the
requirement that the integral of $\rho_n$ over a fixed phase space region
containing the chaotic saddle remains finite in the limit $n \rightarrow \infty$.
The limit distribution $\rho_{\infty}$ is the density $\rho_c$ of the
conditionally invariant measure of the true-time map. 
 Equation~(\ref{FP2true}) is consistent with the properties of the flow and map measures in billiards discussed in
 Sec.~\ref{sec.averagecollision2}. A nonzero stationary c-density of the map can only exist if we compensate the exponential loss
 of the densities in time $t$ (and coordinate $r$) . This can be achieved by applying an instantaneous ``kick'' to the flow density in the
 form of a multiplicative factor  
$K({\bf x},{\bf x'})>1$ applied when the billiard wall is reached at $({\bf x},r=t_{\text{coll}}({\bf x}))\equiv({\bf x'},r=0)$. The flow
 density right after the collision can then be written as 
$\rho_{Fc}({\bf x'},r=0)=A_c \rho_c({\bf x'})=K({\bf x},{\bf x'})\rho_{Fc}({\bf x},r=t_{\text{coll}}({\bf x})) =A_c K({\bf x},{\bf
 x'}) \rho_c({\bf x}) e^{-\kappa  t_{\text{coll}}({\bf x})}.$
 This is consistent with (\ref{FP2true}) applied to the limit distribution
of billiards $(J\equiv1)$ if $K({\bf x},{\bf x'})=e^{\kappa(t_{\text{coll}}({\bf x})+t_{\text{coll}}({\bf x}'))}$.

The problem of maps with partial leaks can be treated as a further 
generalization. Since the dynamics of trajectories is then closed, we write the map as
${\bf f}_{\text{closed}}$. The energy escape rate $\tilde{\kappa}$ for partial leaks follows from an  iteration
scheme in which the reflection coefficient $R({\bf x})$ also appears, in
a similar spirit as in Tanner's work on ray dynamics with transmission and reflection in periodically driven problems~\cite{Tanner:1998,Tanner:2013,Chappell:new} . In our notation, $R$ shows up 
on the right-hand side 
since there is an immediate loss of density wherever $R$ is different from unity. 
If collision times are also taken into account, we find 
\begin{equation}
e^{-\tilde{\kappa} t_{\mathrm{coll}}({\bf x'})} \tilde{\rho}_{n+1}({\bf x}') = 
                \frac{R({\bf x}) \tilde{\rho}_{n}({\bf x})}{\mid \mathcal{J}({\bf x})\mid}_{\vert_{{\bf x} \in {\bf f}_{\text{closed}}^{-1}({\bf x}')}}.
\label{FPleak}
\end{equation}
The limit distribution $\tilde{\rho}_{\infty}$ is the density $\tilde{\rho}_c$ of the
conditionally invariant measure in the true-time map of the partially leaking system.
(One might also have leaks in a naturally open system, in which case the open 
map ${\bf f}$ should be used in the relation above.) 
Full leaks can also be seen as partial leaks with $R({\bf x})=0 \;\;\; \mbox{for} \;\;\; {\bf x} \in{I}.$
For traditional maps with leaks, $t_{\text{coll}}({\bf x})=1$, Eq.~(\ref{FPleak}) yields the escape rate $\tilde{\gamma}$ of such maps:
$\tilde{\kappa} \mapsto \tilde{\gamma}$. It is interesting to see from~(\ref{FP2true}) and~(\ref{FPleak}) that in closed systems ($R\equiv
1, \kappa=0$) the Perron-Frobenius operator coincides with the classical form~(\ref{eq.FP}). There is then no essential difference between the
true-time and the map picture due to the simple proportionality of the flow's and the map's measure as expressed by~(\ref{eq.A}).

The operator formalism for true-time maps
with partial leaks developed above is a new result of our paper.
It unifies and generalizes all previous approaches. In fact,  
Eq.~(\ref{FPleak}) suggests that the physically rather different phenomena
of collision times and of reflection (without taking into account collision times)
are described by essentially the same mathematical mechanism: the density $\tilde{\rho}({\bf x})$  in Eq.~(\ref{FPleak}) should be multiplied by a function
of the phase-space 
coordinates~${\bf x}$. As a consequence, the inclusion of the collision time distribution has effects on the iteration similar to those of a reflection
coefficient larger than unity.  We note that Eq.~(\ref{FPleak}) is a particular case of the generalized 
operators considered in the mathematical literature (see, e.g.,~\cite{Faure:new}),
augmented here with a well defined physical interpretation.

It is now straightforward to obtain an exact relation for~$\tilde{\kappa}$.  Consider the c-density $\tilde{\rho}_c({\bf x})$
of a system with full or partial leaks, i.e., the limit distribution of Eq.~(\ref{FPleak}). By integrating both sides over the full phase space, we obtain
\begin{equation}\label{eq.kappaderivation}
\begin{array}{lllll}
 \langle e^{-\tilde{\kappa} t_{\text{coll}}} \rangle_{\tilde{c}} & \equiv \int_\Omega d{\bf x}' e^{-\tilde{\kappa} t_{\mathrm{coll}}({\bf x'})} \tilde{\rho}_{c}({\bf x}') \\ \\
     & = \int_\Omega d{\bf x}' \dfrac{R({\bf x}) \tilde{\rho}_{c}({\bf x})}{\mid \mathcal{J}({\bf x})\mid}_{\vert_{{\bf x} \in {\bf
        f}_{\text{closed}}^{-1}({\bf x}')}} \\ \\
& = \int_\Omega d{\bf x} R({\bf x}) \tilde{\rho_{c}}({\bf x}) = \langle R \rangle_{\tilde{c}},
\end{array}
\end{equation}
where we used  $| \mathcal{J}({\bf x}) |=| d{\bf x'}|/|d{\bf x}|$, and $f_{\text{closed}}(\Omega)=\Omega$. This provides a proof of
our exact formula~(\ref{eq.kappacorrect})\footnote{Consider defining the collision time as a function of the initial coordinate
${\bf f}^{-1}({\bf x})$ as $\hat{t}_{\text{coll}}({\bf f}^{-1}({\bf x}))= t_{\text{coll}}({\bf x})$,
or $\hat{t}_{\text{coll}}(({\bf x}))= t_{\text{coll}}({\bf f}({\bf x}))$.
It is then natural to shift the factor $e^{-\kappa \hat{t}_{\text{coll}}({\bf x})}$
to the right hand side of Eq.~(\ref{FPleak}), and the same argument that led to Eq.~(\ref{eq.kappaderivation}) leads to
$\langle R e^{\kappa \hat{t}_{\text{coll}}} \rangle_{\hat{c}}=1$. For the full leak case, Eq.~(\ref{eq.kappacorrect}) becomes $\langle e^{\kappa
\hat{t}_{\text{coll}}} \rangle_{\hat{c}}=1+\mu_c(I)$, as it will be shown in~\cite{APT:2013}.
The distance from ${\bf f}^{-1}({\bf x})$ to ${\bf x}$ (i.e.,
${t}_{\text{coll}}({\bf x})$) is the same as from ${\bf x}$ to ${\bf f}^{-1}({\bf x})$.
The latter distance is obtained along a trajectory that starts at $(s,-p)$ and ends at ${\bf f}(s,-p)$. By introducing the operator $A({\bf
  x})=A(s,p)=(s,-p)$, this distance is $\hat{t}_{\text{coll}}(A({\bf x}))$. We thus find the simple relation 
$\hat{t}_{\text{coll}}({\bf x})={t}_{\text{coll}}(A({\bf x}))$ and, since ${\bf f}^{-1}=A{\bf f}A$ and $A^2=1$, 
${t}_{\text{coll}}({\bf f}({\bf x})) = {t}_{\text{coll}}(A({\bf x}))$ and ${t}_{\text{coll}}(A({\bf f}^{-1}({\bf x}))) =
{t}_{\text{coll}}({\bf x})$.
}.
Although Eqs.~(\ref{eq.kappacorrect}) and~(\ref{eq.kappaderivation}) are
identical, it is worth emphasizing that~(\ref{eq.kappacorrect}) was obtained from a qualitative argument based on properties of billiard
dynamics. Since Eqs.~(\ref{FP2true}) and~(\ref{FPleak}) are valid for any true-time map, the derivation presented here
shows that Eq.~(\ref{eq.kappaderivation}) is not restricted to billiards, it holds for leaky systems in general.

\subsection{Examples in leaky baker maps}\label{ssec.baker}

In order to illustrate our formalism in simple examples, we consider area-preserving baker maps.
This is motivated not only by
the possibility of an analytic treatment, but also by the fact 
that several previous publications~\cite{KNPS:2006,Nonnenmacher2005,Novaes2009,Pedrosa:2009,Ermann:2012,Pedrosa:2011}
use leaky versions of these maps to investigate quantum systems (see Section~\ref{sssec.eigenstates}).

First we consider the triadic area-preserving baker map defined on 
the unit square $(x,y)\in [0,1] \times [0,1]$: 
\begin{equation} \label{eq:ld_baker3}
\begin{array}{lll}
(x_{n+1},y_{n+1})  = (\frac{1}{3} x_n,{3}y_n) & \text{for } y_n \le 1/3 ,\\
(x_{n+1},y_{n+1})  = (\frac{1}{3}(x_n+1),3y_n-1) & \text{for } 1/3 \le y_n<2/3,\\ 
(x_{n+1},y_{n+1})  = (\frac{2}{3}(x_n+2),3y_n-2) & \text{for } y_n \ge 2/3, 
\end{array}
\end{equation}
with a  simple choice of the leak: $I$ is a band of height $1/3$ in the 
expanding ($y$) direction in the middle of the square. In this model there are no partial leaks present. The collision times take on two
values only: $t_{\mathrm{coll}}=\tau_1$ if the point is mapped into the column $x<1/3$, and $t_{\mathrm{coll}}=\tau_2$ if the particle is
mapped into the column $x>2/3$.
  
We start from a constant distribution~$\rho_0\equiv 1$ on the
unit square. Since the Jacobian is unity, Eq. (\ref{FP2true}) tells us 
that the measure (under $\rho_0$) of the lower band $y<1/3$ is mapped on the column $x<1/3$ with the measure on it multiplied by  
$e^{\kappa \tau_1}$. Similarly, the measure of the right
column $2/3<x<1$ will be $e^{\kappa \tau_2}/3$. The measure from the mid band is not mapped anywhere because it is in the leak. 

This construction extends to finer scales in a self-similar manner due to the simple choice of the leak. The measure converges to the c-measure,
which therefore has to remain (conditionally) invariant under the above iteration.
The value of the escape rate therefore follows by prescribing the invariance of the c-measure. In this case it is sufficient to
consider the measure projected on the $x$ axis because the density is constant along $y$.
By construction, this measure is originally $1$, and after one step it is the sum of the two values just determined, and thus
\begin{equation}\label{eq.simpkappa}
e^{\kappa \tau_1} 
+e^{\kappa \tau_2} =3.
\end{equation}
This is an (irrational) equation for the escape rate $\kappa$. The validity
of the new formula (\ref{eq.escaperate}) can be easily verified: 
$$\ec= e^{-\kappa \tau_1}\frac{e^{\kappa \tau_1}}{3}+e^{-\kappa \tau_2}\frac{e^{\kappa \tau_2}}{3}=2/3,$$
which corresponds to $1-\mu_c(I)$ since the leak has height $1/3$ and $\rho_c$ is independent of $y$.

In the limit of traditional maps, $\tau_1=\tau_2=1$, Eq.~(\ref{eq.simpkappa}) yields $\kappa \rightarrow \gamma=\ln{(3/2)}$. Since the
stretching rate is $3$ everywhere in the phase space, the maps average Lyapunov exponent is $\bar{\lambda}=\ln{3}$. The Kantz-Grassberger
formula (\ref{KG1}) then yields $D_1^{(1)}(I)=\ln{2}/\ln{3}$, i.e., the unstable manifold of this leaky baker map
carries the structure of the classical triadic Cantor set. 

\begin{figure}[!ht]
\centering
\includegraphics[width=0.7\columnwidth]{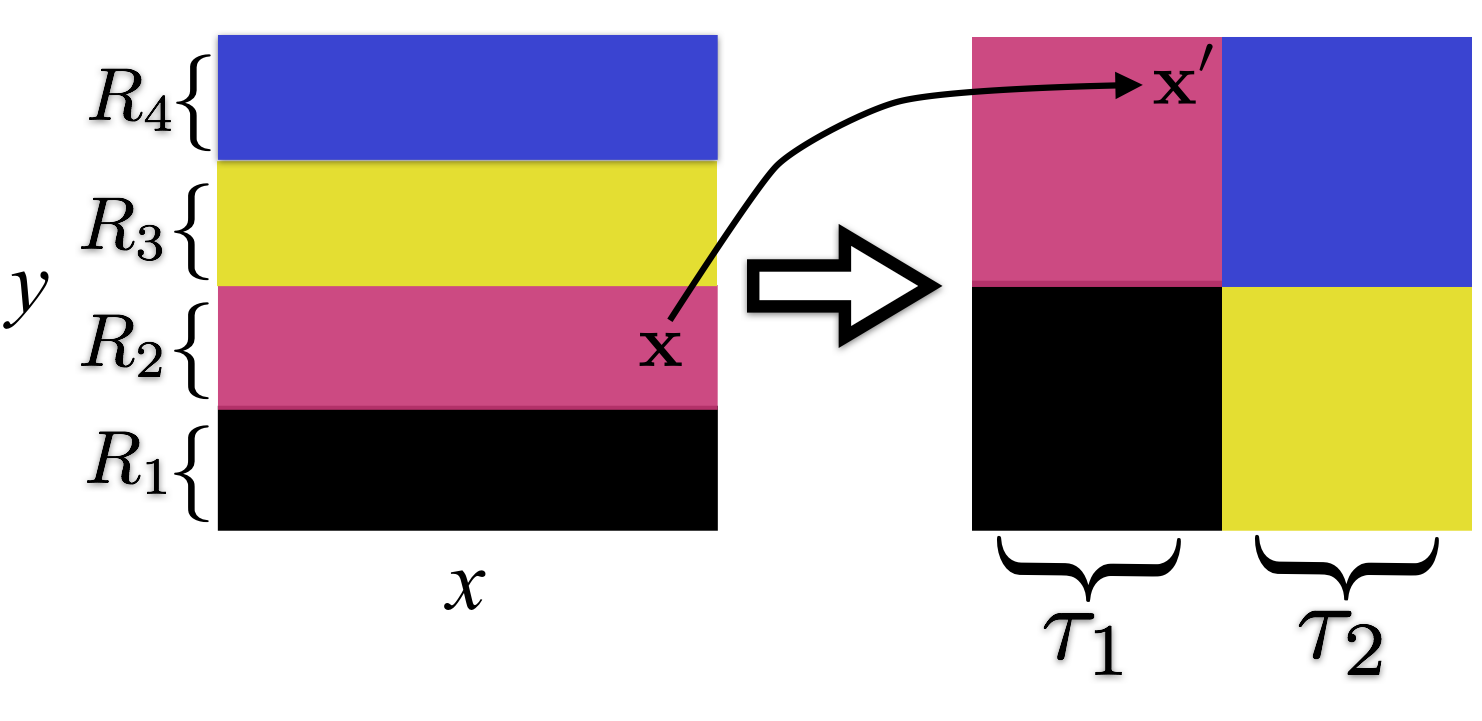}
\caption{(Color online) Illustration of the dynamics of a dyadic true-time area-preserving baker map with partial leaks. The right square is
obtained by applying 
baker map~(\ref{eq:ld_baker1}) to the left panel.}
\label{fig.baker}
\end{figure}

As a more complex example, consider the dyadic baker map
\begin{equation} \label{eq:ld_baker1}
\begin{array}{ll}
(x_{n+1},y_{n+1}) = (\dfrac{1}{2}x_n,2y_n) & \text{for } y_n \le 0.5 ,\\
(x_{n+1},y_{n+1}) = (1-\frac{(1-x_n)}{2},1-2(1-y_n)) & \text{for } y_n>0.5,
\end{array}
\end{equation}
with collision times taking on again two values only: $t_{\mathrm{coll}}=\tau_1$ if for the image point $x'<1/2$, and $\tau_2$ otherwise,
but with partial leaks. These leaks
are introduced with reflection coefficients $R_1$, ..., $R_4$ on four horizontal strips of height $1/4$, as
Fig.~\ref{fig.baker} illustrates. 

We again consider  $\rho_0 \equiv 1$.
After one iteration one finds
that 4 different c-densities appear in the four quadrants of the unit square. 
A next iteration refines the picture, but one feature remains: the c-measure coarse grained on these four quadrants has 4 different densities. The values on the 4 rectangles change as iteration goes on, 
but the four-value structure remains unchanged. 

In order to find an analytic expression for the energy escape rate, it proves to be 
sufficient to deal with the course grained c-density $\tilde{\rho}_c(y)$
projected on the expanding ($y$) axis. As follows from above, 
after a large number of iterations, this c-density is piecewise constant and has a jump at $y=1/2$. We therefore assume its form as

\begin{equation} \label{eq.rhocy}
\tilde{\rho}_c(y) =
 \left \{ \begin{array}{ll} 2-c & \mbox{ for } 0< y< 0.5  \\
                                      c & \mbox{for } 0.5 \ge y\le 1 ,\\
                                       \end{array} \right.
\end{equation}
which fulfills normalization $\int_0^1 \tilde{\rho}_c(y) dy=1$ for any $c$. 
The values of $c$ and $\tilde{\kappa}$ follow from the requirement that this 
projected measure remains invariant after one more time step.  

The lowest horizontal strip of height $1/4$ (and of c-measure $(2-c)/4$)
is mapped, in view of (\ref{FPleak}), on the bottom left quarter of the square with a new measure
$\exp(\tilde{\kappa} \tau_1)R_1(2-c)/4$ (note that the Jacobian is unity).    
The second horizontal strip is mapped on the top left quarter 
with c-measure $\exp(\tilde{\kappa} \tau_1)R_2(2-c)/4$. The third and fourth 
strips come into the remaining quarters with weights  
$\exp(\tilde{\kappa} \tau_2)R_3c/4$ and $\exp(\tilde{\kappa} \tau_2)R_4 c/4$, respectively
(the last one representing the top right quarter). 

The criterion of the invariance of the projected measure is that the 
total c-measure in the two bottom (top) quarters is the same as the
integral of (\ref{eq.rhocy}) over $0<y \le 1/2$ ($1/2<y \le 1$).
Thus, we find two equations 
$$1-\frac{c}{2} = \frac{1}{4}[e^{\tilde{\kappa} \tau_1} R_1(2-c)+e^{\tilde{\kappa} \tau_2} R_3 c],$$
\begin{equation}\label{eq.newseteq} 
\frac{c}{2} = \frac{1}{4} [e^{\tilde{\kappa} \tau_1} R_2(2-c)+e^{\tilde{\kappa} \tau_2} R_4 c].
\end{equation}
After rearrangement, we obtain
\begin{equation}\label{eq.newkappa}
4-2e^{\tilde{\kappa} \tau_1} R_1-2e^{\tilde{\kappa} \tau_2} R_4 
+e^{\tilde{\kappa} (\tau_1+\tau_2)}(R_1R_4-R_2R_3) =0,
\end{equation}
and 
\begin{equation}\label{eq.newc}
c=\frac{4-2e^{\tilde{\kappa} \tau_1} R_1}{2-e^{\tilde{\kappa} \tau_1} R_1 
+e^{\tilde{\kappa} \tau_2}R_3} =
\frac{2e^{\tilde{\kappa} \tau_1} R_2}{2-e^{\tilde{\kappa} \tau_2} R_4 
+e^{\tilde{\kappa} \tau_1}R_2}.
\end{equation}
Equation (\ref{eq.newkappa}) is an implicit equation for
the escape rate, while (\ref{eq.newc}) provides the value of $c$
determining the jump in the c-measure (\ref{eq.rhocy}) projected
onto the $y$ axis.     
We verified that $\tilde{\kappa}$ obtained from such implicit relations
agree with direct numerical simulations of the baker map including 
the intensity~$J_n$ and real time~$t_n$.

To have more analytic insight, now we focus on the particular case of $\tau_1=\tau_2=1$
(map with partial leaks). Equation~(\ref{eq.newkappa}) leads then to a quadratic
expression for escape rate $\tilde{\gamma}$:
\begin{equation}\label{eq.newgamma}
4-2(R_1+R_4)e^{\tilde{\gamma}} 
+(R_1R_4-R_2R_3) e^{2\tilde{\gamma}} =0.
\end{equation}%

\begin{figure}[!ht]
\centering
\includegraphics[width=1\columnwidth]{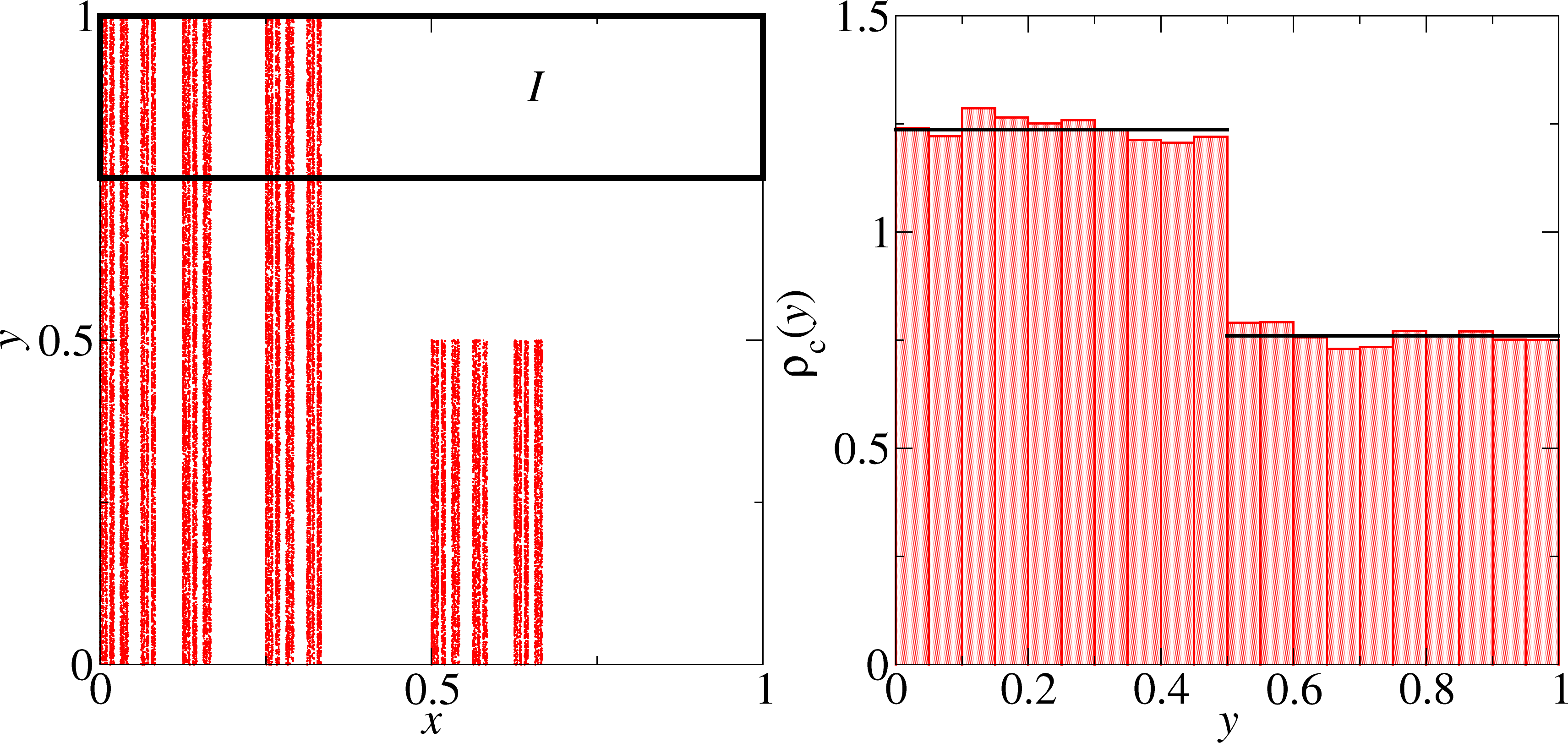}
\caption{(Color online) Numerical results for the dyadic baker map with a full leak $I$ over the uppermost strip of height $1/4$: $R_4=0,R_1=R_2=R_3=1,\tau_1=\tau_2=1$. 
Left panel: Unstable manifold obtained as the endpoint of particles surviving up to $n=30$ iterations. Right panel: Distribution of the c-density projected
on the $y$ axis $\rho_c(y)$. The straight lines are the analytical results: $2-c=\sqrt{5}-1$ for $y<0.5$ and $c=3-\sqrt{5}$ for $y>0.5$.}
\label{fig.R4-1}
\end{figure}

As a simple particular case, we assume that there is a full leak over the 
uppermost horizontal band, and no partial leak anywhere:
$ R_4=0, R_1=R_2=R_3=1$. From Eq.~(\ref{eq.newgamma}) and (\ref{eq.newc})
we obtain $e^{\gamma}=\sqrt{5}-1, \;\;\;  c=3-\sqrt{5}$. Since the average Lyapunov exponent in this uniform baker map
is $\bar{\lambda}=\ln{2}$, the stable manifold's information dimension is, in view of Eqs.~(\ref{sumDu}) and (\ref{KG1})
$$
D_1^{(s)}=2-\frac{\ln{(\sqrt{5}-1)}}{\ln2},
$$
clearly below $2$.   
The numerically generated unstable manifold of this map can be seen in the left panel of
Fig.~\ref{fig.R4-1}, and confirms the fractal property. The right panel shows the projected $c$-measure which clearly exhibits a jump at
$y=1/2$. The plateau values agree well with the theoretical predictions.

Had we taken the leak with $R_1=0$ ($R_2=R_3=R_4=1$), the same result would have been obtained.
The situation is different, however, for $R_2=0$ or $R_3=0.$
In the first case, we see from (\ref{eq.newc}) that $c=0$, implying vanishing
c-measure for the entire upper half-square. The escape rate is then 
$\gamma=\ln{2}$ which implies $D_1^{(s)}=1$. Fractality is then lost, 
the stable and unstable manifolds are one-dimensional. The dynamics is fully leaked,
there is no chaos, 
a single unstable fixed point, the one at $(0,0)$, governs the escape dynamics 
(hence $\gamma=\bar{\lambda}$).  For $R_3=0$, $c=2$, the c-measure vanishes in the
bottom half square, the situation is otherwise the same. 
These observations again illustrate that the location of
a leak of the same area is very important, even chaos can be lost if they do
not overlap with a period-1 orbit. 
  
Now, we consider the more general case of a partial leak of arbitrary reflection 
coefficient  
over the uppermost strip: $R_4\le 1$ ($R_1=R_2=R_3$=1). The solution of (\ref{eq.newgamma}), (\ref{eq.newc}) leads to
$$e^{\tilde{\gamma}}=2-c, \;\;\; c= \frac{R_4-3\pm \sqrt{R_4^2-2R_4+5}}{R_4-1}.$$
We took here the $+$ root because for the $-$ root the density $2-c$ for $y\in[0,0.5]$ would be negative. 
{Using the result above and  Eq.~(\ref{eq.newgamma}) with $R_1=R_2=R_3=1$ it is not difficult to confirm the validity of Eq.~(\ref{eq.kappacorrect}) for this partial leak case.}
As expected, for $R_4\rightarrow1$ we obtain $c=2-c=1$ and $\tilde{\gamma}=0$.

\section{Implications in strongly chaotic systems}\label{sec.fullchaos}\label{sec.4}
\begin{figure*}[ht!]
\includegraphics[width=1.8\columnwidth]{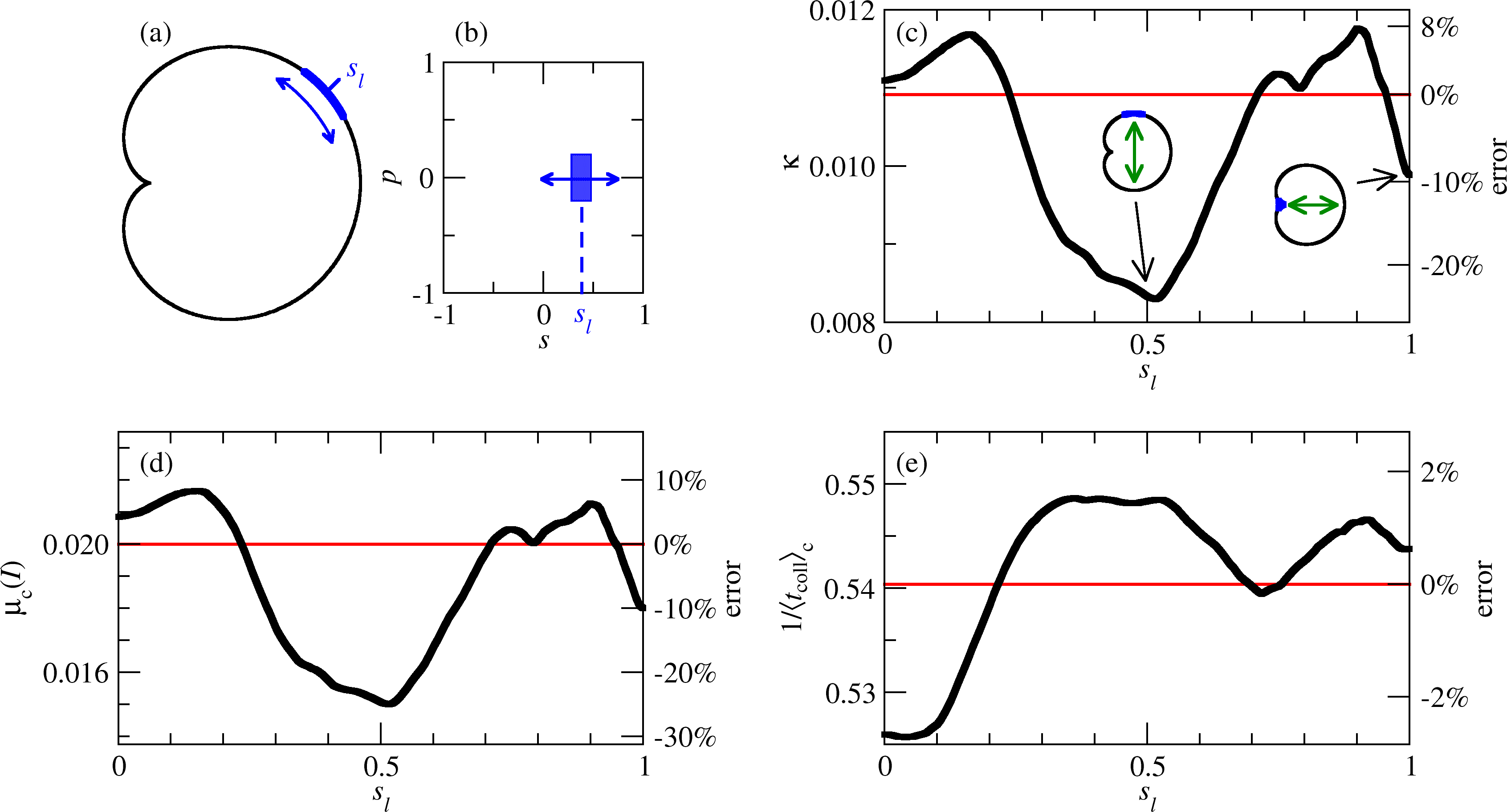}
\caption{(Color online) Dependence of the escape rate on the position of the leak. (a) Cardioid billiard with a leak centered at some
  position $s_l$ with $p_l=0$. (b) Phase
  space representation showing the leak $I=[s_l-\Delta s, s_l+\Delta s] \times [p_l -\Delta p,p_l+\Delta p]$ with $\Delta p=0.2,
  \Delta s=0.1, p_l=0,$ and $s_l\in[0,1]$. (c) The escape rate~$\kappa$ obtained 
  from numerical simulations as the ones in Fig.~\ref{fig.robnik1}. (d) The c-measure of the leak~$\mu_c(I)$ and (e) the inverse of the mean
  collision time. Results in (d) and (e) were calculated from $\rho_c(s,p)$ as described in Appendix~\ref{sec.ap-algorithm}. Using these
  values to compute $\kappa$ through Eq.~(\ref{eq.kappa}) leads to results undistinguishable from those of  (c).  
The estimates based on the closed billiard theory of Sec.~\ref{ssec.closed} are shown as horizontal lines and correspond to: in (c)
the naive estimate $\kappa^*$ given by
  Eq.~(\ref{eq.kappastar}), in (d) $\mu(I)=\Delta s \Delta p$, and in (e) $\tcoll$ given by Eq.~(\ref{eq.tau2d3d}). The y axis on the right edges 
  of  (c)-(e) indicate the relative deviation between these values and the actual data.}
\label{fig.kappasl}
\end{figure*}

\begin{figure}[h!]
\includegraphics[width=0.9\columnwidth]{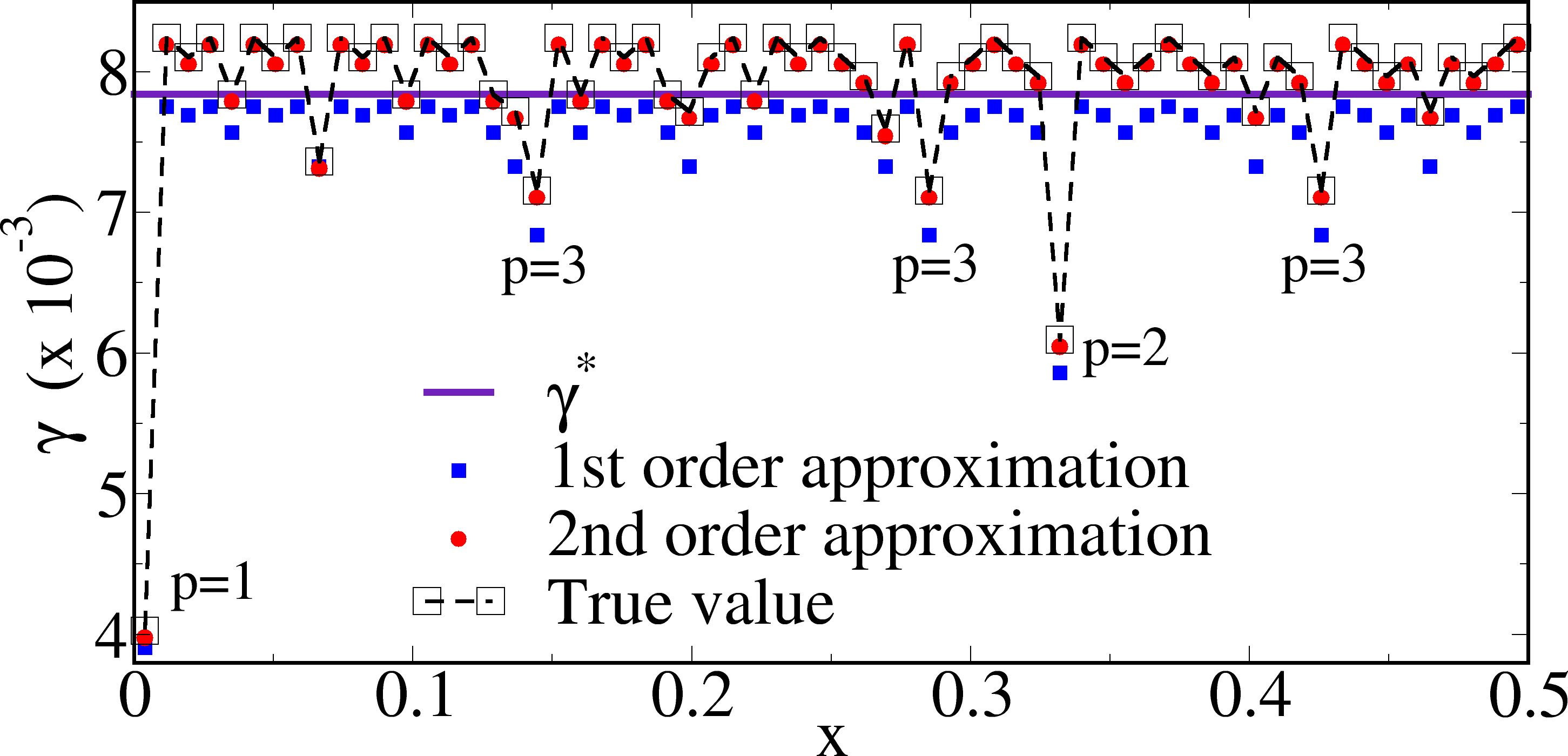}
\caption{Dependence of the escape rate~$\gamma$ on the location of the leak in the strongly chaotic doubling map: $x_{n+1}= 2 x_n (\mod 
   1)$. Leaks have $\mu(I)=2^{-7}$ and are placed in non-overlapping positions starting at $x=0$ (Markov partitions). 
    The period $p$ of the shortest unstable periodic orbits are marked at the positions $x$ of leaks containing these orbits. The
    approximations to the true~$\gamma$ (squares) correspond to: $\gamma^*=-\ln(1-\mu(I))$ as in~(\ref{eq.kappastar}) (horizontal line),
    Eq.~(\ref{eq.periodic}) (squares, first order), and the corrected results obtained by~\cite{Orestis2012} (circles, second order). (Data by
    O. Georgiou). }
\label{fig.doubling}
\end{figure}

\subsection{Dependence of the escape rate on the leak}\label{ssec.dependence}

The main message of Sec.~\ref{sec.2}  is that for systems with finite leaks 
the naive theory based on the close
dynamics differs from the correct theory based on open systems.
One of the most 
striking and best studied effects arising due to this difference is the dependence of the
escape rate on the position of a fixed-size leak~\cite{Paar1997,Schneider2002,Altmann2004,Bunimovich2007,Afraimovich2010,Bunimovich:2011,Demers:2011}. This result is shown in Fig.~\ref{fig.kappasl}
for the cardioid billiard with a leak~$I=[s_l-\Delta s,s_l+\Delta s] \times [p_l-\Delta
p,p_l+\Delta p]$ with fixed size~$\Delta s=0.1,\Delta p=0.2, p_l=0,$ and different positions~$s_l$. Physically this type of leak, illustrated in Fig.~\ref{fig.kappasl}(a,b), could be realized in optical systems by
replacing the perfect mirror boundaries in the  region $[s_l-\Delta s,s_l+\Delta s]$ by  dielectric
material with refraction index~$n=1/\sin(\Delta p)$. The results in Fig.~\ref{fig.kappasl}(c) confirm the nontrivial dependence of~$\kappa$ on~$s_l$, which can take
values both smaller and larger than the naive estimation~$\kappa^*=-\ln(1-\Delta s \Delta p)/\tcoll$ given by Eq.~(\ref{eq.kappastar}).
The theory developed in
Sec.~\ref{ssec.transient} tells us that $\kappa$ depends on $\mu_c(I)$ and
$\tcoll_c$ 
through the improved formula Eq.~(\ref{eq.kappa}).
Figure~\ref{fig.kappasl}(d,e) show that both factors $\mu_c(I)$ and
$\tcoll_c$ contribute to~$\kappa \neq \kappa^*$  but  the
variation of~$\mu_c(I)$ is the stronger factor in the dependence of~$\kappa$ on $s_l$ (at
least for this size of the leak). The results also indicate that the improved formula~(\ref{eq.kappa}) provides an excellent approximation of the
  numerical results [and to (\ref{eq.escaperate})] in this example.

\begin{figure*}[!ht]
\includegraphics[width=1.8\columnwidth]{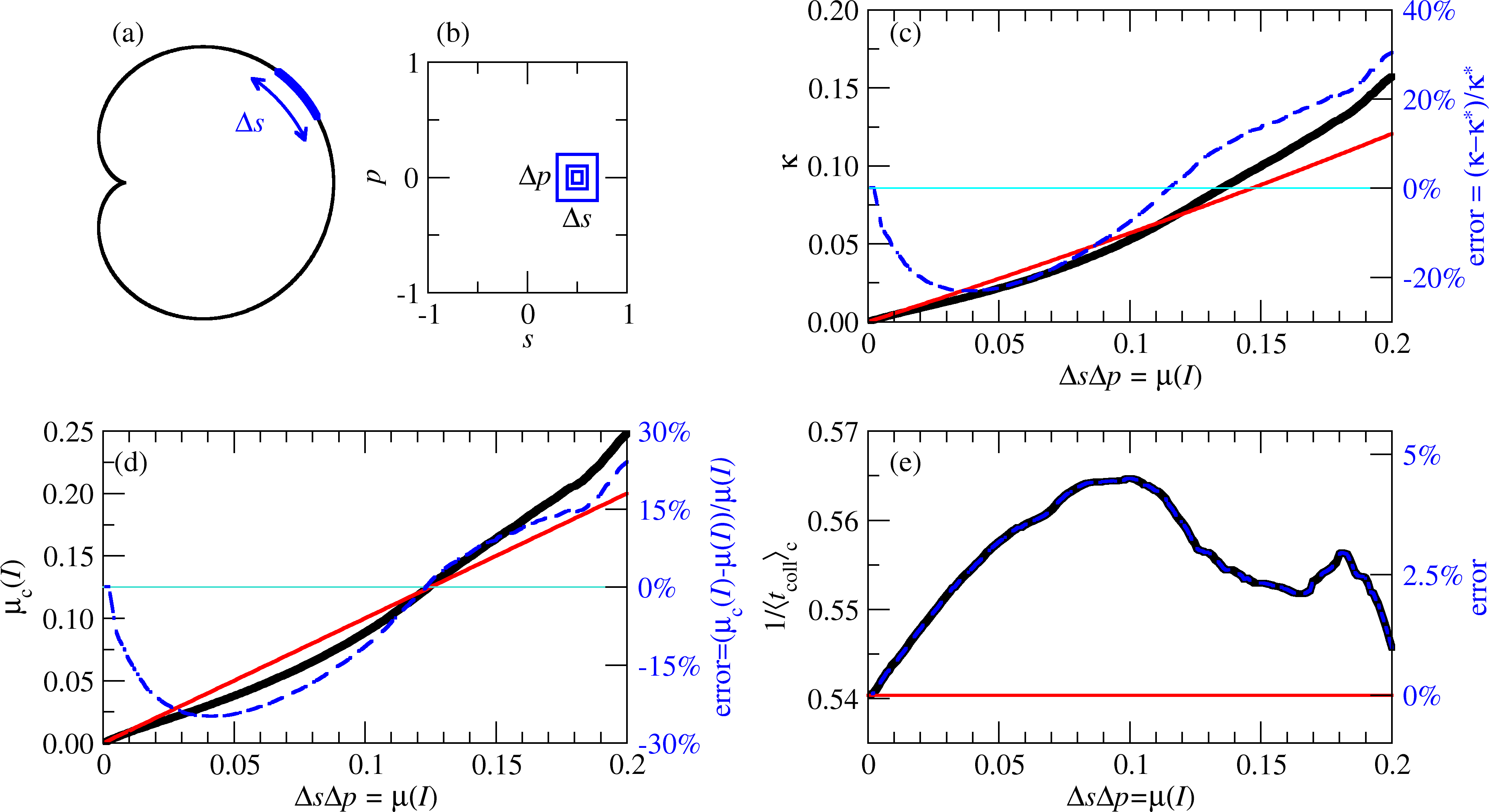}
\caption{(Color online) Dependence of the escape rate on the size $\Delta s \Delta p$ of the leak. (a) Cardioid billiard with a leak centered at $s_l=0.4, p_l=0$
  and variable size. (b) Phase
  space representation showing the leak $I=[s_l-\Delta s, s_l+\Delta s] \times [-\Delta p,+\Delta p]$, with $\Delta p=
  \Delta s \in [0,1/\sqrt{5}]$. (c) Escape rate~$\kappa$ as a function of $\Delta s\Delta p$. The improved estimate~(\ref{eq.kappa}) again provides a
    good approximation to the numerical results.
  (d) The c-measure of the leak~$\mu_c(I)$ and (e) the inverse of the mean
  collision time. Results from (d) and (e) were calculated from $\rho_c(s,p)$ as described in Appendix~\ref{sec.ap-algorithm}. 
    Black symbols: numerical results.
  Thin (red) lines: estimates based on the closed billiard theory of Sec.~\ref{ssec.closed}: in (c) the naive estimate $\kappa^*$ given by
  Eq.~(\ref{eq.kappastar}), in (d) $\mu(I)=\Delta s \Delta p$, and in (e) $\tcoll$ given by Eq.~(\ref{eq.tau2d3d}). Blue dashed line:
  relative deviation between thin and thick (black and red) lines, see the $y$ axis at the right edge of panels (c)-(d).
 }
\label{fig.kappadeltas}
\end{figure*}

Apparently the first to report the dependence of the maps escape rate~$\gamma$ on leak position were 
\cite{Paar1997}. The most pronounced effect shown 
for the doubling map, see Fig.~\ref{fig.doubling}, was the relationship between small values of~$\gamma$ and positions of the
leak around short periodic orbits of the system (see also~\cite{Altmann2004,Bunimovich:2011}). This effect is less pronounced but also visible in
Fig.~\ref{fig.kappasl}, where local minima of the~$\kappa$ vs. $s_l$ curve are obtained when the leak is
placed around the lowest periodic orbits of the billiard: the horizontal orbit at
$(s=0,p=0) \mapsto (s=-1,p=0)$  and the vertical orbit $(s=-0.5,p=0)\mapsto(s=0.5,p=0)$.

An intuitive explanation of these results is found by looking at the images and preimages of leak~$I$~\cite{Paar1997,Buljan2001}. The surviving
trajectories at iteration~$n$ correspond to all trajectories that are not in any of the $n$
preimages of~$I$. Now, if preimages overlap repeatedly, there are more surviving trajectories for increasing~$n$, and therefore the escape rate is smaller.
The overlap between the leak and its images (or preimages) is obviously increased when the leak is around periodic orbits.
We already learned in Sec.~\ref{ssec.transient} that the escape rate can be calculated in terms of the periodic
orbits inside the leak through Eq.~(\ref{eq.upo3}). In this formalism, orbits with low
period play an important role, with weights inversely proportional to their instability.
If periodic orbits with low period (in particular those with
small expansion rates) are in the leak, they appear in the sum~(\ref{eq.upo3}) and reduce~$\gamma$.

This qualitative argument has been rigorously extended for different classes of strongly chaotic systems, such as $1D$ expanding linear
maps~\cite{Afraimovich2010,Bakhtin:2011,Bunimovich:2011,Bunimovich:2012,Keller:2009,Fergusson:2012}. An important feature in these approaches is that all
systems investigated admit a Markov partition and the holes are {\it chosen} to coincide with one element of the partition. For small leak sizes, a fractal
dependence of the escape rate~$\gamma$ has been observed~\cite{Bunimovich:2011,Knight:2011} (see also~\cite{Altmann2010}). As noticed
by~\cite{Bunimovich:2011,Bakhtin:2011}, the position dependence of~$\gamma$ might be so strong that 
 there are cases in which a hole two times {\em larger} than another one can have a {\em smaller} escape rate.
These results can also be understood in the exact expansions developed and applied to strongly chaotic
billiards with finite but small leaks
in~\cite{Bunimovich2007,DettmannSmall}. 

Beyond the Markovian approach, an alternative explanation based on recurrence times and Kac's lemma appears in~\cite{Altmann2004}. 
While a great number of analytical results can be obtained in $1$D Markov systems, the results shown above in the cardioid billiard
(area-preserving true-time map)
appear to show a smoother dependence on position.

We  mention that strong dependences are observed also by varying other parameters of the leaks such as the \emph{orientation} of (asymmetric) leaks, as observed for Hamiltonian
systems in~\cite{Schneider2002}. Interestingly, similarly complicated
even fractal dependences with the position of the leak have also been observed in the diffusion coefficient~\cite{Klages:book,Knight:2011}.
Exponential decay and leaks with different shapes were
investigated in detail in the periodic Lorentz gas by~\cite{Demers:2010}

Complementary to the position-dependence of~$\kappa$, the
dependence on the leak size is illustrated in
Fig.~\ref{fig.kappadeltas}~\cite{Schneider2002,Altmann2004,Bunimovich2007}.
 We change the size of the leak
$I=[s_l-\Delta s,s_l+\Delta s] \times [p_l-\Delta p,p_l+\Delta p]$ by changing~$\Delta p=\Delta
s$ at a fixed $s_l=0.4, p_l=0$, as illustrated in Fig.~\ref{fig.kappadeltas}(a,b). The dependence of~$\kappa$
on~$\Delta s$ depicted in Fig.~\ref{fig.kappadeltas}(c) follows roughly the dependence
of~$\kappa^*$ (red line in c) but a nontrivial behavior is observed apart from this trend. Again, both the
measure~$\mu_c(I)$ shown in Fig.~\ref{fig.kappadeltas}(d) and the mean collision
time~$\tcoll_c$ shown in Fig.~\ref{fig.kappadeltas}(e) are clearly
different from the closed system prediction (red lines).  The oscillations in $\mu_c(I)$
are stronger than the ones in  $\tcoll_c$ (at least for the 
leak sizes considered here) and provide the strongest contribution to~$\kappa \ne \kappa^*$. 
In 1-D piecewise-linear chaotic maps it was shown that the main properties (escape rate, entropy, fractal
  dimensions) of the leaky map vary nonsmoothly with the leak size (as a Devil's stair-case)~\cite{Zyczkowski:1999,Lai:1999,Demers:2011} and position~\cite{Orestis2012}. Note that in view of the
  dependence of $\kappa$ on the leak position discussed above, one can easily find situations in which {\em larger} leaks have {\em smaller}
  escape rates not only in the case of Markov leaks.

One particularly important limit is the case of vanishingly small leaks~$\Delta s, \Delta
p \rightarrow 0$. This is the traditional limit mathematicians are interested in (e.g.,
in the context of Poincar\'e recurrences)~\cite{Haydn:2005}. 
It is interesting to see what happens with the position dependence of $\gamma$ in the limit $\mu(I)\rightarrow 0$. For the
  case of the doubling map with Markov leaks, the escape rate~$\gamma$ depends only on the periodic orbit of lowest period $p$ inside the leak and is
  given in leading order by (see e.g.~\cite{Keller:2009,Bunimovich:2012})
\begin{equation}\label{eq.periodic}
\gamma = \mu(I)(1-2^{-p}),
\end{equation}
for arbitrary small $\mu(I)$. In this limit almost every leak position will have $p\rightarrow\infty$ so that
Sabine's result is recovered in $\mu$-almost every case. Indeed, also the results  in
Fig.~\ref{fig.kappadeltas}(c-e) show that all quantities converge to the closed systems
prediction (red lines) in this limit. The dashed line in Fig.~\ref{fig.kappadeltas}(c) shows
that the relative difference $(\kappa-\kappa^*)/\kappa^*$ is of the order of~$10\%$ for
$\mu(I)\approx0.1$ but that it also consistently decays for small~$\Delta s$. Altogether, these convergences are physically relevant,
particular manifestations of the more general convergence~$\rho_c(s,p) \rightarrow \rho_\mu(s,p)$ for
$\mu(I)\rightarrow 0$, discussed in Sec.~\ref{ssec.transient}. 
For finite but small leaks, an approximation which improves~(\ref{eq.periodic}) was obtained by~\cite{Orestis2012} for maps with complete
symbolic dynamics, see Fig.~\ref{fig.doubling}. It considers not only the period $p$ but also the full symbolic dynamics of the periodic orbits inside $I$ and shows that
the average of~$\gamma$ taken over all (Markovian) leak positions is {\it larger} than $\mu(I)$ (and $\gamma^*$), contrary to 
what~(\ref{eq.periodic}) suggests. 
An alternative approach which leads to an optimal expansion of~$\kappa$ versus the leak size was developed by~\cite{Bunimovich2007} and
applied to billiards. See also~\cite{BunimovichWebb:2012,Cristadoro:2012} for recent alternative approaches.

\subsection{Multiple leaks and basins of escape}\label{ssec.basins}

The idea of introducing more than one leak into the system is very natural in numerous circumstances~\cite{Bleher1988}. Quantum systems often have more than one
leak due to input/output/transmission/reflection/antennas, and multiple leaks in chaotic systems have also been considered~\cite{Buljan2001,Bunimovich2007,Bunimovich2005,Orestis2011,Portela2007}. The results of the previous sections show that the
c-measure depends sensitively on the position of the leak. As a consequence, the results for multiple leaks are not only different from
those obtained in the closed
system approximation, but also cannot be easily obtained from the results for each leak alone. 
For instance, with
two leaks $I_1$ and $I_2$, the escape rate is, in general, different from the sum of  
the single leak case:
\begin{equation} \label{eq:ld_2kappa}
\kappa(I_1 + I_2)\neq \kappa(I_1) + \kappa(I_2).
\end{equation}
This results is not surprising in view of the nontrivial dependence of $\kappa$ on leak size, reported in Fig.~\ref{fig.kappadeltas}, in
which case it was clear that doubling the leak size does not imply doubling $\kappa$. 
Indeed, the difference reported in Eq.~(\ref{eq:ld_2kappa}) has been understood in terms of the overlap of the preimages of the
two leaks \cite{Buljan2001,PP:2003}, and has been
developed more systematically by Bunimovich and Dettmann. {They found that $\kappa(I_1+I_2)$ can be expressed as $\kappa(I_1)+\kappa(I_2)$ plus
a series of correlation terms (with decreasing importance) 
\cite{Bunimovich2007}.}

\begin{figure}[!hb]
  \includegraphics[width=0.8\columnwidth]{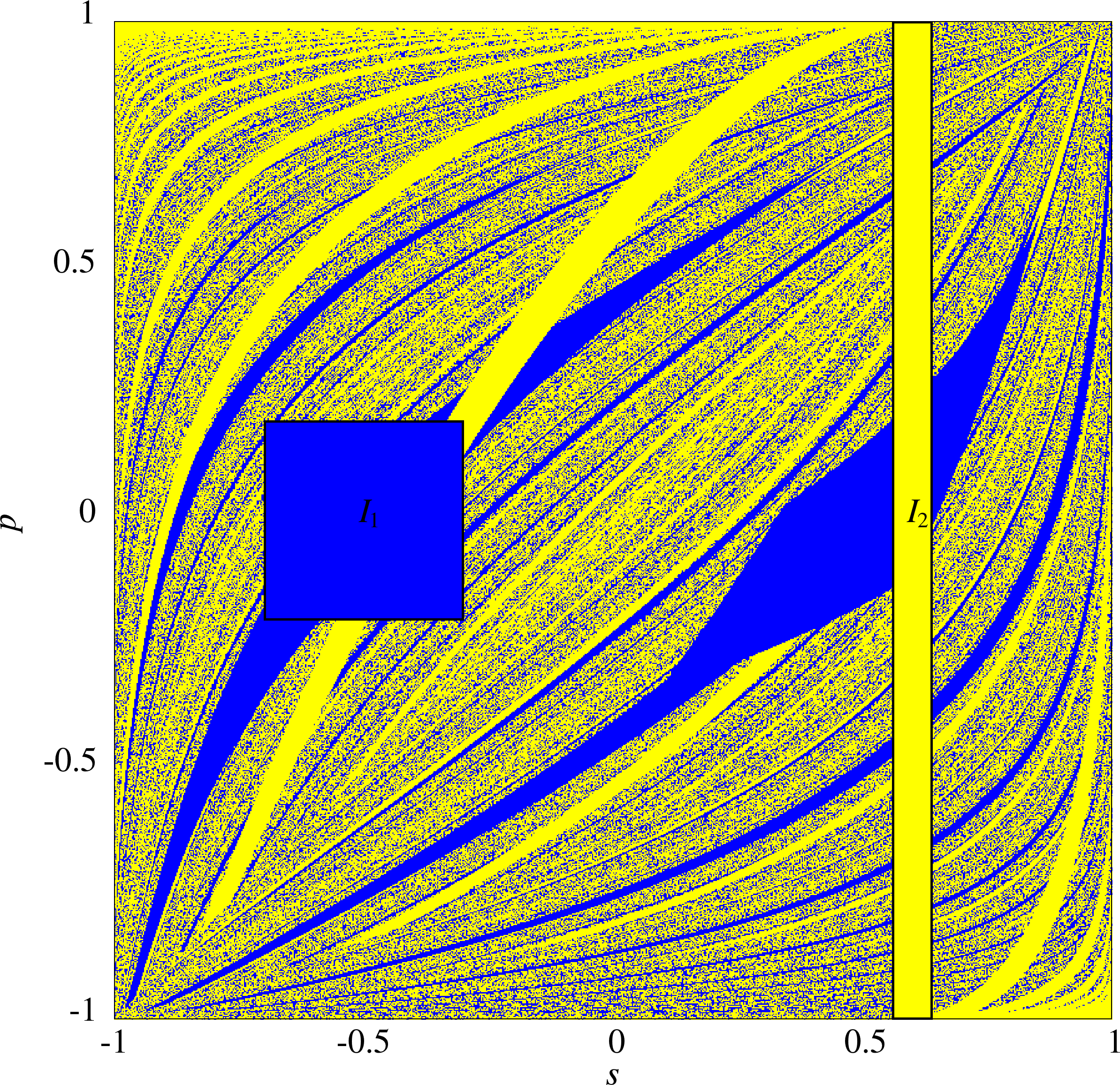}
\caption{(Color online) Escape basins for two leaks in the cardioid billiard. 
Leak~$I_1$ is centered at $s_l=-0.5$, $p_l=0$ with $\Delta s=\Delta p=0.2$, and $I_2$ at
$s_l=0.6, p_l=0$ with $\Delta s=0.04$ and no restriction in the collision angle is applied
(i.e., $\Delta p=1$). Both leaks have the same area in the phase space, nevertheless, $41\%$ of the trajectories escape through leak~$I_1$
(blue/dark gray region) while
$59\%$ of trajectories escape through leak~$I_2$ (yellow region) when both leaks are open. Different characteristic 
values obtained for this system are reported in Tab.~\ref{tab.2l}. }
\label{fig.basin2}
\end{figure}

\begin{table*}
\begin{center}
\begin{tabular}{|c||c || c || c | c | c|}
\hline
{Leaks open during experiment:}
&{Only $I_1$}
&{Only $I_2$}
&\multicolumn{3}{|c|}{Both $I_1$ and $I_2$}
\\\hline
{Measurement ($\downarrow$) applied in leak $I_j=$ ($\rightarrow$)}
&{$I_1$}
&{$I_2$}
&{$I_1+I_2$}
&{$I_1$}
&{$I_2$}
\\\hline
{$\mu(I_j)$}
&{0.04}
&{0.04}
&{0.08}
&{0.04}
&{0.04}
\\
{$\kappa$}
&{0.01607}
&{0.02360}
&{0.03776}
&{0.03780}
&{0.03773}
\\\hline
{$\mu_c(I_j)$}
&{0.02856}
&{0.04332}
&{0.06697}
&{0.02766}
&{0.03931}
\\
{$\tcoll_c$}
&{1.8059}
&{1.8799}
&{1.8424}
&{1.8424}
&{1.8424}
\\\hline
{$\mu_c^{\dagger,j}(I_j)$}
&{0.02856}
&{0.04332}
&{0.06697}
&{0.06999}
&{0.06500}
\\
{$\tcoll_c^{\dagger,j}$}
&{1.8059}
&{1.8799}
&{1.8424}
&{1.9273}
&{1.7869}
\\\hline
\end{tabular}
\caption{ Measurements in the system depicted in Fig.~\ref{fig.basin2}.  
The superscript $\dagger,j$ in the last two rows indicates that the c-measure was restricted to the set of points~{\bf $x$} which escape
through leak~$I_j$. The $\kappa$ in the three last columns are consistent with each other and with the values obtained from 
  $\kappa_1 = -\ln(1-\mu_c)/\tcoll_c$ with 
 $\mu_c^{\dagger,j}(I_j)$,  $\tcoll_c^{\dagger,j}$
  (the values in the last two rows of these columns). 
 Obviously, $\mu_c^{\dagger,j}=\mu_c$
    whenever only $I_j$ is open ($I_j=\{I_1,I_2,I_1+I_2\}$, compare rows $5,6$ to $7,8$.
For the computation 
  procedures, see Fig.~\ref{fig.convergence} (for the $\dagger,j$ cases the set $S(t^*)$ was divided in two subsets according to the leak through
  which trajectories escape). Error bars are of order $5$ in the last digit.
}\label{tab.2l}
\end{center}
\end{table*}

In contrast to the escape rate, the dimensions of the invariant sets of the system opened with multiple leaks can be estimated from the 
dimensions of the invariant sets with single leaks. Consider the case of two leaks~$I_1$ and~$I_2$. The saddle (or the manifolds) obtained for
the case in which $I_1$ and~$I_2$ are simultaneously opened corresponds to the intersection of the saddle (manifold) when
only~$I_1$ is opened with
the saddle (manifold) obtained when only~$I_2$ is opened. Very generally, fractal dimensions $D(I_1+I_2)$ of the intersection is~\cite{Falconer-book}
$$ D(I_1+I_2) = D(I_1)+D(I_2)-D_{\text{embedding}},$$
where $D(I_1)$ ($D(I_2)$) is the dimension of the set 
obtained when only $I_1$ ($I_2$) is opened and
$D_{\text{embedding}}$ is the dimension of the embedding space. For full leaks, the dimension and the escape rate are related to the Lyapunov
exponent through relation~(\ref{KG1}).
The argument above can trivially be extended to more then two leaks. It is not valid, however, in cases when the saddles of the two leaks are trivially
connected to each other (e.g., $I_2\subset I_1$ or $I_2$ is an image of~$I_1$).

The interesting phenomenology of multiple leaks is better illustrated through an example. We consider the cardioid billiard with two leaks, $I_1$ and
$I_2$, of the same size $\mu(I_1)=\mu(I_2)=0.04$ but at different positions, as depicted in Fig.~\ref{fig.basin2}. We first consider the effect of each
of these leaks separately. Following the 
procedures described in Fig.~\ref{fig.kappasl}, we obtain the results reported in the first two columns of Tab.~\ref{tab.2l}.

We now consider the case when both leaks are simultaneously open. The result reported in Tab.~\ref{tab.2l} confirms
inequality~(\ref{eq:ld_2kappa}). Similarly, we  compute $\mu_c(I_i)$, the c-measure for both leaks open calculated at $I_i$.
 The mean collision time is not defined with respect to a given leak, as it depends on~$\rho_c$ in the whole phase space, hence the
 identical values in Tab.~\ref{tab.2l}. 
A natural question is that of the nature of the sets of initial conditions which
escape through each $I_i$, i.e., of the properties of the escape basins $B_i$, as the ones depicted in Fig.~\ref{fig.basin2}. 
 The border between the two escape regions contains 
the stable manifold of the chaotic saddle~\cite{LaiTel-book}.
Notice that the total area of the basins correspond to the amount of initial conditions that escape through each leak, while $\mu_c(I_1)$
and $\mu_c(I_2)$ are proportional to the rate of escape through each leak for large times. 
Note that the escape rate should be independent of the leak through which the particle flux is monitored when both leaks are opened
  simultaneously. The obtained $\kappa$ values are indeed the same within the numerical precision
(three columns on the right). 
Apart from the standard c-measure (in $\mu_c$ and $\tcoll_c$), which considers
  normalization in the full phase space, we have also computed the c-measure restricted to the set of points $({\bf x})$ which escape
  through leak~$I_j$. These results appear in the last two rows of Tab.~\ref{tab.2l} and carry a superscript $\dagger,j$.
In addition, it is interesting to note that 
all values are consistent with those obtained from
  $\kappa \simeq -\ln(1-\mu_c)/\tcoll_c$ with 
 $\mu_c^{\dagger,j}(I_j)$,  $\tcoll_c^{\dagger,j}$.

It is shown in \cite{Bunimovich:2011} that it is possible to construct examples in which arbitrarily small escape rates are achieved even in the
presence of leaks of arbitrarily large sizes. This surprising claim can be understood intuitively from the results of this section. Starting
from a system with an arbitrarily small leak, consider expanding the leak in such a way that the new leak also contains many images (and/or
preimages) of the original leak. 
This new leak can take an arbitrarily
large proportion of the phase space without affecting the saddle and thus the escape dynamics which is by construction slow.

\subsection{Emission}\label{emission}\label{ssec.emission}

\begin{figure}[!ht]
\includegraphics[width=0.9\columnwidth]{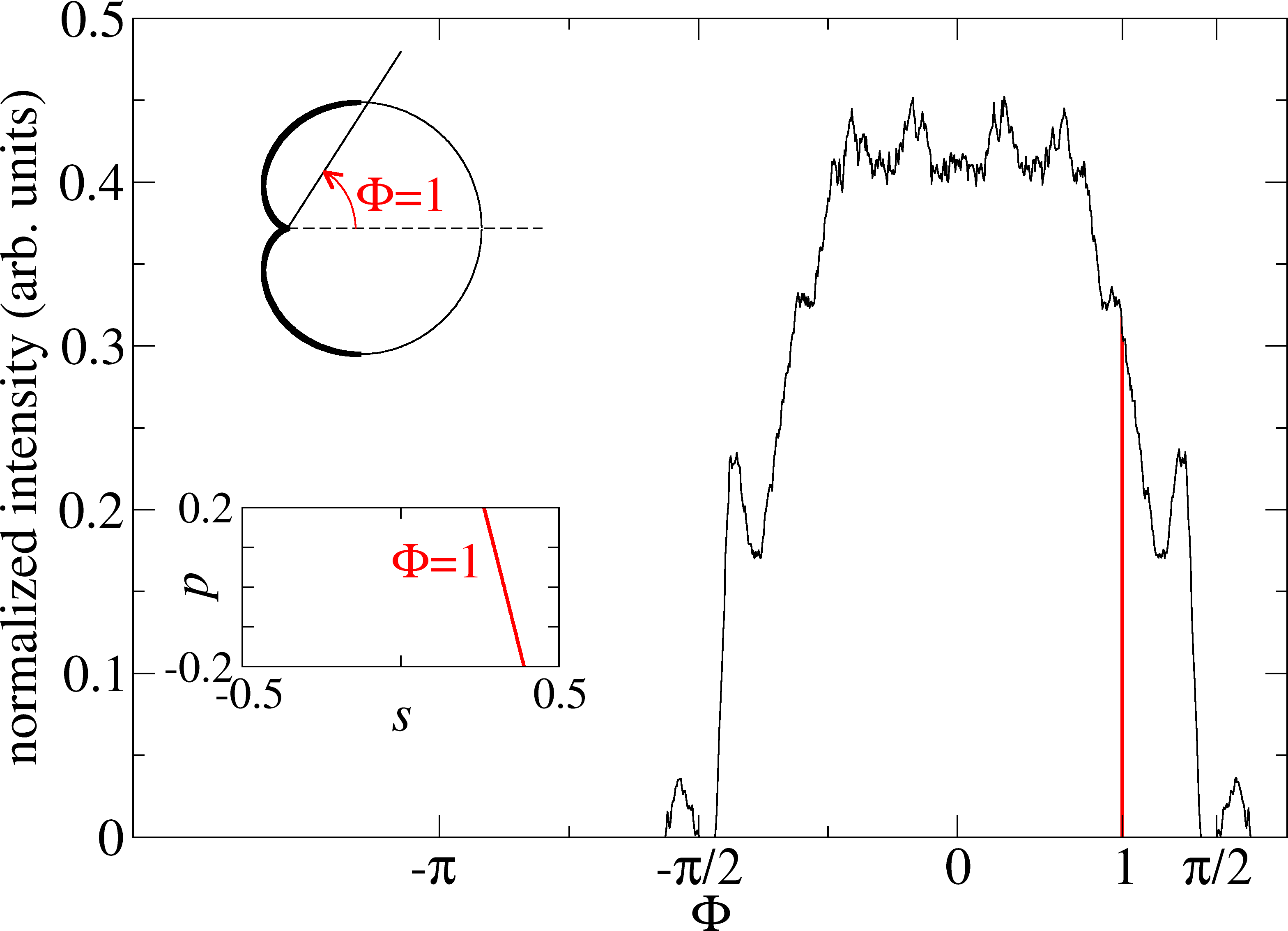}
\caption{(Color online) Far-field emission for the cardioid billiard with a leak $I$ centered
  at   $(s_l,p_l)=(0,0)$, with $\Delta s = 0.5$ and $\Delta p = 0.2$. The far-field intensity distribution is computed by
  collecting the number of trajectories emitted in the asymptotic direction given by the angle~$\Phi$ (see upper inset), measured over the time
  interval $(t_0 \ldots \infty)$ with $t_0 \gg 1$. The lower inset displays the leak with the phase-space positions corresponding to a fixed emission angle
  $\Phi$ (gray/red curve).  The initial density was uniform in the full phase space~$\rho_0({\bf x}) = \rho_\mu({\bf
    x})$, $1000$ bins in $\Phi\in[-\pi,\pi]$ were used. }
\label{fig.emission1}
\end{figure}

\begin{figure}[!ht]
\includegraphics[width=0.8\columnwidth]{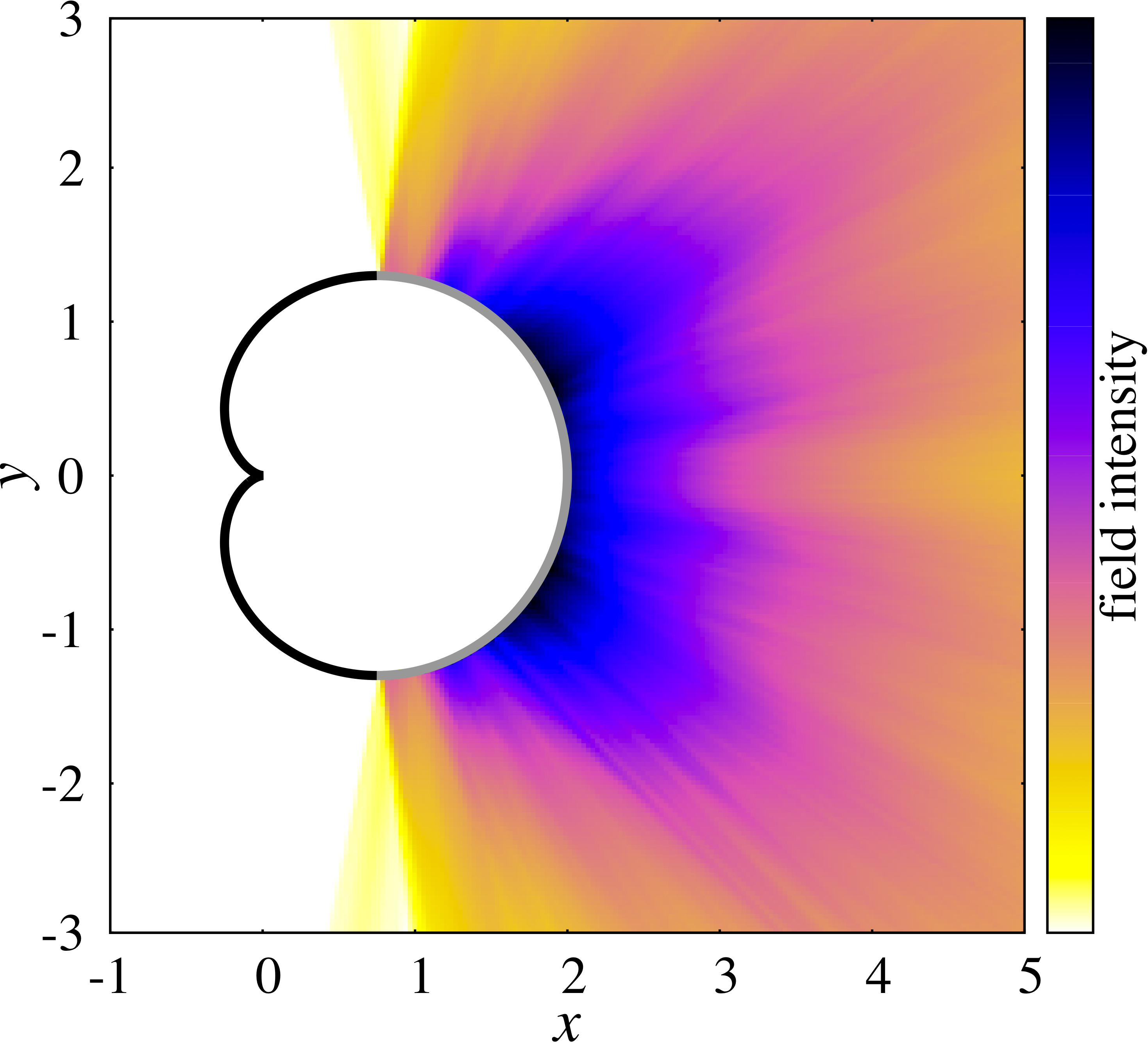}
\caption{(Color online) Emission from the cardioid billiard described in Fig.~\ref{fig.emission1}
over the full configuration space.  The color code indicates the density of trajectories outside the
  billiard at large times~$t>50$. Fig.\ref{fig.emission1} is obtained by summing up all the intensities in a given direction. $10^8$ initial
  conditions were used and data are presented on  a grid of $200 \times 200$. }
\label{fig.emission2}
\end{figure}

The most natural observable quantity in the configuration space of leaking systems
is
the emission of trajectories through the leak. Here we provide a representative configuration in which emission plays an important role, and
connect the observed quantities to our theoretical formulation. 

Consider that detectors are placed around a circle far away from the cardioid billiard,
as usually considered in optical microcavities.
We introduce a leak on the right-hand side of the cardioid billiard 
with $s_l=0, p_l=0$, $\Delta s=0.5, \Delta p=0.2$.
The detectors collect the intensity of light emitted through the leak under different polar angles~$\Phi$ measured from the center of the 
billiard. 
Numerically, we distribute a large number of trajectories 
with an initial density $\rho_0({\bf x})$ 
in the phase space and record the emission angle $\Phi$ of escaped trajectories in the configuration
 space. The proportion of trajectories escaping in a small interval 
 around $\Phi$ is measured. For short times, the distributions are
very irregular and depend strongly on the initial distribution 
$\rho_0({\bf x})$ 
of the trajectories inside the cavity. 
After a transient period of the time, $\rho({\bf x},t) \rightarrow \rho_c({\bf x})$, the total intensity decays exponentially and the shape of the spatial 
distribution remains the same. Figure~\ref{fig.emission1} shows the spatial distribution as a function of $\Phi$, the so-called far-field
emission. A multipeaked nonuniform emission is observed. The  spatial density of light rays for 
$t \gg 1$ is shown in Fig.~\ref{fig.emission2}.

\begin{figure}[!h]
\includegraphics[width=1\columnwidth]{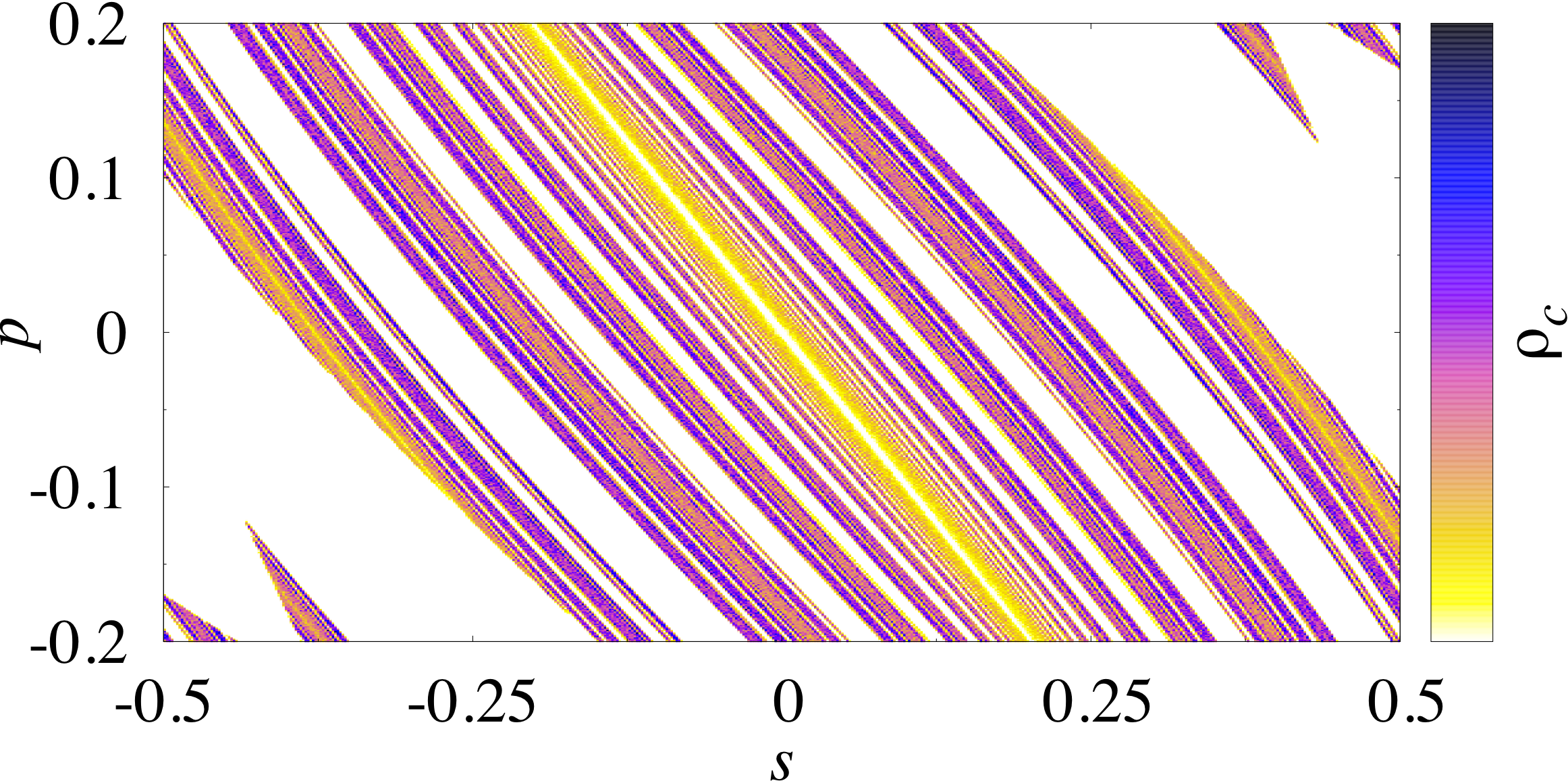}
\caption{(Color online) Invariant density $\rho_c(s,p)$ inside the leak of Figs.~\ref{fig.emission1}, \ref{fig.emission2}. The emission patterns are ultimately determined by $\rho_c(s,p)$, e.g., the zero density around $(s,p)=(\pm0.5,0)$ leads to the zero emission around $\phi=\pm\pi/2$ in Fig.~\ref{fig.emission1}.}
\label{fig.emission3}
\end{figure}

For long times, the emission is fully described by the c-density $\rho_c({\bf x})$ inside the leak (${\bf x} \in I$). For instance, the
emission angle~$\Phi$ is a geometrical function of ${\bf x}=(s,p)$ in the leak, $\Phi=\phi({\bf x})$, and the far-field intensity
$\varrho(\Phi)$ is given by
\begin{equation}\label{eq.projection}
\varrho(\Phi)\sim \int_I \rho_c({\bf x}) \delta(\Phi-\phi({\bf x})) d{\bf x}. 
\end{equation}
Figure~\ref{fig.emission3} shows $\rho_c({\bf x})$ in $I$ for our example.
The leak placed only on the right side of the billiard $s\in[-0.5,0.5]$ and the nonuniform distribution of $\rho_c({\bf x})$ are
responsible for the peaked and nonuniform emission in Figs.~\ref{fig.emission1} and~\ref{fig.emission2}.  
 The filamentary structure of $\rho_c({\bf x})$ inside $I$ seen in Fig.~\ref{fig.emission3}
reflects the filamentary pattern of the unstable manifold of the chaotic saddle, 
as discussed in
Sec.~\ref{ssec.transient}. 
After performing the projection by $\rho_c(s,p)$ as indicated in Eq.~(\ref{eq.projection}), these filaments give rise to the zigzaged far-field
emission shown in Fig.~\ref{fig.emission1}.
 In Sec.~\ref{ssec.microcavities} we shall see that similar emission properties can be 
experimentally observed in lasing microcavities, {in which case the factor $(1-R({\bf x}))$ has also to be included in
  Eq.~(\ref{eq.projection}) in order to account for the partial 
  reflection/transmission property of the leak (see also Sec.~\ref{ssec.partial}).}

\section{Extension to weakly chaotic systems}\label{sec.weak}\label{sec.5}
\subsection{Closed-system phase space}

\begin{figure}[ht!]
\centering
\includegraphics[width=1\columnwidth]{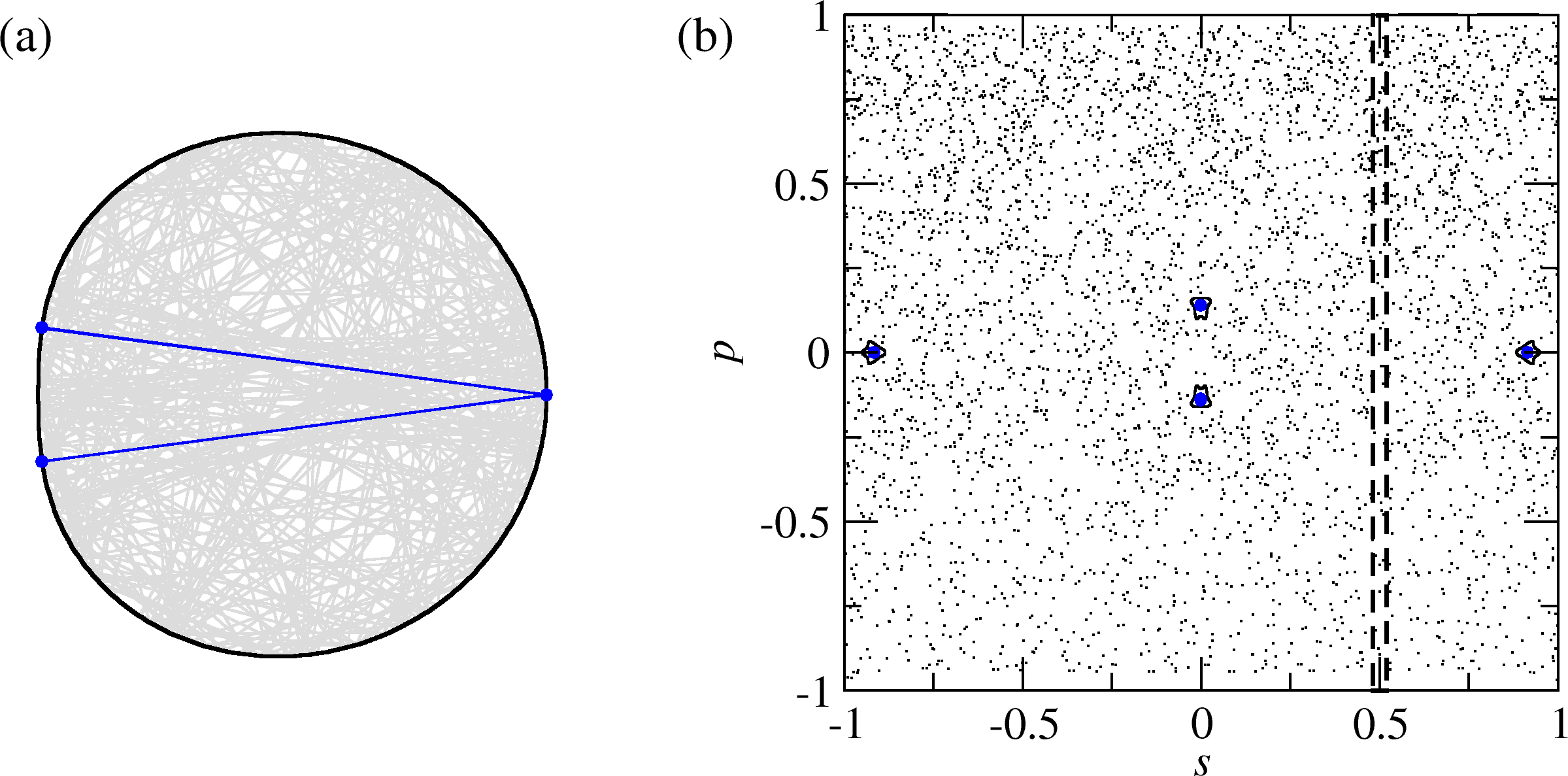}
\caption{(Color online) Lima\c{c}on billiard defined by Eq.~(\ref{eq.limacon}) with $\varepsilon=0.46$. (a) Configuration space and (b) phase space of the
  closed billiard. The blue trajectory (V-shaped curve and $\bullet$) is a stable period-4 orbit around which a KAM island exists. A chaotic trajectory is shown
  as gray lines in (a) and as black dots in (b). The dashed line in (b)
  indicates the leak used in Figs.~\ref{fig.nonhyperbolic2} and~\ref{fig.nonhyperbolic3}.}
\label{fig.nonhyperbolic1}
\end{figure}

So far we have focused on the case of strongly chaotic systems and used the cardioid billiard to illustrate the theory. While
these results apply to a broad class of (Hamiltonian and dissipative) systems, there is an evident need to expand them to the
larger class of weakly chaotic systems. For instance, in closed Hamiltonian systems stable periodic orbits and, around them, quasiperiodic
Kolmogorov-Arnold-Moser (KAM) tori form regions of regular motion in the phase space. These regions coexist with regions of chaotic motion.
The situation is illustrated in Fig.~\ref{fig.nonhyperbolic1} for the lima\c{c}on billiard (\ref{eq.limacon}) with $\varepsilon=0.46$. Such mixed-phase-space systems are generic among all Hamiltonian systems and we focus on this type of nonlinearity.  This is the typical case for billiards whose 
boundaries are defined by arbitrary (smooth) curves. Famous examples are the \limacon{}~\cite{Robnik1983} billiard for any  $\varepsilon \neq 1$ in
Eq.~(\ref{eq.limacon}) and the annular billiard~\cite{Saito:1982}.

The arguments above emphasize that the closed system is not ergodic: the phase-space is divided in multiple independent components. One
can imagine that the dynamics inside any chaotic component are described by the theory above. In reality, the situation is more 
involved because KAM tori are sticky surfaces that can be thought to affect trajectories in the surrounding chaotic component. Because of
the smoothness of the dynamics, the local finite-time Lyapunov exponent close to the border of the KAM islands approaches zero (the Lyapunov
exponent of the tori). A trajectory in the chaotic region that come near some KAM surface, wanders close to that surface for a long
time before leaving it and showing intermittent bursts of chaos.
This effect is called {\em stickiness} and the
dynamics is said to be weakly chaotic (references are given in the next Section). 
Stickiness is typical in nonhyperbolic dynamical systems and can be thought as a consequence of the vanishing local Lyapunov exponent 
in the sticky region. 
The independence of this effect from nonergodicity becomes evident
by noting that even zero measure sets can lead to {\em stickiness} and weak chaos (e.g., the bouncing ball orbits discussed below).

When a leak is introduced in a mixed-phase-space system, only the ergodic components that intersect the leak will be
affected. Here we focus on leaks placed in chaotic components and, accordingly, the estimations of the measure of the leak {$\mu(I)$ has
  to be normalized by the measure of the chaotic sea~$\mu(\Omega_{\text{chaos}})$. This usually involves estimating the measure of the
  regions of regular motion $\mu(\Omega_{\text{regular}})$ and subtracting from $\mu(\Omega)\equiv1$ because typically $\mu(\Omega)=\mu(\Omega_{\text{chaos}})+\mu(\Omega_{\text{regular}})$}. Beyond this trivial correction accounting for the nonergodicity of the system, weak chaos and stickiness is
  manifested in 
the survival probability~$P(t)$. 

\begin{figure*}[!ht]
\centering
\includegraphics[width=1.8\columnwidth]{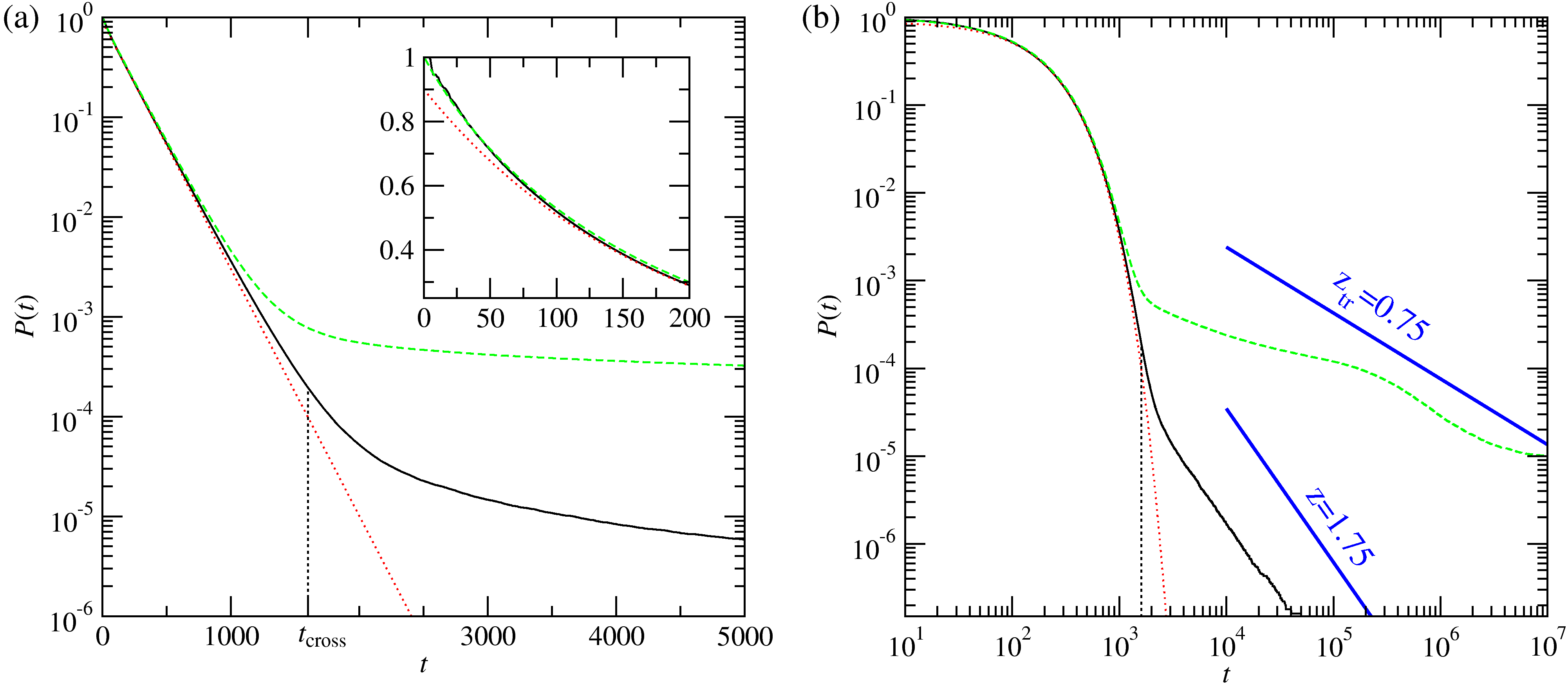}
\caption{(Color online) Survival probability~$P(t)$ inside the \limacon{} billiard depicted in Fig.~\ref{fig.nonhyperbolic1} with a
  leak~$I$ centered around $s_l=0.5,p_l=0$ with~$\Delta s=0.05,\Delta p=1$. Two different initial ensembles $\rho_0({\bf x})$ are taken:
  $\rho_r({\bf x})$
  on the   image of~$I$ (continuous, black); and $\rho_\mu({\bf x})$ uniform in the chaotic component (the regions outside KAM tori) of the closed billiard (dashed,
  green). The exponential curve (dotted, red) fits $P(t)$ between times $1/\kappa$ and $2/\kappa$. 
(a) Log-linear representation; Inset: magnification for short times. (b) Log-log representation. 
$P(t)$ decays 
with intermediate escape rate $\kappa \approx0.0057(3)$,  and with a power-law $P(t)\sim t^{-z}$
for  $t>t_{\text{cross}} \approx 1600$ (see (\ref{eq:scattering_completo})).
Thick blue lines with scalings $z=1.75$ and $z_{tr}=0.75$ are shown for comparison.
  }
\label{fig.nonhyperbolic2}
\end{figure*}

\subsection{Decay of the survival probability in open systems}\label{ssec.weaksurvival}

The most important effect of the presence of sticky regions on the survival probability~$P(t)$ is that it modifies the asymptotic decay from
exponential to power-law $P(t)\sim t^{-z}$, where $z$ is the algebraic decay exponent related to the properties of chaotic regions
close to regular ones.  
This power-law scaling of~$P(t)$ can be related to other observables such as Poincar\'e recurrences~\cite{Chirikov1984,Zaslavsky:2002} (as
in Sec.~\ref{ssec.poincare}), long-term 
correlations~\cite{karney,Chirikov1984,Chirikov:1999}, $1/f$ spectrum~\cite{Geisel:1987}, Lyapunov exponents~\cite{Kantz:1987,Artuso:2009}, and anomalous
transport~\cite{karney,Geisel:1988}. 
The value $z=2$ can be obtained analytically in leaking billiards with bouncing ball or marginally unstable (parabolic) periodic
orbits~\cite{Gaspard:1995,Altmann:MUPOS}, such as 
the Sinai~\cite{Bauer1990,Legrand:1990,Fendrik:1995,Kokshenev:2000},  
the stadium~\cite{Vivaldi:1983,Dumont:new,Alt1996,Armstead:2004,Nagler2007,OrestisStadium},
the mushroom~\cite{Altmann:mushroom,Tanaka:2006,Miyaguchi:2007,OrestisMushroom}, and other 
billiards~\cite{Fendrik:1997,Altmann:MUPOS}, and also for area-preserving maps with sharply divided 
phase-space~\cite{Fendrik:1997,AMK:2006,Akaishi2009}. 
Even a single marginally unstable point in area-preserving map can lead to stickiness~\cite{Artuso:1998}, allowing for a direct connection
to one-dimensional (Pommeau-Manneville type) intermittent maps~\cite{Artuso:2008}. Stickiness also appears in higher
dimensional Hamiltonian systems~\cite{Ding:1990,Fendrik:1995,Altmann.higherN}. In area-preserving maps different stickiness scenarios can be
distinguished~\cite{Zaslavsky:2002,Zaslavsky-book}, 
but there are examples of billiards with divided phase space for which stickiness is absent~\cite{BunimovichNonlinearity}.
For the  generic KAM scenario  
strong fluctuations are observed due to the presence of {\em Cantori}~\cite{Meiss:rmp} acting as partial barrier to the transport of
particles. The universality of $z$ in the KAM scenario is an old problem that has not been completely solved despite different approaches and
substantial advance in the last 30 years~\cite{Meiss:1985,Meiss:1986,Chirikov1984,Chirikov:1999,ketzmerick.comment,chirikov.reply,cristadoro,venegeroles}. The
most recent results~\cite{cristadoro} indicate the universal exponent to be $z\approx1.57$.

Even if  for long times the decay is expected to be power-law due to the tori, there are interesting preasymptotic regimes of $P(t)$ for
systems with mixed phase space.
For very short times $t<t_s$, initial-condition and system-dependent fluctuations are typical and may play a predominant role in specific
applications~\cite{Dumont:new,DietzMushroom,Tanaka:2006,Grete:2007,Altmann2008}.  
Furthermore, when the leak is far away from any KAM tori and the chaotic component is large~\footnote{This situation is 
  important in high-dimensional systems for which the measure of the regular
regions decreases (but is still different from zero).}, typical trajectories will exit before having the chance of
approaching the KAM islands. 
These trajectories will experience an effective hyperbolic system and
render an intermediate-time exponential decay, as reported in different
systems~\cite{Jung1993,Fendrik:1994,Gaspard:1995,Alt1996,Kokshenev:2000,Zaslavsky:2002,Altmann2008,OrestisStadium,OrestisMushroom}.

Based on the different decay regimes discussed above, we can write the survival probability as~\cite{Altmann2008}:
\begin{equation}\label{eq:scattering_completo}
P(t) \approx \left\{ \begin{array}{lll}
    \text{ irregular } & \mbox{for} \;\;\;\; 0<t<t_s,\\
   a e^{-\kappa t} & \mbox{for} \;\;\;\; t_s < t < t_{\text{z}}, \\
   a e^{-\kappa t} + b (\kappa t)^{-z} & \mbox{for}  \;\;\;\; t_{\text{z}} < t,\\
\end{array} \right. 
\end{equation}
where $t_{\text{z}}$ corresponds to the time needed for the first trajectories to approach the sticky region, $a/b$ is
  proportional to the ratio of the measure of the chaotic and of the regular components of the phase space, and $t_s,t_z$ depend on the initial condition~$\rho_0({\bf x})$.  Even if trajectories are
started in the sticky region (see Sec.~\ref{ssec.weakic}), both power-law and exponential regimes are seen because for { large chaotic components $a
  e^{-\kappa} \gg b \kappa^{-z}$}. These regimes are clearly observable in the example of the \limacon{} billiard, 
as Fig.~\ref{fig.nonhyperbolic2} illustrates.

An actual crossover time $t_{\text{cross}}$ between the exponential and the algebraic 
decay can be defined as~\cite{Altmann2008,Akaishi2009}
\begin{equation}
a e^{-\kappa t_{\text{cross}}}= b (\kappa t_{\text{cross}})^{-z},
\label{eq:nckappa}
\end{equation}  
which is the time when the contributions from the hyperbolic and the nonhyperbolic 
components are of {\em equal} importance. 

We can easily estimate the dependence of the crossover time on the size of the leak~$\mu(I)$ assuming that the ratio $b/a$ 
depends at most weakly on $\kappa$ for small $\mu(I)$.  
From Eq.~(\ref{eq:nckappa}) we obtain that~\cite{Altmann2008} 
\begin{equation} \label{eq:scattering_ncmu}
t_{\text{cross}} \sim 1/\kappa \sim 1/\mu(I).
\end{equation} 
A logarithmic correction to this relation was found by~\cite{Akaishi2009}  as
$t_{\text{cross}} \sim 1/\mu(I)-(\log(\mu(I))+1)/\mu(I)$. For exact
calculations and simulations in specific systems see also~\cite{Kokshenev:2000,OrestisStadium,OrestisChaos,Altmann2012}. Taking this correction into account, after the usual rescaling of time $t\mapsto t \kappa$ 
and in the limit
$\mu(I)\rightarrow0$, we find that in the rescaled units $\kappa \rightarrow 1$ and $t_{\text{cross}} \rightarrow \infty$, i.e., the exponential
decay always {\em dominates} $P(t)$. This provides an example of precise statements for infinitely small leaks that mask
an interesting dynamical  phenomenon (the power-law) 
because for any finite leak a transition to
power-law decay exists. Instead, here we do not apply any rescaling and discuss the dependence of the intermediate-time regimes in
Eq.~(\ref{eq:scattering_completo}) on leak size and other parameters. Alternatively, one could take a different rescaling of time, e.g.  $t \mapsto t \mu(I)/\log
\mu(I)$, that would
not suppress the crossover.
 
\subsection{Dependence on the initial distribution}\label{ssec.weakic}

The connection between escape time in open systems and recurrence time in closed system described in Sec.~\ref{ssec.poincare} remains valid
for weakly chaotic system as well. In particular, the intermediate-time decay of $P(t)$ in Eq.~(\ref{eq:scattering_completo}) appears in the
(cumulative) distribution $P_r(T)$ of the Poincar\'e recurrence times $T$: $P_r(T) \sim P(t)_{t=T}$. With initial condition  $\rho_r({\bf x})$ 
[in the image of the leak $I$ taken with respect to
${\bf f}_{\text{closed}}$  as in Sec.~\ref{ssec.poincare})] 
the entire distributions coincide $P_r(T)=P(t)_{t=T}$. In
area-preserving cases $\rho_r$  is a uniform distribution over the image  ${\bf  f}_{\text{closed}}(I)$ of the leak.

The power-law exponent shows an important dependence on the initial condition~$\rho_0({\bf x})$.
In fact, $z$ in Eq.~(\ref{eq:scattering_completo}) characterizes ``scattering'' cases, i.e., situations in which the support of $\rho_0({\bf x})$ is {\em far away} from KAM
surfaces. For~$\rho_0({\bf x})$ inside the sticky region, arbitrarily close to KAM islands,  the escape process -- called {\em transient chaos} situation~\cite{Pik:1992} --  can be shown to be characterized by a survival probability $P_{tr}$ for which
\begin{equation} \label{eq:algebraic_decaytr}
P_{tr}(t) \sim t^{-z_{tr}}, \;\;\;\; 
\mbox{for large}\ t,
\end{equation}
with a different decay exponent $z_{tr}$. As shown by~\cite{Pik:1992,Meiss1997} and explained below,
$z_{tr}$ is smaller than $z$ with unit difference:
\begin{equation} \label{eq:algebraic_zdiff}
z_{tr}=z-1.
\end{equation}
The impact of such a slower decay of the survival probability can be clearly seen in Fig.~\ref{fig.nonhyperbolic2} since the support of
$\rho_r$ remains far away from KAM islands, but this is not the case for $\rho_\mu$.

The similarity between a properly opened up dynamics and the Poincar\'e recurrences
in the closed system  can be used
to explain the difference between the algebraic decay exponents $z$ and $z_r$ \cite{AMK:2006}. 
Consider first initial conditions touching the sticky
region, the case of transient chaos, and examine the time  
a trajectory takes to escape to a region far away from the sticky region. The survival probability distribution $P_{tr}(\tau)$ 
is  proportional to the natural measure $\mu(\tau)$ of the region of the phase space 
to which the trajectories stick for a time longer than $\tau$. Because of ergodicity,
we can write 
\begin{equation} \label{eq:scattering_Ptau}
P_{tr}(\tau) \sim \mu(\tau) = \frac{t_{\tau}}{t_{\text{total}}},
\end{equation}
where $t_\tau$ is the total time spent in the sticky regions (in events with recurrences time $T>\tau$) within  the total observation time
$t_{\text{total}}$, see Fig.~\ref{fig.poincare}).

For the recurrence problem with a single trajectory
of length $t_{\text{total}}$ initialized far away from the sticky region, the cumulative probability
$P_r(\tau)$ to find recurrence times $T$ larger than $\tau$ can be expressed as
\begin{equation} \label{eq:scattering_Prtau}
P_r(\tau) = \frac{N_{\tau}}{N} \sim P(\tau),
\end{equation}
where $N_{\tau}$ is the number of recurrences with recurrence times larger than $\tau$ 
and $N$ is the total number of recurrences observed in the time interval $t_{total}$. The right-hand side expresses the above mentioned
connection between recurrence and escape times.
Since the total observation time can be estimated as $N$ times the
mean recurrence time $\bar{T}$ [Eq. \ref{eq.kac})], 
we have $t_{\text{total}} \sim N \bar{T}$. Similarly, the total time~$t_\tau$ spent inside the sticky region 
is approximately the number $N_{\tau}$ of recurrences {with} times longer than $\tau$ 
multiplied by $\tau$: $t_{\tau} \sim N_{\tau} \tau$.  As $\bar{T}$ is a constant, independent of $\tau$,
these allow us to write
\begin{equation} \label{eq:scattering_PtauP}
P(\tau) \sim \frac{N_{\tau}}{N} \sim \frac{t_{\tau}/\tau}{t_{\text{total}}/\bar{T}}=\frac{t_{\tau} \bar{T}}{t_{\text{total}} \tau}
\sim \frac{P_{tr}(\tau)}{\tau},
\end{equation}
from which the shift of algebraic decay exponent by 1, Eq.~(\ref{eq:algebraic_zdiff}),
immediately follows from the asymptotics of Eqs. (\ref{eq:scattering_completo}) and (\ref{eq:algebraic_decaytr}).

\subsection{Hyperbolic and nonhyperbolic components of chaotic saddles}\label{ssec.saddles}

A theory based on invariant sets has to consider that in this case the chaotic saddle responsible for transient cases  encircle KAM tori, moreover, it comes
arbitrarily close to the tori~\cite{Lau:1991,Christiansen:1993}.
The survival probability in Eq.~(\ref{eq:scattering_completo}) suggests that the invariant sets governing the temporal decay of weakly
chaotic systems can be divided in hyperbolic and nonhyperbolic components~\cite{Jung1993,Fendrik:1994}.
The intermediate-time ($t_s < t < t_z$) exponential decay is the manifestation of the hyperbolic component of the chaotic saddle, and the 
power-law decay, becoming observable for long times $ t > t_{\text{cross}}$, of the nonhyperbolic component.

\begin{figure}[!b]
\includegraphics[width=1\columnwidth]{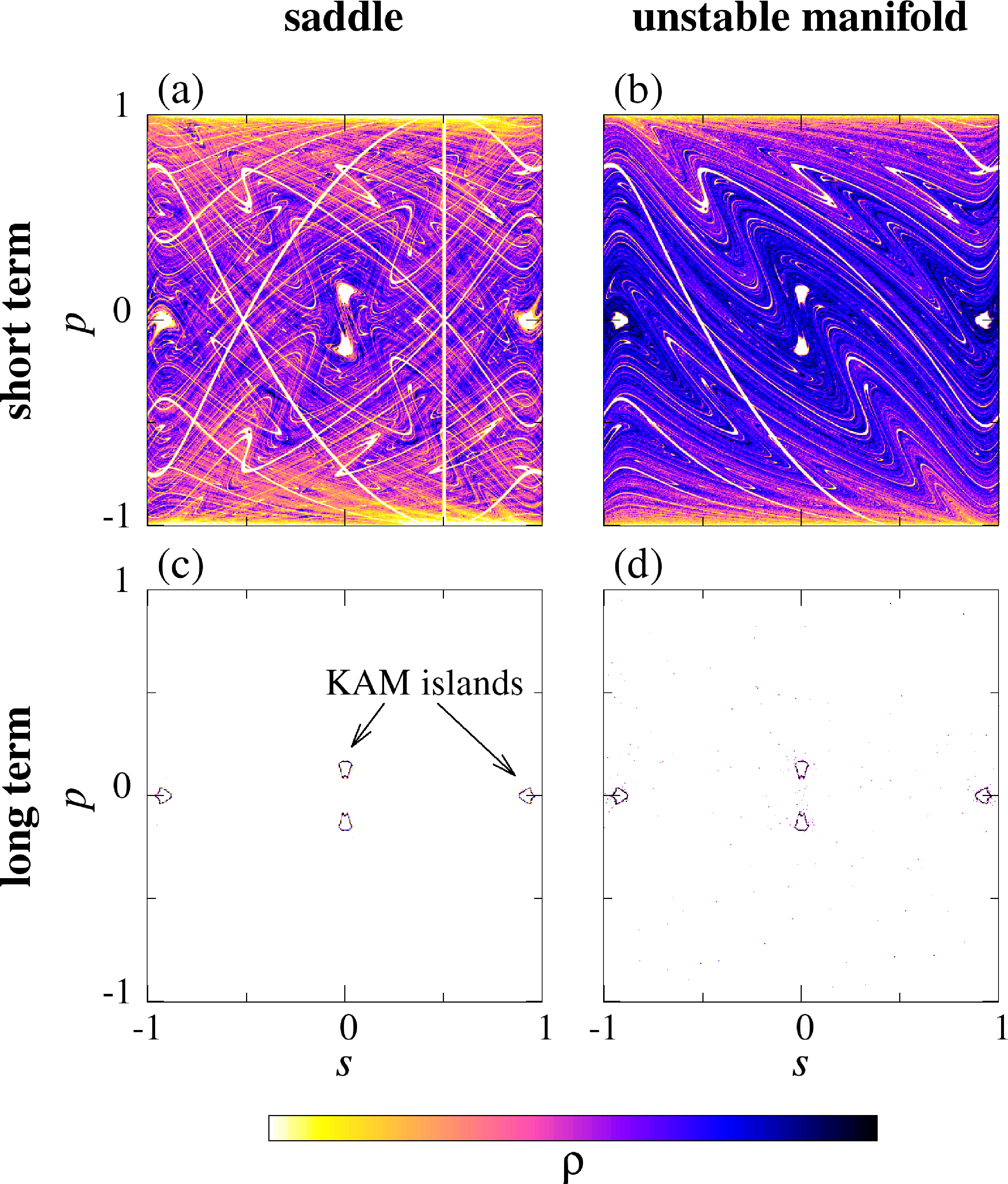}
\caption{(Color online) (a,c) Chaotic saddles and (b,d) the corresponding unstable manifolds computed at different times~$t^*$ for
    the leaky billiard in Fig.~\ref{fig.nonhyperbolic1}.  The color code indicates the phase-space density~$\rho(s,p)$ and is valid for all panels. (a,b) For short times  $t^*\approx 1/\kappa\approx
    175<t_{\text{cross}}\approx1600$ the hyperbolic component of the saddle is dominant and $\rho$ exhibits the characteristic
    fractal/filamentary patterns (compare with Fig.~\ref{fig.saddle2}).  (c,d) For long times $t^*\approx
    2t_{\text{cross}}$ the nonhyperbolic component is dominant and the densities stick to the KAM
    islands responsible for the power-law decay in (\ref{eq:scattering_completo}).  
See Sec.~\ref{ssec.appendixnew} for details on the simulation.
} 
\label{fig.nonhyperbolic3}
\end{figure}

In Fig.~\ref{fig.nonhyperbolic3} we show numerical approximations of the chaotic saddle and its unstable manifold in the hyperbolic and
nonhyperbolic regimes. We  employed the same procedures as used before for the case of strongly chaotic systems, see
Appendix~\ref{ssec.appendixnew}, but varied the effective time $t^*$ used in the simulation. For intermediate values of $t^*$ the exponential decay
dominates $P(t)$ in Eq.~(\ref{eq:scattering_completo}), and the obtained saddle and its unstable manifolds are disjoint from the region containing the KAM island
and show the structure typical of hyperbolic systems. For large times $t^*>t_{\text{cross}}$ the asymptotic dynamics is governed by the
nonhyperbolic regions, and the saddle and unstable manifold concentrate around the islands.

We assumed, as in Sec.~\ref{ssec.transient}, that a single chaotic saddle exists. An interesting and
nontrivial example that violates this assumption was reported by Dettmann and Georgiou in the Stadium billiard with two full
leaks~$I_i$ at the boundary: $I_1$ in the flat and $I_2$ in the circular component of the billiard's boundary~\cite{Orestis2011}. With only
one of these leaks, $P(t) \sim t^{-2}$ as reviewed in Sec.~\ref{ssec.weaksurvival}. Consider now the survival probability $P_i^j(t)$ of
particles that start in leak $i$ and leave through leak $j$ ($i=j$ corresponds to a reflection survival probability and 
$i\neq j$ to a transmission survival probability). From the four $P_i^j(t)$ in the example above, only $P_1^1(t)$ shows the expected power-law tail.
$P_1^2(t),P_2^1(t),$ and $P_2^2(t)$ decay exponentially. This surprising result is a consequence of the sticky region due to the parallel walls in the
Stadium billiard, which cannot be approached from all positions of the phase space if a leak is placed along these parallel walls. 
More generally, this example shows that the ergodicity of the closed system is not automatically transferred to the leaky system.

An interesting effect beyond the results of this paper was reported by~\cite{beims2010}: the shape of the 
leak (rounded, squared, etc.) introduced at the border of the billiard can modify the dynamics of the billiard (e.g., create sticky or
chaotic motion). Similarly, openness in optical systems can lead to a modification of the dynamics of the reflected rays due to non-specular reflection
close to the angle of total internal reflection~\cite{Schomerus2006,AltmannEPL2008,Song:2010}. Here we consider only leaks for which the
trajectory either escapes or is reflected as in the closed system.

Finally, we discuss further examples which go beyond the setup considered in this section. 
In integrable cases,  
 leaks (or recurrence regions) in billiards~\cite{Bauer1990,Vicentini:2001,Bunimovich2005} and
 maps~\cite{Buric:2003} lead to a power-law decay of $P(t)$ with $z=1$. This is due to stickiness caused by families of marginally unstable
 periodic orbits (e.g., parallel walls) and is consistent with 
 Eq.~(\ref{eq:algebraic_zdiff}) since there is no chaotic region and hence all initial conditions are close to sticky regions.
 Leaks placed inside regular islands will lead to similar observations. Leaks (or recurrence regions)
 centered in periodic orbits have been considered by~\cite{Hu:2004}. For leaks centered at 
the border of regular and chaotic regions, a weighted sum of exponential and power-law escapes 
was found by~\cite{Buric:2003}.
A similar composition was also reported by~\cite{Vicentini:2001,Kokshenev:2003} for the case of
leaking polygonal billiards, which are neither integrable nor chaotic. \cite{Dettmann:2012,Leonel:2012} report a decay of survival
probability with  a stretched exponential $P(t) \sim \exp(-\alpha t^\beta)$  when a leak is introduced in the open
 bouncer model (vibrating billiard), which exhibits  mixed phase  space and Fermi acceleration in the closed version.

\section{Applications} \label{sec.applications}\label{sec.6}

We now present different applications for which the results of the previous sections are relevant to understand the specific
phenomena and observations. Accordingly, the models appearing below intend to capture the essential dynamical aspects of the system but do
not intend to  describe all details of a given experimental configuration.
The extent to which dynamical systems provide an appropriate description of reality has been subject of discussion in the context of wave
dynamics for a long time. For instance, in acoustics the dynamical systems description corresponds to the ray approach, which has been repeatedly debated~\cite{Joyce:1975} but systematically used
nevertheless. Two 
recent advances that confirm the validity of the ray picture are: (i) the derivation of Sabine's law in the wave
picture~\cite{Legrand:1991,Dennis} and (ii) the recognition of the relevance of classical periodic orbits in room acoustics~\cite{BerryFor}. 
Similar considerations involving ray and wave pictures apply to the quantum and optics applications below, while the astronomy, fluids,
and plasma applications also have simplifying assumptions and regimes of validity of their own.

\begin{figure*}[!bt]
\includegraphics[width=2\columnwidth]{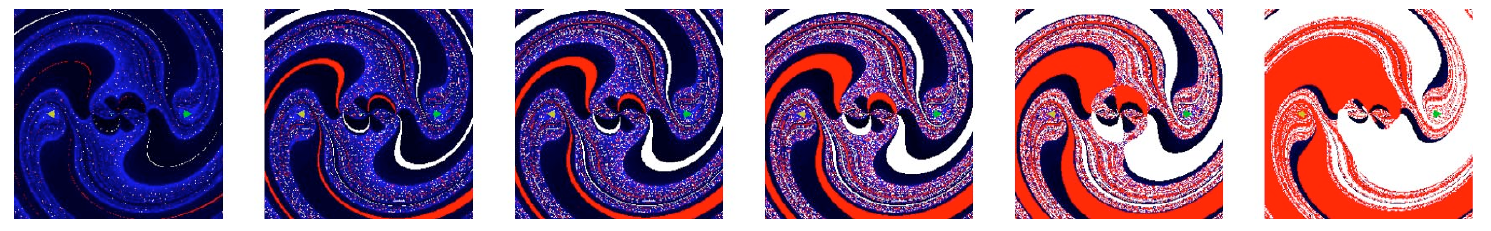}
\caption{ (Color online) The configuration space of the small body in the restricted three-body problem 
in a frame corotating with the main bodies. This section of the phase space is obtained at a fixed
energy under the condition that the particle's velocity toward the origin vanishes, and the angular velocity
is  negative. The colors correspond to trajectories colliding with main body $1$
  (white), colliding with main body $2$ (gray, red), and escape to infinity (dark - black to light blue, from short to long escape times).
 From left to right the diameter of the two main bodies is increased, what corresponds to increasing the size of the leaks (not visible in this section) 
and the collision  probability, from~\cite{Nagler:2004}.}
\label{fig.astronomy2}
\end{figure*}

\subsection{Planetary astronomy}\label{ssec.astronomy}

Chaotic systems with leaks might play a role in one of the most traditional problem of celestial mechanics, the three-body problem (three masses interacting
through gravitational forces). This has been noted by~\cite{Bleher1988} and explored in detail for particular cases by~\cite{Nagler:2002,Nagler:2004,Nagler:2005}. 
The most natural leakage mechanism corresponds to collisions of the finite-size celestial bodies, a realistic possibility even in our Solar System~\cite{Laskar:2009}.
   
The simplest case discussed by~\cite{Nagler:2004} corresponds to two main bodies with equal mass $M$ moving along the same
circle around their center of mass, and a test body of mass $m \ll M$
(circular restricted three-body problem). In the corotating reference frame centered at the center of
mass, this problem can be described by a four-dimensional time-independent system. By using the conservation of energy and employing a
suitable Poincar\'e surface of section, one can reduce the dynamics to a two-dimensional discrete-time problem, as the billiard systems considered here. 
For a fixed energy, 
confined trajectories coexist with trajectories that escape to infinity. 
The system is nonintegrable, with
regular and chaotic trajectories coexisting in the phase space. 

Leakage is introduced by considering that the masses of the two main bodies are not concentrated
in a point and therefore collisions occur whenever the pointlike test particle approaches one of the bodies to a distance~$r<R$, where $R$ is the
radius of the main bodies (the velocity can be arbitrary). The size of the leak 
in the configuration space is proportional to the radius $R$ of the main bodies, and the position of the leak is given by the
positions of the main bodies, which is fixed in the corotating frame. This situation corresponds to the problem of systems with more than one coexisting
leaks discussed in Sec.~\ref{ssec.basins}. 
In fact, at a given energy, particles can also go to infinity; therefore here
there are two leaks in an otherwise open system.
Figure~\ref{fig.astronomy2} illustrates how the basin of escape to infinity and the basin of collision with each of the main bodies varies as a function of the leak size~$R$. 
The typical fractal-like structures observed in Fig.~\ref{fig.basin2}
are clearly seen and coexist with smooth boundaries~\cite{Nagler:2004}.
\cite{Nagler:2005} extended this analysis to the case of different masses and different circular orbits for the
main bodies.

\subsection{Hydrodynamical flows}\label{ssec.fluids}

The advection of tracer particles in hydrodynamical flows represents an 
important application of dynamical-system's theory.
Since molecular diffusion is typically negligible on the relevant time and length scales,
the equation of motion for an idealized particle of zero size and zero mass expresses
the fact that the particle velocity, $\dot{\bf r}$, 
coincides, at any instant of time,  with the flow velocity ${\bf u}({\bf r},t)$. 
The velocity field is assumed to be known and the advective dynamics
is thus described by
\begin{equation} \label{eq:fluid_drdt}
\dot{\bf r}(t)={\bf u}[{\bf r}(t),t].
\end{equation}
The solution to this differential equation is the path ${\bf r}(t)$ of the particle.

Chaos is typical in two-dimensional time-dependent flows and in any kind of three-dimensional flows, leading to the phenomenon of {\em chaotic advection} \cite{Aref:1984}.
Indeed, the main physical mechanism for fluid stirring is advection, 
whose efficiency can be greatly enhanced by chaotic dynamics. The 
spreading of pollutants on large scales is also dominated by advection. 
Potential applications of chaotic advection thus range from laboratory
investigations of fluid dynamics to the study of large-scale environmental flows, and these 
aspects are well reviewed in the literature~\cite{Ottino:book,Aref:book,LaiTel-book}.

The existence of a leak in a  flow implies a sink for fluid elements.
In such cases the amount of fluid in a finite container is decreasing in time,
and this leads to a qualitative change of the dynamics sooner or later.
Here we focus on a less evident, but physically more appealing, realization of leaking dynamics, which is related to 
advected particles and to the first arrival to certain regions of the flow. The problem of reactions in fluid flows provides an
  interesting example.
Reactive particles have typically no considerable influence on flow, 
therefore the velocity field remains the same ${\bf u}({\bf r},t)$ as without reactions.
The reactive dynamics, represented, e.g., by a temporal change
of certain particle properties, is thus superimposed on the advection problem
(\ref{eq:fluid_drdt}).  The change might happen upon entering a region of the flow. {The most important examples of activities in flows involve chemical  (e.g., $A+B \rightarrow C+D$) and 
biological (e.g., $A+B \rightarrow 2A$) reactions. The combination of reaction and advection provides a realistic model for a plethora of 
applications~\cite{TMGG:2005, Neufeld:book}. Next we discuss two particular problems in 
which leaking is introduced in a closed hydrodynamical flow by allowing particles to enter or also react in preassigned fluid regions (that
play the role of leaks).}

\subsubsection{Spreading of pollutants in the environment}\label{sssec.spreading}

Imagine  that a pollutant is released in an observation region within a water basin. A typical problem in the prevention of environmental pollution is to
determine which coastal region the pollutant will be advected to so that one can estimate which parts are most likely to
be affected  by the pollution release. This is one particular example of the general problem of partitioning the 
initial conditions in an observation region of a closed
flow according to the first arrival of advected particles to predetermined subregions of interest.
The boundary between the different partitions typically shows fractal patterns that correspond to 
the {\em stable manifold} of the chaotic saddle formed by tracers that never reach any
of the target regions. The 
target regions act as leaks for
the dynamics of the tracers [following Eq.~(\ref{eq:fluid_drdt})], but not for the velocity field ${\bf u}({\bf r},t)$. 

\begin{figure}[!ht]
\includegraphics[width=1\columnwidth]{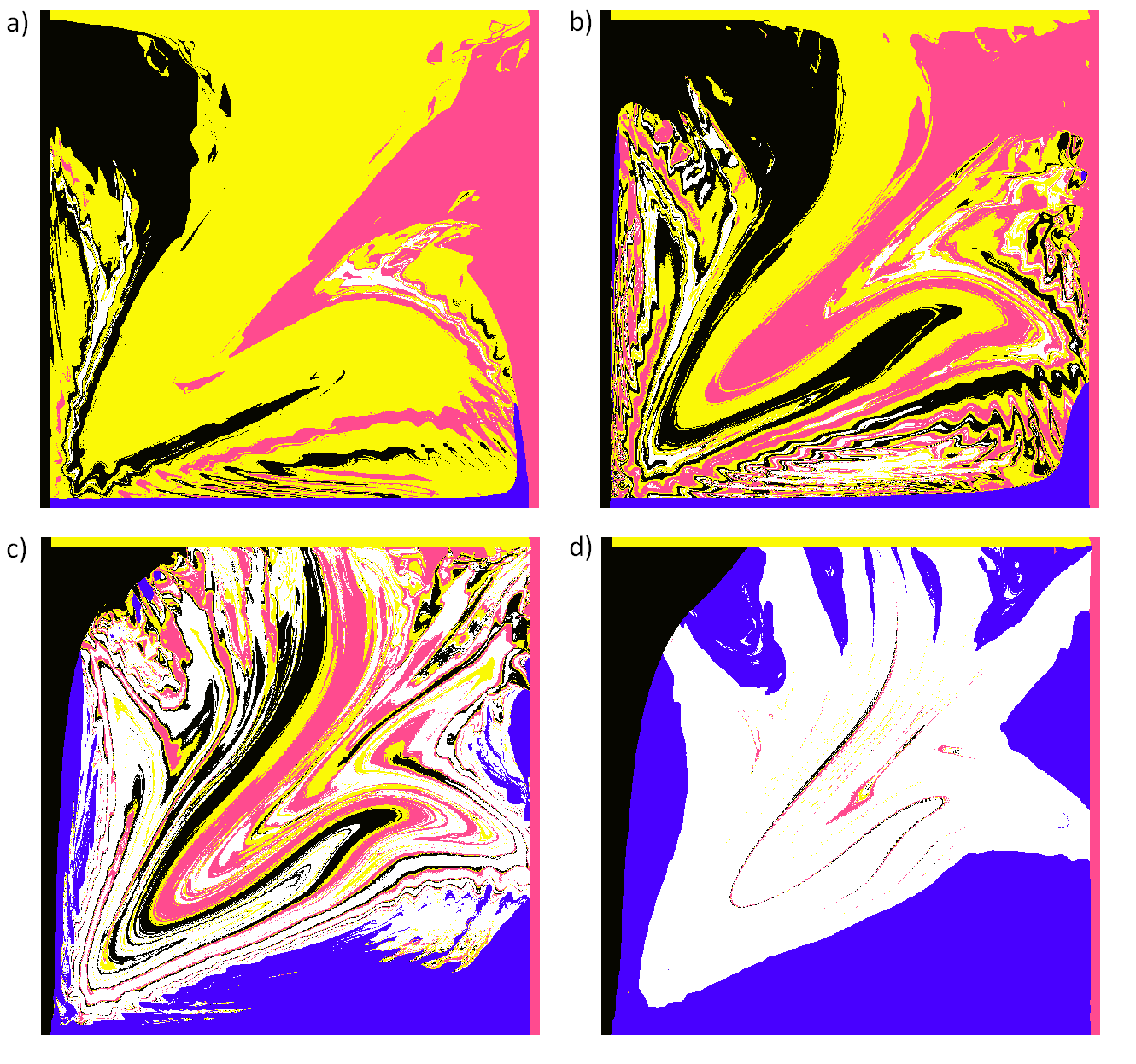}
\caption{(Color online) Spreading of pollutants in a model of a wind-driven lake. The wind field is periodic with period $T$.
Positions $x,y$ are colored according to 
  which of the four coastal regions (bands along the boundaries, also colored) a tracer starting with initial condition $x,y$ [at time $t=0$ (mod $T$)] is advected to. Tracers that do not reach  any of the bands along the shores 
over $60$ h of observation are colored white.  a)-d) represent layers at depth
$0.7, 1.0, 1.3,$ and $1.6$ m. (Picture by S. Szanyi.)
}
\label{fig:to4}
\end{figure}

\begin{figure*}[!ht]
\includegraphics[width=2\columnwidth]{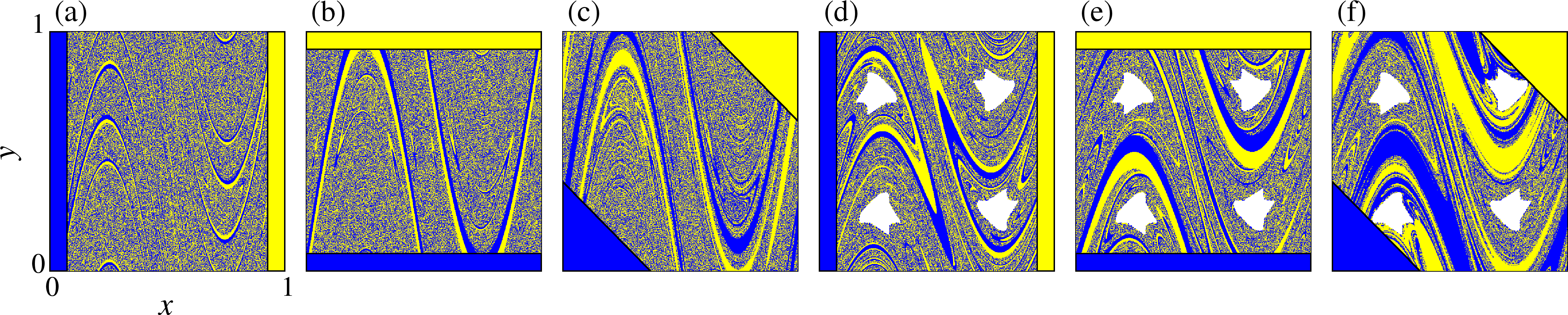}
\caption{(Color online) Tracer distributions with resetting
in a closed time-periodic flow. The particles' color (concentration) is set to blue (dark) and yellow (light) at two resetting bands with total area $0.14$
placed (a,d) vertically, 
(b,e) horizontally, and (c,f) as a triangle at the corners.
In the full phase space the two colors come arbitrarily close to each other along a filamentary pattern. 
In (a)-(c) the bands foliate a strongly chaotic flow, while in (d)-(f) the fluid flow is slower and  KAM islands (white) appear outside the
resetting regions. (Data by G. Dr\'otos).}
\label{fig:reset}
\end{figure*}

Figure \ref{fig:to4} shows an example in a square-shaped wind-driven lake of $2$km by $2$km. 
The boundaries are vertical walls of height $2$m, and below this depth, the lake 
has a pyramidal form with the deepest point at a depth of $2.5$m in the middle.  
The water flow is generated by wind stress.
A similar problem was studied in \cite{Karolyi2010} by applying a shallow-water
approximation in which layers of different depths are assumed to move in a synchronized 
manner. In Fig.~\ref{fig:to4} 
the flow is
obtained from a numerical solution of the hydrodynamical equations in three-dimensions~\cite{Cioffi2005}.
A wind of strength $12$m/s changes periodically in time, with a period of
$T=8$ h. It blows from the south west (lower left corner) for $T/2$, changes abruptly 
to the south east, and after an interval of length $T/2$ it changes back again, etc.

The region of observation is the full lake area outside narrow bands along the coasts. 
Layers at different depths are investigated. Each point in a layer is colored  
according to which of the coastal
bands (of width $100$m and of height $2$m) along the four vertical walls at the shores will be reached by the tracer
first (the vertical coordinate of the first arrival is not recorded).
The results in Fig.~\ref{fig:to4} indicate a strong height-dependence. In the uppermost
layer, pollutants released in the lake are most dangerous for the northern shore~\cite{Szanyi:2012}. The western shore is somewhat less polluted
than the eastern one. 
The southern coast is {hardly} affected. This is consistent with the fact that the average wind direction is southerly, and particles are thus pushed mainly northward.
Considerable deviations from this pattern occur when going to deeper and deeper layers. 
The pollution of the northern coast is becoming less and less strong, while
the southern one becomes heavily polluted. 
This is due to the development of an overturning circulation that has an overall southward component in the deeper regions.
The hazard for the four shores thus strongly depends on the level in which   
pollution is released.  
The boundaries between different 
colors contain the {stable manifold} of a three-dimensional chaotic saddle residing in the
region of observation. The main result obtained from this example is that a two-dimensional shallow-water approximation might lead to an
oversimplification of the three-dimensional flow and also of the advection patterns
since the observed height-dependence in the pollutant distribution cannot appear then.

\subsubsection{Reactivity in flows, resetting}

Perhaps the simplest model of reactions is provided by what was invented and called 
by Pierrehumbert \cite{Pierrehumbert:1994,Ngan2000}
as the {\em resetting mechanism}.  
Whenever a tracer enters a preselected region of the flow, a given property of the tracer such as
concentration (or color) is reset to a value associated with that region, regardless of its
previous value \cite{Gaspard:2005,Neufeld2000}. This mimics a situation where dye is introduced by diffusion from  
a solid surface and is maintained at the saturation concentration in a diffusive 
boundary layer.  An atmospheric  example of the resetting mechanism is provided by the dynamics of water vapor.
Water vapor is removed from an air parcel whenever it enters a region
where the local humidity is lower than that of the parcel.
Its humidity is then reset to this lower value and the difference is rained out.  
When a parcel of low humidity comes into a humid region (typically close
to the Earth's surface), its water vapor content is reset to a high value.
The water vapor distribution in the atmosphere is similar indeed 
to the ones obtained from resetting models~\cite{Pierrehumbert2007}.
As in the example of Sec.~\ref{sssec.spreading}, the leaking mechanism always refers to the particles of a given type (color), and not to
the fluid, as fluid is never lost from the system.

In two-dimensional incompressible flows~${\bf \nabla  u}$=0, and the advective dynamics 
(\ref{eq:fluid_drdt}) is area-preserving. The resetting problem provides thus
a close analogue of a leaky $2D$ billiard.   
Chaotic advection in closed flows is characterized by space-filling chaos, and correspondingly by space-filling stable and unstable
manifolds. Nonhyperbolic regions might also exist around KAM tori. 
The resetting mechanism reveals the foliations of the dynamics, in a similar spirit as a leak, because in both cases a particle can be
considered to be lost after entering a preselected region. The   
advective dynamics with the leak is typically transiently chaotic. Particles never escaping the complement of the leak(s),
both forward and backward in time, form a chaotic saddle. The manifolds of this saddle 
are subsets of the closed system's manifolds since the advective dynamics outside the leak 
is exactly the same as in the closed system.

\begin{figure*}[!ht]
\includegraphics[width=2\columnwidth]{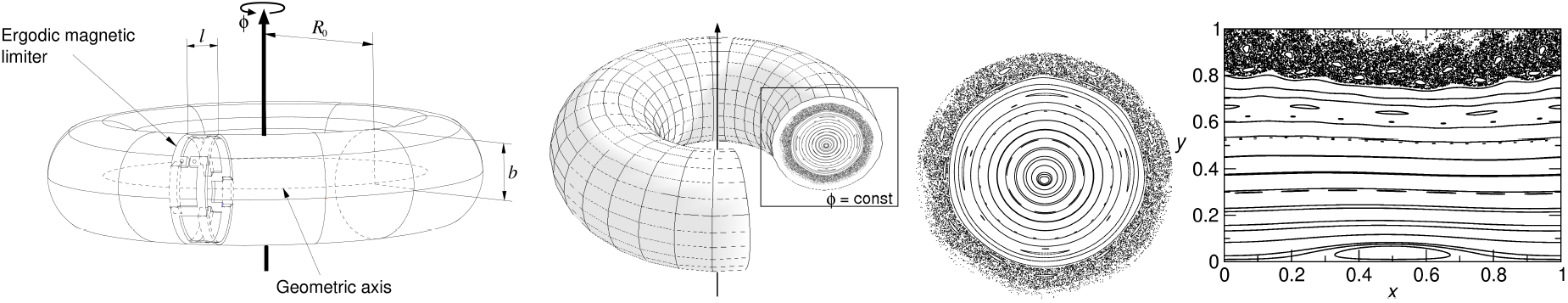}
\caption{Construction of an area-preserving map of the magnetic field lines in a tokamak. From left to right: main geometrical parameters of
  a tokamak with an ergodic magnetic limiter, a torus in which the magnetic field lines are confined, and the surface of section at
  fixed  toroidal angle~$\Phi$ used to construct the discrete-time map.  Black dots on this map represent field line trajectories with many
  iterates confined to the torus. They correspond to long-lived charged particles.  Trajectories are shown both in polar
  $(\mathfrak{r},\theta)$ and  in rectangular coordinates $(x=\theta/2\pi, y=\mathfrak{r}/b)$. }
\label{fig.plasma2}
\end{figure*}

With two or more resetting regions, tracers of different
concentrations or different colors come close to each other along fractal-like boundaries. 
An example is shown in Fig. \ref{fig:reset}
{where the flow is chosen as the alternating sinusoidal 
shear flow model~\cite{Pierrehumbert:1994}. The dark (light)
dots are obtained as initial conditions of trajectories reaching the dark (light)
region when iterated {\em backward} in time. 
Particles close to the boundary have long lifetimes outside the resetting regions and therefore
they come close to the saddle in the time-reversed dynamics. The boundaries trace thus out here the {\em unstable}   
manifold of the chaotic saddle.
The resetting pattern is a fingerprint of the chaotic saddle underlying
the leaking advection  dynamics, where the leak is the
union of the resetting regions. 
Figure~\ref{fig:reset} show that resetting regions of the same area lead to different patterns and also to different 
average lifetimes  in full agreement with what we saw for strongly chaotic billiards
in Sec.~\ref{sec.fullchaos}. Panels d)-e) represent a case where the average
fluid velocity is smaller 
and four elliptic islands appear, with a nonhyperbolic component of
the chaotic saddle around the outermost KAM tori.

Resetting-like methods have been applied to visualizing the foliations of 
three-dimensional model flows \cite{Tuval2004} and they also have applications in 
geophysics \cite{SFH:2005,Schneider2007}. 

\subsection{Magnetic confinement of plasma}\label{ssec.tokamaks}

The research on plasma confinement devices in the last $60$ years has been driven by the expectation of designing controlled fusion
reactors. One of the best studied and most promising strategies is to use magnetic confinement machines with toroidal shape, as illustrated in
Fig.~\ref{fig.plasma2}. These machines are called
tokamaks.
Here we describe the dynamics in lowest order approximation only, when the charged particles follow the magnetic field lines
while being within the tokamak~\cite{book-tokamak}.
A chaotic layer at the border of the plasma is long known to enhance the confinement of particles to the core. This enhancement is 
accomplished by controlling plasma wall interactions in the tokamak~\cite{Engelhardt:1978}.
Chaos is present whenever the ideal toroidal symmetry is broken, which can be achieved by
generating weak electric currents along the toroidal vessel via the so-called {\em
ergodic  magnetic limiters} shown in Fig.~\ref{fig.plasma2} (left).
In any specific tokamak the wall
is fixed and the efforts to control plasma wall interactions concentrate on  manipulating the 
  magnetic structure at the plasma edge~\cite{Schmitz:2012}.

\begin{figure}[!h]
\includegraphics[width=1\columnwidth]{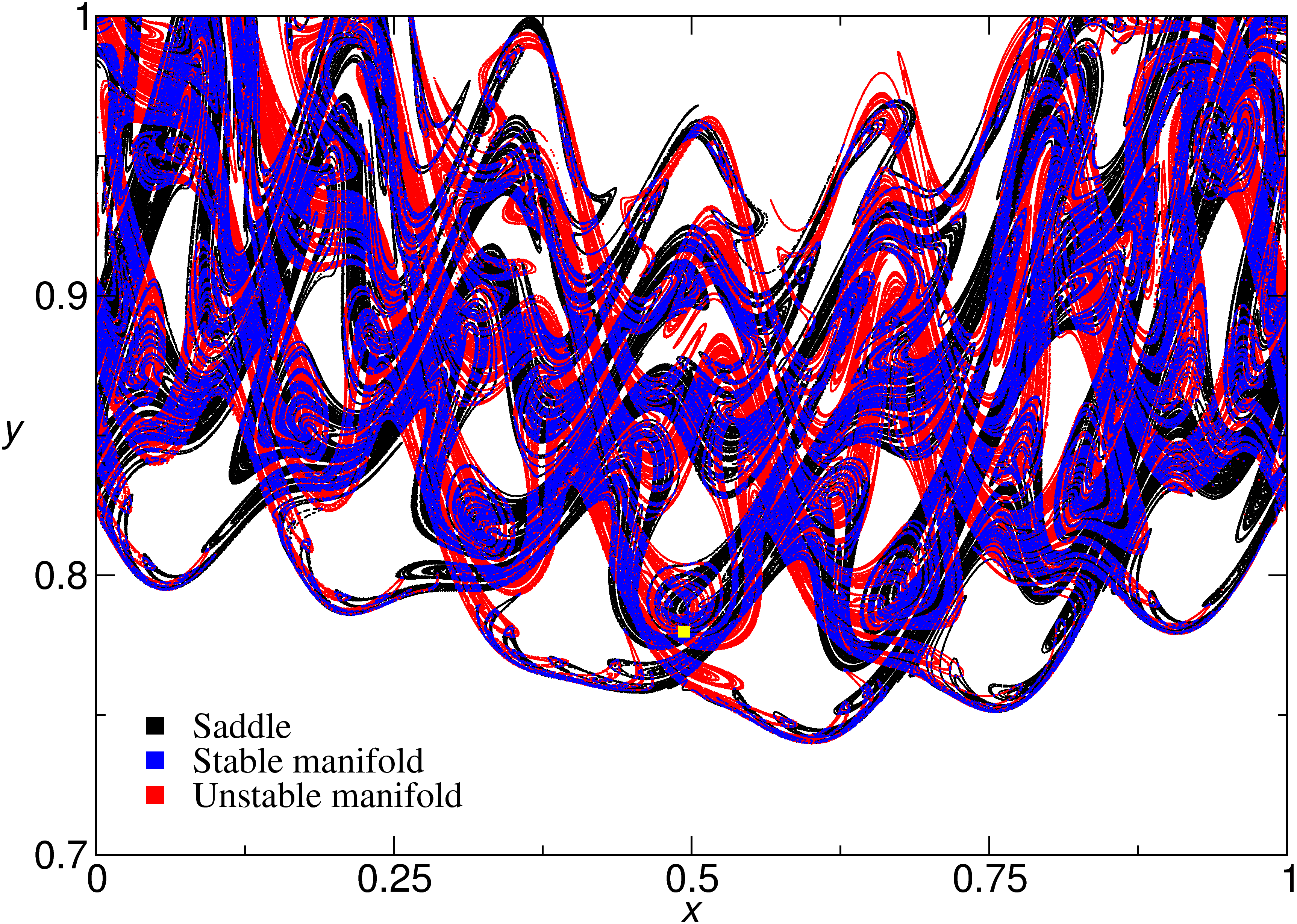}
\caption{(Color online) The saddle (blue), stable (black), and unstable (red) manifolds in the Ullmann-Caldas tokamak map.
The coordinates $x,y$ are as in Fig.~\ref{fig.plasma2}.
The (stable) unstable manifold was obtained as the \mbox{80-th} (pre-)image of a small ball of initial conditions around the unstable
periodic point indicated as a small square at $\sim(0.49, 0.78)$. The intersection of both manifolds approximates the chaotic
saddle for the leaky particle dynamics.
The intersection of the manifolds with the line $y=1$ can be considered as
"footprints" on the wall~\cite{Portela2007}. }
\label{fig.plasma3}
\end{figure}

 Here we are interested in the dynamics of the magnetic field lines and of the charged particles. 
These lines are divergence free and thus correspond to trajectories of volume preserving flows. They can therefore be described by
time-dependent Hamiltonian systems where, for the tokamak geometry, the toroidal angle $\Phi$ of the trajectories corresponds to the time
variable of the flow and slices of them can be described by area-preserving maps~\cite{Morrison:2000}. 
Figure~\ref{fig.plasma2} illustrates the steps that map the
 magnetic field line dynamics into a discrete
map between  two successive intersections with the
 Poincar\'e surface of section with fixed toroidal 
 angle $\Phi$, see~\cite{Portela.review} for a recent review. 
Iterations in these maps correspond
to toroidal turns and are thus proportional to the magnetic field line length.
The connection to systems with leaks is established 
by noting that the evolution of the particles is interrupted at the vessel's wall and at any obstacle
inside the tokamak chamber, like probes and antennas.  
Here 
we concentrate on the main loss of particles in tokamaks, which is due to collisions with
the chamber's wall located at $y=1$. 
Although being ideally mainly confined to the torus, the magnetic field
extends outside the tokamak wall as well. The chaos in the field lines
is thus not limited to $y\le 1$. 
For the charged particles, it is therefore essential to consider a modified  map in which the
wall acts as a leak of the field line map. 
Since the wall can be considered to not affect the magnetic field lines, the particle
map is identical to the field line map but is restricted to $y \le 1$.
 Even if the position of the leak in a map is fixed at $y=1$, it can effectively be controlled by changing other control parameters of the map. For instance, in Fig. 26 this could be achieved by choosing a different wall position (parameter $b$) or current in the ergodic magnetic limiter.

The asymptotic particle dynamics of the leaky system is governed by a chaotic saddle and its manifolds. 
Figure~\ref{fig.plasma3} shows an approximation of the corresponding invariant sets for the so-called Ullmann-Caldas map~\cite{Ullmann:2000}, based on the manifolds 
of a single fixed point embedded into the chaotic saddle~\cite{Portela2007}. 
From the results of Sec.~\ref{sec.4} we conclude that for more general situations the invariant sets depend sensitively on the particular choice of the
  leak.
\cite{Viana:2011} provided a detailed review of the role of the chaotic saddle and fractal
structures in plasma confinement devices. We emphasize that the chaotic saddle is nonhyperbolic due to the presence of KAM islands, as
discussed in Sec.~\ref{sec.weak}. 

\begin{figure}[!h]
\includegraphics[width=0.85\columnwidth]{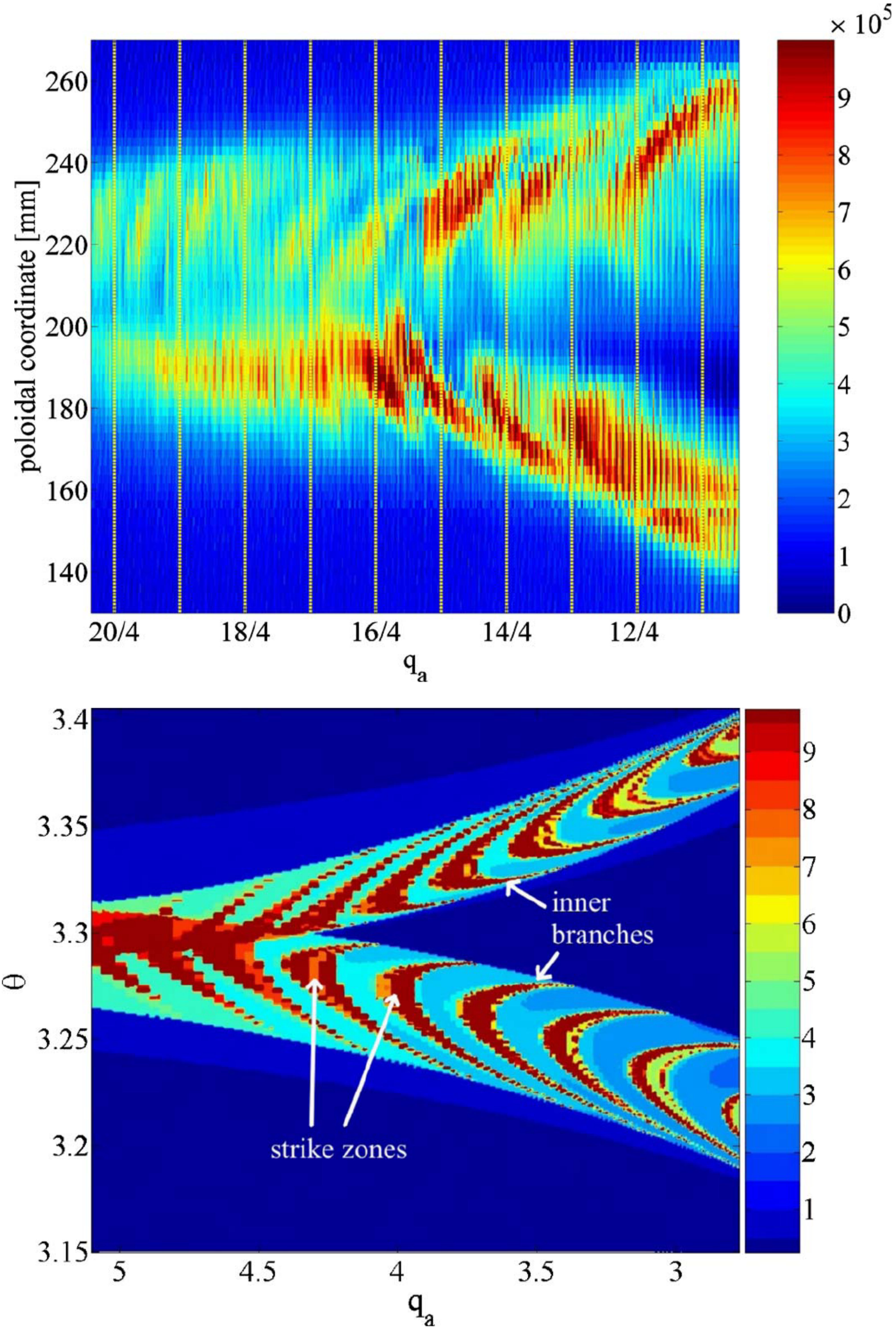}
\caption{(Color online) Comparison between experimental and numerical results for the heat flux pattern in tokamaks.  Upper panel:
  Experimental measurements of the heat flux (color scale) through different poloidal coordinates (proportional to $x$ in Fig.~\ref{fig.plasma2}) 
  shown for different edge safety factors $q_a$.  Lower panel: Numerical simulations indicating the length (in toroidal turns) of the field lines that hit the
  tokamak wall vs $q_a$. From~\cite{Wingen:2007}.} 
\label{fig.plasma4}
\end{figure}

Experimental signatures of the invariant sets of the leaking dynamics appear in the heat flux into 
the tokamak's wall~\cite{Evans:2002,Silva:2002,Wingen:2007,Viana:2011}. It was noticed that the heat flux is not uniform in the poloidal
angle (proportional to $x$ in Fig.~\ref{fig.plasma2})~\cite{Takamura:1989,Shen:1989}.  Apart from the dispersion caused by collisional
effects~\cite{Schelin:2011}, the charged particles leave the system closely following the unstable manifold of the
chaotic saddle of the leaky system and are deposited at the positions where this manifold intersects the wall.
Such lines survive for a long time and reach deeper inside the torus (hot plasma) \cite{Abdullaev:2001}.
Note that the plasma current in the $\Phi$ direction implies that particles with different electrical charge follow the magnetic
field lines in opposite directions. Recalling that time corresponds to the length of the magnetic field lines, we see that this implies that
while the positively charged particle escape following the unstable manifold, the negatively charged particles follow the stable manifold of
the same chaotic saddle. Altogether, this reasoning predicts that the heat flux at the wall should be enhanced at the position of the
intersection of these manifolds with the wall and thus be proportional to the length of the magnetic field lines within the tokamak. 
Figure~\ref{fig.plasma4} shows experimental and numerical results that confirm this prediction. In the upper panel the 
concentration of the heat flux in specific poloidal angles is shown for different edge safety factors $q_a$, a control parameter that is
inversely proportional to the plasma current.
In the lower panel
we observe that the same patterns are observed in the length of the field lines. 

Here, again, similar to the advection and resetting problem,
we have a chance to see a direct fingerprint of transient chaos
in the configuration space, since the phase space and
the configuration space coincide in both cases.

\subsection{Optical microcavities}\label{ssec.microcavities}

\begin{figure}[!h]
\includegraphics[width=0.5\columnwidth]{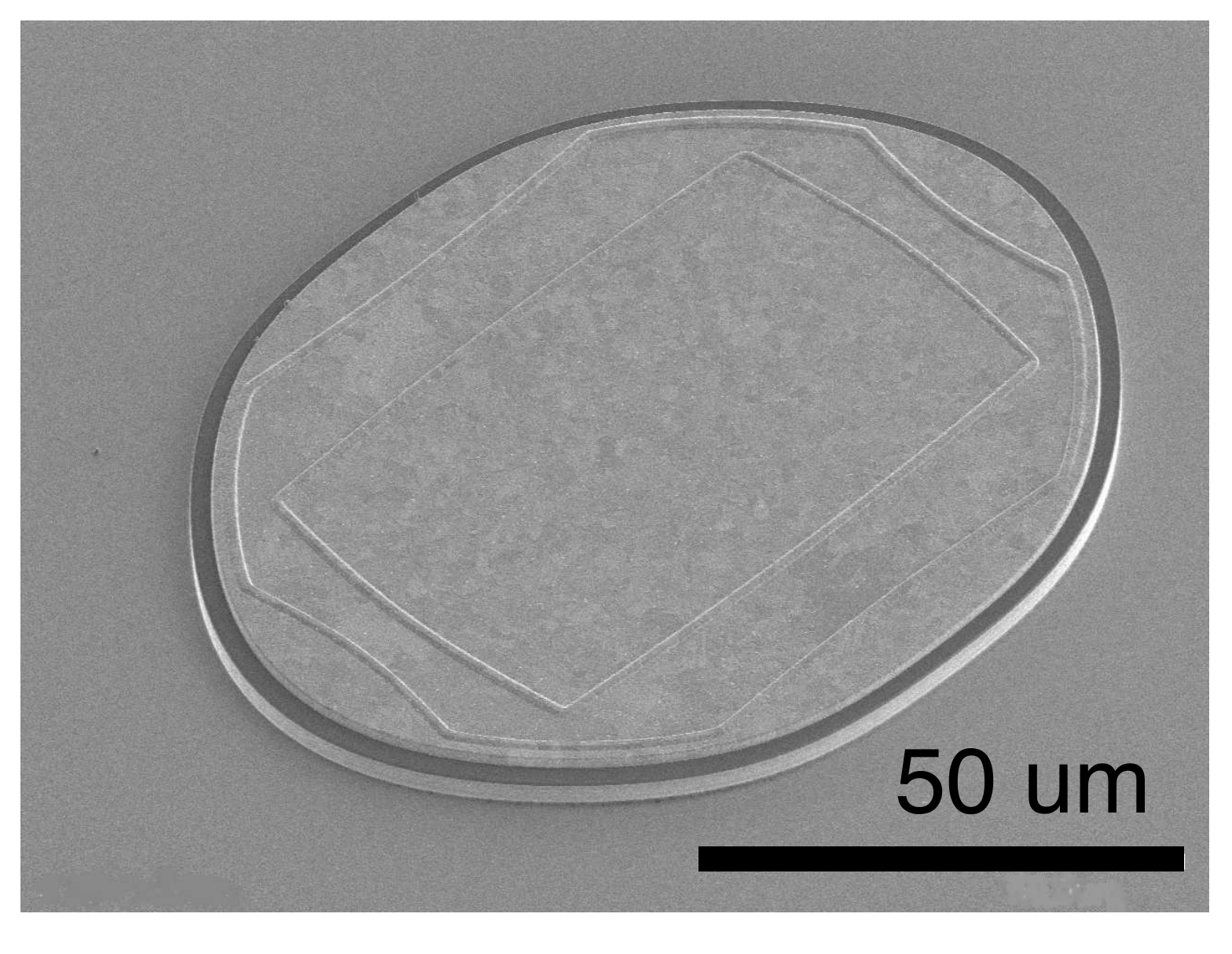}
\caption{Scanning electron microscope image of a microlaser with deformed-disk shape. From~\cite{Shinohara2010}.}
\label{fig.optics1}
\end{figure}

\begin{figure}[!h]
\includegraphics[width=0.8\columnwidth]{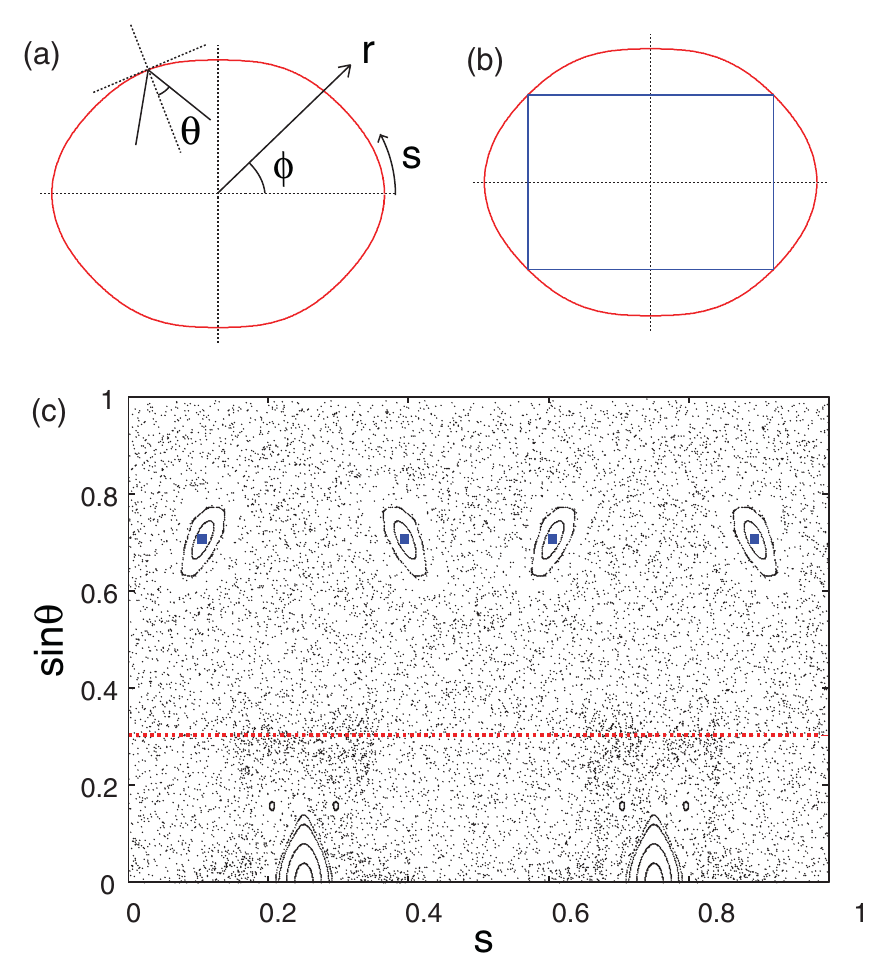}
\caption{ Ray model of the microlaser depicted in Fig.~\ref{fig.optics1}. (a) Definitions of the coordinates. (b) A stable 
periodic  orbit. (c) Phase space of the closed billiard. The upper
edge of the partial leak is indicated by the dashed (horizontal/red) line in
  panel (c). From~\cite{Shinohara2011}.} 
\label{fig.optics2}
\end{figure}

\begin{figure}[!ht]
\includegraphics[width=0.8\columnwidth]{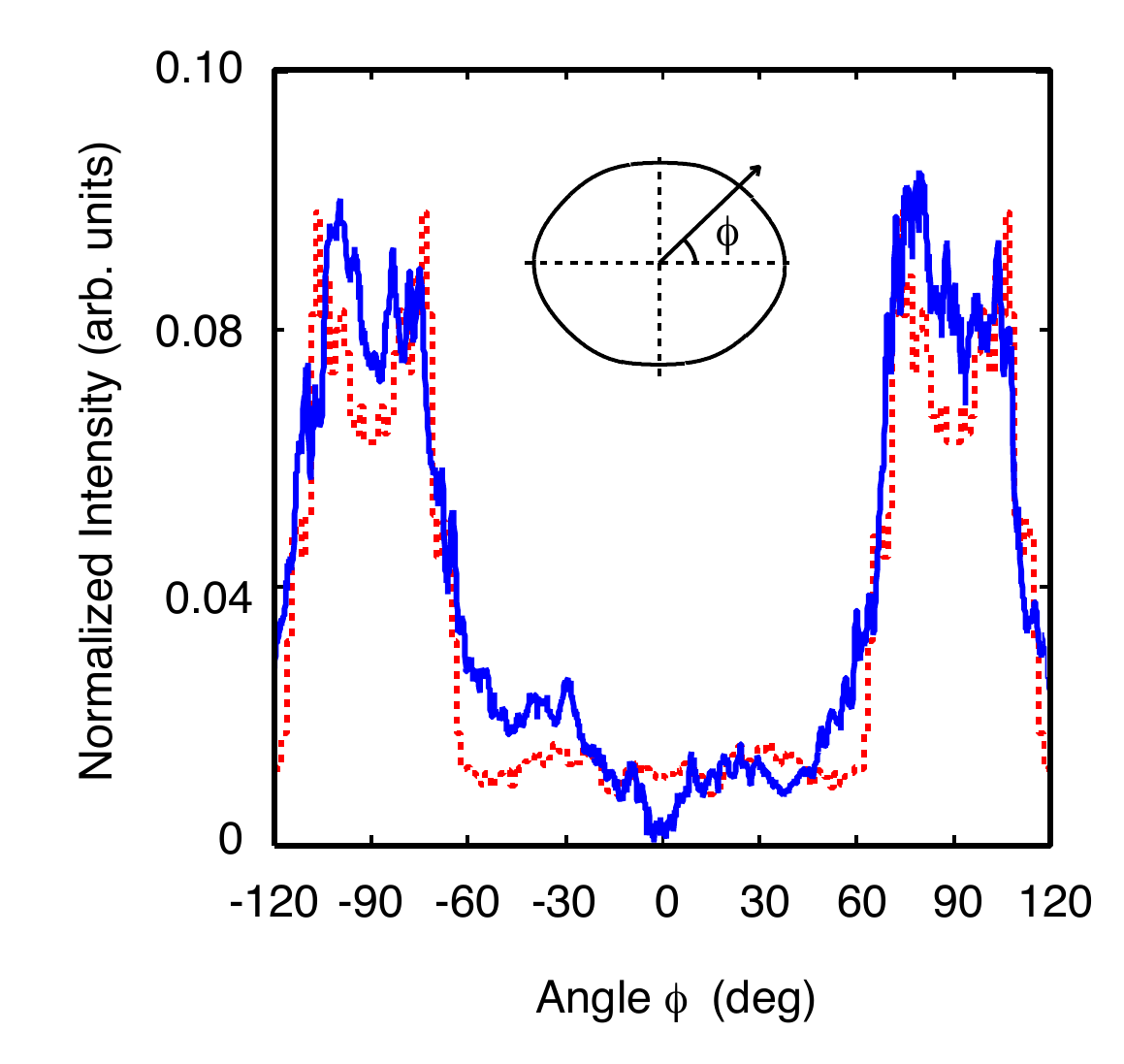}
\caption{Far-field emission observed experimentally (solid line) and predicted by the ray-model (dashed line). 
From~\cite{Shinohara2011}.}
\label{fig.optics4}
\end{figure}

Optical microcavities are used in applications ranging from dynamic filters in optical communications to quantum electrodynamics 
(see~\cite{Vahala2003} for a review). Here we focus on the case of optical microcavities used as lasers. Beside the practical
applications, these cavities allow for fundamental scientific investigations (e.g., in quantum chaos) in systems in which only partial leaks are present. 
An example of an optical microcavity is given in Fig.~\ref{fig.optics1} from~\cite{Shinohara2010,Shinohara2011},
a system used here to illustrate the achievements of the last $15$ years of intense
research~\cite{NS:1997,Schwefel:2004,Lee:2004,Ryu2006,Tanaka:2007,Wiersig2008,Yan2009,DMSW:2009,AltmannPRA,Harayama2011}. 

The connection to leaking chaotic systems is based on the observation that the shape of the microlaser matters: while the regular trajectories present in
spherical and circular shaped cavities provide the good confinement necessary for lasing (high Q modes), deformation of these geometries leading to
chaotic dynamics~\cite{MNCSC:1995,NS:1997} can add to the good confinement other desired properties such as the directionality of the
emission~\cite{NS:1997,GCNNSFSC:1998,Liu:2002,Schwefel:2004,Tanaka:2007,Lebental:2007,Wiersig2008,Yan2009,DMSW:2009,Song:2009}. Even if 
these systems are often far from the limit of vanishing wave-length (geometrical optics), the so-called ray-wave correspondence applies
in a surprising large number of cases, see~\cite{Harayama2011} for a recent review. A description based on rays makes robust
  numerical simulations possible and provides intuitive
 interpretations of 
  experimental observations that can be 
directly connected to orbits and the geometry of the corresponding billiard.

The ray-model in optical cavities is equivalent to the dynamics in closed
billiards with partial leaks. Figure~\ref{fig.optics2} shows the phase space of the system used by~\cite{Shinohara2010,Shinohara2011}. As mentioned in  
Secs.~\ref{ssec.leak} and~\ref{ssec.partial}, the partial
leak has a natural physical origin in the collisions inside the cavity (with refractive index $n_{\text{in}}> n_{\text{out}} \equiv 1$).
Rays with angles smaller than the critical angle of total internal reflection ($p < p_c=\sin \theta_c = 1/n_{\text{in}}$, dashed line in
Fig.~\ref{fig.optics2}c) are partially transmitted (with angles given by Snell's law) and partially reflected.  The
intensities of the reflected and transmitted rays are given by Fresnel's
law~\cite{NS:1997,Schwefel:2004,Lee:2004,Ryu2006,Tanaka:2007,AltmannPRA,Harayama2011}. For rays with $p > p_c$ no
transmission takes place. 

Interestingly, \limacon{} shaped microlasers as the ones 
used in the main part of this review have also been
used in experiments and simulations~\cite{Wiersig2008,Yan2009,Shinohara2009,Song:2009}. An important additional aspect of 
lasing  cavities is the gain medium that continuously pumps energy into the system enhancing the intensity of long-living trajectories
(a precise modeling of this effect is beyond the scope of this paper). For corrections to the ray model due to wave  
effects see~\cite{Schomerus2006,AltmannEPL2008}.

The main relevant observable is the far-field emission
intensity, as computed in Fig.~\ref{fig.emission1}
(see also Sec.~\ref{ssec.emission}). A comparison between experimental results and ray-dynamics simulations is presented in
Fig.~\ref{fig.optics4}. The agreement is 
remarkable in particular if one takes into account that the system is far from the formal semiclassical limit of small wavelength, see~\cite{Shinohara2010,Shinohara2011,Redding:2012} for further evidence of the robustness of ray results.
The important and somehow surprising aspect of the emission in
Fig.~\ref{fig.optics4} is the concentration of the emission in specific directions, even if the leak lies in a region where the closed system
is more or less uniformly chaotic. 

The specific directions of emission have been explained using the unstable manifold of a single unstable periodic
orbits close to the border of the leak region, first by~\cite{Schwefel:2004}. Another important theoretical development was the proposal of
the existence of a so-called quasistationary energy distribution in the classical ray dynamics, see~\cite{Lee:2004,Ryu2006}. 
As noticed by~\cite{AltmannPRA}, and better explained in Secs.~\ref{ssec.partial} and \ref{sec.3}, these two important concepts can more generally be expressed,
respectively, as: the unstable manifold of the chaotic saddle (which aligns with the unstable manifold of specific periodic orbits) and the
c-measure distributed along it. 
 In as much the ray-model provides a good description of
  the laser application, these two key concepts of the theory of dynamical systems have direct observable consequences for the
far-field emission of lasing microcavities. Here, the results of Sec.~\ref{sec.weak} have to be taken into account because 
microlasing cavities are typically not strongly chaotic~\cite{Liu:2002}. The main effect of weak chaos is that the long-living modes (high-Q factors)
are usually concentrated inside KAM islands and regions of regular motion. It remains to be seen to what extent the effect of weak chaos
described in  Sec.~\ref{sec.weak} can be detected experimentally.

Recent developments show links to many other results discussed in this review.
For instance, while dielectric microcavities have 
leaks restricted in the $p$ direction, recent experiments use waveguides that restrict the leak also in the other
  direction. Additional leaks are
  constructed around positions $s_l$
 that showed to be efficient in {\em channeling} the rays out of the cavity~\cite{Song:2012,Redding:2012}. Such coexistence of different
 leaks was discussed in  Sec.~\ref{ssec.basins}. Another example is the periodic-orbit formula for the resonance spectrum of dielectric 
  cavities proposed by~\cite{Bogomolny:2008} and experimentally verified  in regular~\cite{Bittner:2010} and  chaotic~\cite{Bittner:2012}
  systems. The main novelty in this formula is the incorporation of partial leakage as a multiplicative term proportional to the 
  product of the reflection coefficients at collisions along periodic orbits, in a similar manner as the reflection coefficient appears in
  the operator~(\ref{FPleak}).

\subsection{Quantum and wave chaos in systems with leaks}\label{ssec.quantum}

The quantum and wave analogues of classically chaotic systems with leaks appear in theoretical analyses, controlled experiments, and real-world
systems. 
Quantum mechanically, the major differences to closed systems originate from the non-Hermitian Hamilton operator 
(nonunitarian Scattering matrix). The intrinsic openness of any quantum mechanical experiment, highlighted in a
quote from the early $1990$'s in Sec.~\ref{sec.1}, has already been fully discussed~\cite{Stockmann:book}. Still, an important question influencing the research in quantum and wave chaos in the last decade was the effect of
openness on results known for closed systems. A driving force are experiments in quantum
dots, microwaves, optics, acoustics etc. These more recent developments appear in essays on open quantum systems:
\cite{Kuhl2005} reviews spectral and scattering properties with focus on classical waves (microwaves and sound
waves); \cite{Fyodorov:2005} uses a formalism based on correlation 
and distribution functions that goes beyond random matrices; \cite{Rotter:2009} 
uses a Feshbach projection operator formalism to non-Hermitian Hamiltonian operators; \cite{Nonnenmacher:2011} reviews the mathematical
results and methods of semiclassical theories of wave operators in scattering systems; and \cite{Novaes:2012} gave an overview of the
properties and recent research on the eigenstates of leaky quantum-chaotic maps. 

The results of these review papers are to a great extent valid for scattering systems in general, with the main distinction
being between, as they call, ``weak and strong absorption'' (coupling to the environment).
Throughout we  distinguished leaking from genuinely open systems based on the key elements: (i) the possibility of comparing
the results to a closed-system and (ii) the control over properties of the leak (see
Sec.~\ref{ssec.motiv}). 
These two elements affect spectral properties, which were recently observed in experiments~\cite{Barthelemy:2005,Dietz:2006,Xeridat:2009}
and fully analyzed theoretically~\cite{Savin:2006,Poli:2009}. Point (i) is used, for instance, when the universal distribution of nearest
neighbor levels observed in strongly chaotic closed systems~\cite{Stockmann:book} is compared to the results obtained in leaking systems~\cite{Poli:2012}.  Point (ii) is related to 
the importance of the localization of the leak or perturbation in the phase space, a recent observation that is  summarized by Savin, Legrand, 
and Mortessagne as: 
\begin{quote}
{\em Open wave-chaotic systems in the presence of energy losses (absorption) are nowadays under intense experimental and theoretical
investigations \ldots
Most of the works concern the case of uniform absorption which is responsible for homogeneous broadening $\Gamma_{\text{hom}}$ of all the modes (resonance states). However, in some
experimentally relevant situations \ldots one should take
into account also localized-in-space losses  which lead to an inhomogeneous part $\Gamma_{\text{inh}}$ of the widths which varies from mode
to mode. 
}\cite{Savin:2006} 
\end{quote}
The use of antennas and measurement devices typically fulfills properties (i) and (ii) mentioned above.

Next we focus on three 
measurable properties of quantum systems which can be directly connected
to classical properties and the main results of this paper.

\subsubsection{Loschmidt echo (fidelity)}\label{echo}\label{ssec.Loschmidt}

The absence of well-defined trajectories in quantum mechanics make it difficult to precisely define 
chaos~\cite{Berry1987}, a concept usually based on the exponential instability of classical trajectories under small perturbations of initial
conditions (one positive Lyapunov exponent). 
The effects of classical chaos in quantum and wave systems can be observed following an idea by
Peres~\cite{Peres1984}, who proposed to compare the evolution of a wave packet with its evolution in a perturbed Hamiltonian.
A central concept here is the Loschmidt echo, also known as fidelity.
It is defined as the overlap between two quantum states: the first state is obtained from an initial
state $|\Phi\rangle$ in the course of its evolution up to time $t$ 
under a Hamiltonian $H$; and the second state results from the same initial
state by evolving up to the same time with a perturbed Hamiltonian $H'$.
The measure of this overlap is
$$ M(t) = |\langle \Phi| e^{iH't/\hbar} e^{-iHt/\hbar} |\Phi\rangle|^2,$$
where $\hbar$ is Planck's constant.
This quantity can also be interpreted as the overlap of the initial state and the state 
obtained by first propagating this state up to time $t$ with $H$, and then
backward in time, up to $-t$ under $H'$. 
$M(t)$ equals unity at $t=0$, and typically decays in time. 
      
The term {\em Loschmidt echo} refers to the debates in the 19th century about the ireversibility of thermodynamical systems and the
foundations of statistical mechanics (in which Loschmidt participated). Similar {\em echo} concepts are important in classical
systems~\cite{Eckhardt:2003}, with applications in sensing techniques~\cite{Taddese2010}, and have recently been extended to relativistic quantum systems~\cite{Sadurni:2008}.
The term fidelity is used mainly in the field of quantum information.  
The overlap $M(t)$ is known to quantify the robustness of systems to perturbations, 
and has nowadays numerous applications in quantum information,
statistical physics, and quantum chaos~\cite{Gorin:2006}.

The first and most natural investigations in quantum chaos considered {\em global} perturbations (e.g., $H'=H+\varepsilon V$ with $V$ acting
globally). Global perturbations affect all (or a dominant part of) the phase-space accessible to the quantum particle in the course of its time evolution. It was shown that the decay of $M(t)$ with $t$ has a variety of regimes, many of them reflecting classical properties of the
closed system (e.g., Lyapunov exponents). See~\cite{Gorin:2006,Jacquod:2009} for reviews, and~\cite{Garcia-Mata:2011} for
limitations of the Lyapunov regime.  

More recently, experimental realizations in
billiards~\cite{Hohmann:2008} emphasized the importance of the case 
of quantum mechanically {\em local} perturbations~\cite{GR:2007,Goussev:2008,Ares:2009,Kober:2011}. Next we review recent results on local
perturbations that show that in this case the decay of $M(t)$ has an important regime dominated by properties of the corresponding classical system with leak
(e.g., the escape rate). 

Figure~\ref{fig.quantum0} illustrates the concept of local perturbations on the example 
of a chaotic billiard. The perturbation consists of a deformation
of width $w$ localized in a region $\tilde{B}_1$ of the boundary. The rest of the boundary, $B_0$, is unaffected by the perturbation. Three trajectories starting from the same
location ${\bf r}_0$ (and reaching point ${\bf r}$ afterwards) are shown. The two nearby trajectories $S_1$ and $\bar{S}_1$ marked by continuous and dashed lines correspond, respectively, to an evolution with the unperturbed and 
perturbed Hamiltonian. In a semiclassical approximation, the action difference between 
these trajectories should be determined. The third trajectory ($S_0$) hits the 
boundary only at $B_0$. For this case the perturbed and unperturbed trajectories coincide, hence the action difference of this pair is zero.

\begin{figure}[!h]
\includegraphics[width=0.7\columnwidth]{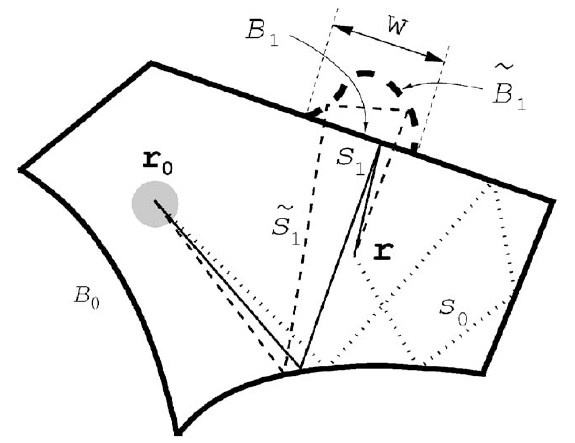}
\caption{Schematic diagram of a billiard with a local perturbation $\tilde{B}_1$
along its boundary and three classical trajectories between  
${\bf r}_0$ and ${\bf r}$, from~\cite{GR:2007}.}
\label{fig.quantum0}
\end{figure}

The quantum mechanical implementation and interpretation of the Loschmidt echo with a localized perturbation is illustrated in 
Fig.~\ref{fig.quantum1}. The evolution of a wave packet 
is shown in the original billiard up to time $t$. At this time the perturbation along the boundary is introduced, and the dynamics is followed in the time-reversed 
evolution up to the same time.
Gray shading represents the probability distribution. The initial
Gaussian wave packet appears thus as a dot in the upper left panel. The arrow in this panel marks the momentum direction of
the wave packet. The lower left panel shows that the original form is only partially recovered 
after the full process, the overlap is thus incomplete~$M(t)<1$.

\begin{figure}[!h]
\includegraphics[width=\columnwidth]{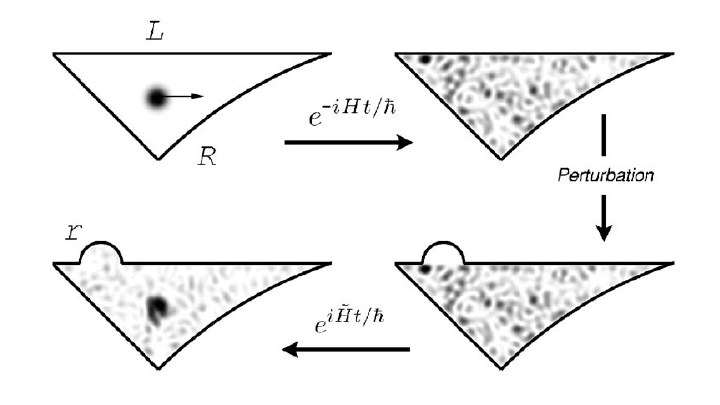}
\caption{ Forward-time wave packet evolution in an unperturbed diamond billiard 
up to a time $t$ corresponding to about $10 \tcoll$, followed by the
time-reversed evolution in the perturbed billiard. The width of the localized perturbation is $w=60$, $15\%$ of the length $L$ of the horizontal edge, and 
the de Broglie wavelength is $\lambdabar=4.8$ in the same units.  From~\cite{GR:2007}.
}
\label{fig.quantum1}
\end{figure}

Goussev and Richter considered a classically weak but quantum mechanically strong perturbation in a region of the boundary of a strongly chaotic
billiard~\cite{GR:2007}. More precisely, the length $w$ and depth~$r$ of the perturbed boundary should be much smaller than the length of the billiard boundary, and the perturbation is considered quantum mechanically strong if $w$ and $r$ are much larger
than the de Broglie wavelength~$\lambdabar$. Their main result is that in such cases there is a long time interval for which the Loschmidt echo decays as:
\begin{equation}\label{eq.2kappa}
M(t) \sim \exp(-2\kappa t),
\end{equation}
where $\kappa$ is the classical escape rate of the billiard with a leak~$I$ extending precisely through the perturbation region of
length~$w$ along the original boundary. 
The time interval over which decay~(\ref{eq.2kappa}) is valid extends  from 
the mean collision time (flight time) $\tcoll$ to a
saturation time $t_s=-1/{(2\kappa)}\ln M_\infty$ where $M_\infty={\lambdabar \sigma}/(2 \pi {A})$, with  $\sigma$ as the wave packet
size at $t=0$, and $A$ as the area of the billiard~\cite{Gutierrez:2009}. Note that $t_s$ can be even larger than the Heisenberg time $t_H \sim 1/\hbar$.
This result
is obtained by showing that, in this regime, the main contribution to~$M(t)$ is due to trajectories that do not collide with the 
perturbation region (collisions there would radically change the dynamics of the trajectories). In our formulation, this means that such trajectories start from the stable manifold of the saddle characterizing the billiard with this leak. Fig.~\ref{fig.quantum2} shows results
of numerical simulations in the billiard depicted in Fig.~\ref{fig.quantum1}.
The quantity $M(t)$ is plotted for different depths $r$, and its decay is governed by the classical escape rate $\kappa$ of the billiard with a leak of length $w$, independent of the value of $r$.  
Time is normalized by the classical mean collision time $\tcoll$. 

\begin{figure}[!h]
\includegraphics[width=\columnwidth]{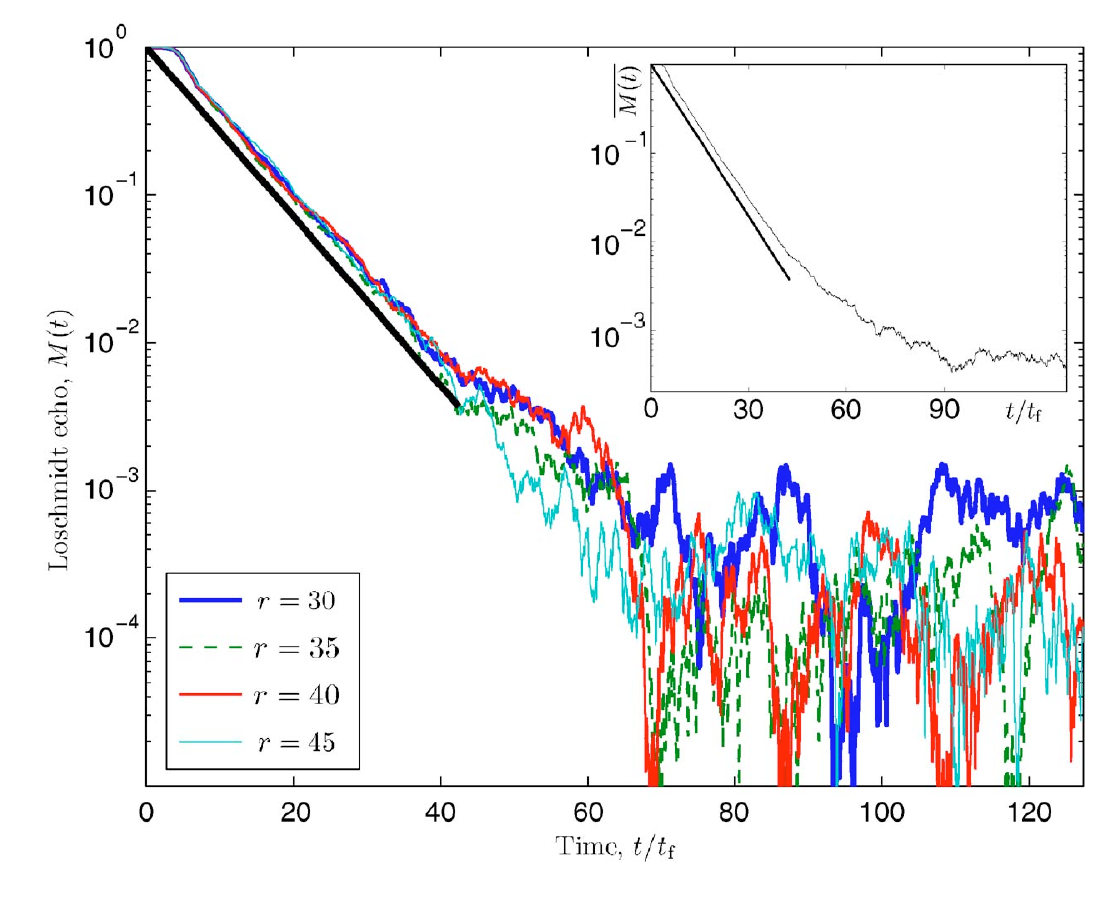}
\caption{(Color online) The decay of $M(t)$ in the diamond billiard of Fig.\ref{fig.quantum1}
for four different values of the depth $r$ at $w=60$. Time is scaled with
$\tcoll\equiv t_f$ as given in (\ref{eq.tau2d3d}). The solid straight line 
corresponds to $\exp{(-2\kappa t)}$ with the escape rate given by (\ref{eq.kappa2d}),  Sabine's result.
The inset presents the decay of $\bar{M}(t)$, the Loschmidt echo averaged over several values of $r$, from
~\cite{GR:2007}.
}
\label{fig.quantum2}
\end{figure}

These results have been refined and generalized to perturbations 
which are quantum mechanically not necessarily strong.
The decay of $M(t)$ then remains exponential, but the decay rate is a factor 
$\alpha$ times the classical escape rate:
\begin{equation}\label{eq.a2kappa}
M(t) \sim \exp(-\alpha \kappa t).
\end{equation}
For pistonlike perturbations, semiclassical expressions have been derived for the
coefficient $\alpha$~\cite{Goussev:2008,Kober:2011}. As a function of the 
perturbation strength, $\alpha$ turns out to be an oscillatory function converging to $2$ for large perturbations.

These results were confirmed in recent experiments in microwave cavities~\cite{Kober:2011}. The
correspondence between the Helmholtz equation (describing 2D microwave cavities) and the Schr\"odinger equation 
make microwave cavities a unique tool for the experimental investigation of quantum chaos~\cite{Stockmann:book,Richter:1999}, including the
exploration of the effect of classical trajectories in chaotic systems with leaks~\cite{Dembowski:2004}. Figure~\ref{fig.quantum3} shows
the configuration used by~\cite{Kober:2011} for the experimental investigation of Loschmidt echo decays. For not very strong perturbations,
good agreement was found with the semiclassical coefficient $\alpha$ in Eq.~(\ref{eq.a2kappa}).
On the experimental side, the investigation of the strong perturbation case remains a challenge since the signals are then expected to be rather weak.

Altogether these results on Loschmidt echoes show that there is
increasing recent interest in systems with leaks, not only in applications but also on the side of fundamental properties of quantum chaos.
In the semiclassical limit such systems reflect the properties of the underlying classical systems, including
semiclassical corrections to the classical escape rate~\cite{Sieber:2001,Waltner:2008,Gutierrez:2009b}. 

\begin{figure}[!h]
\includegraphics[width=\columnwidth]{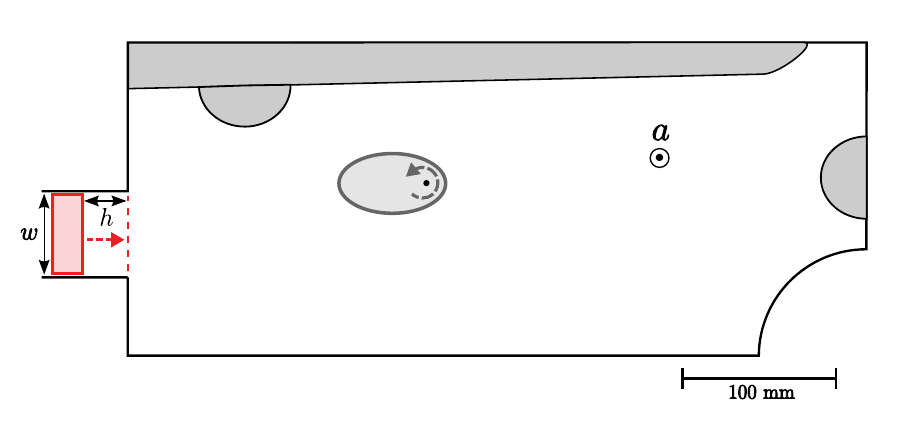}
\caption{(Color online)
Geometry of the chaotic billiard used in the experiments of \cite{Kober:2011}.  
The local perturbation is piston-like, of width $w$ and of depth $h$. 
The measuring antenna is placed at location $a$. The rotatable ellipsis is
used to perform ensemble averages, and the additional elements are introduced to
reduce the influence of bouncing balls.   From~\cite{Kober:2011}.
}
\label{fig.quantum3}
\end{figure}

\subsubsection{Fractal distribution of eigenstates}\label{sssec.eigenstates}

Another signature of chaos in open quantum systems is provided by the fractality of certain invariant sets of the corresponding classical systems. 
The eigenstates of the quantum system turn out to be distributed along the manifolds of the classical chaotic saddle\footnote{ In spite of the fact that 
  repellers (i.e., unstable sets with only expanding   directions) cannot exist at all in Hamiltonian systems, this misleading term is
 often also used in the quantum literature to refer to the chaotic saddle~\cite{LSZ:2003,RPBF:2009,Novaes2009,Eberspracher2010,Pedrosa:2009}.} discussed in
Sec.~\ref{ssec.transient}~\cite{Casati:1999}.
Open quantum systems are characterized by nonunitary evolution operators having a set of right and left decaying nonorthogonal eigenfunctions.   
The left and right eigenstates of the  
nonunitary propagator concentrate in the limit of $\hbar \rightarrow 0$ 
on the stable and unstable manifold of the chaotic saddle, respectively
\cite{KNPS:2006}.
This affects the statistical properties of energy levels (e.g., Weyl's law discussed next). 

Classical maps can be quantized~\cite{Berry1979}.
Quantized area-preserving baker maps with leaks (often called open baker maps in the quantum chaos community) nicely exemplify the results mentioned
above, and can be directly related to the findings reported in Sec.~\ref{ssec.baker}. \cite{KNPS:2006} studied the ternary 
baker map (\ref{eq:ld_baker3}) with a leak,
exactly as in Sec.\ref{ssec.baker}. 
Figure~\ref{fig.quantum4} shows their results: the long-lived (right) eigenstates
concentrate in the semiclassical limit on the unstable manifold (horizontal filaments parallel to the $q$ axis).
Correspondingly, the momentum representation of any long-lived right eigenstate 
is supported on a Cantor set which is the projection of the unstable manifold on the 
momentum $(p)$ axis. This is apparent in Fig. \ref{fig.quantum5}, which also shows a magnification that reveals the self-similar character of 
this distribution.

\begin{figure}[!h]
\includegraphics[width=\columnwidth]{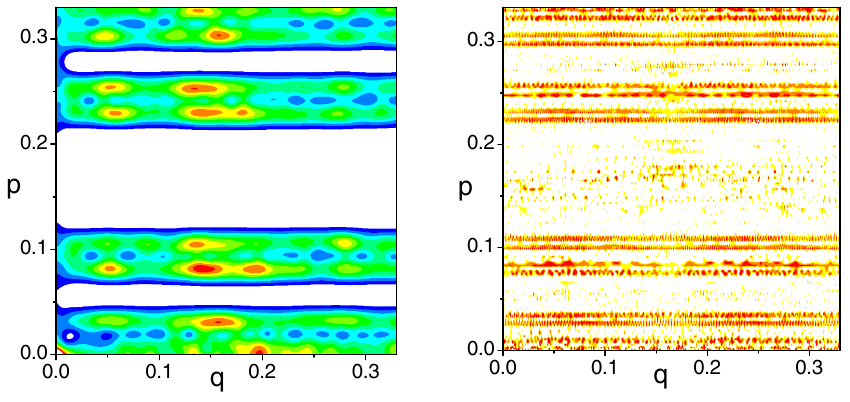}
\caption{(Color online) Representations of the long-lived right eigenstates in the leaky ternary baker map. The classical map is described
    by~(\ref{eq:ld_baker3}) but the variables are here $p\equiv x$ and $q\equiv y$.  The scale of $\hbar$ is $0.0005$.  
Because of symmetry, here only $1/9$th of the phase space is shown, outside the leak
(a central vertical strip of area $1/3$).  
Left panel: The average of the so-called Husimi function. Right panel:
the corresponding Wigner function average. From 
~\cite{KNPS:2006}. 
}
\label{fig.quantum4}
\end{figure}
   
\begin{figure}[!h]
\includegraphics[width=\columnwidth]{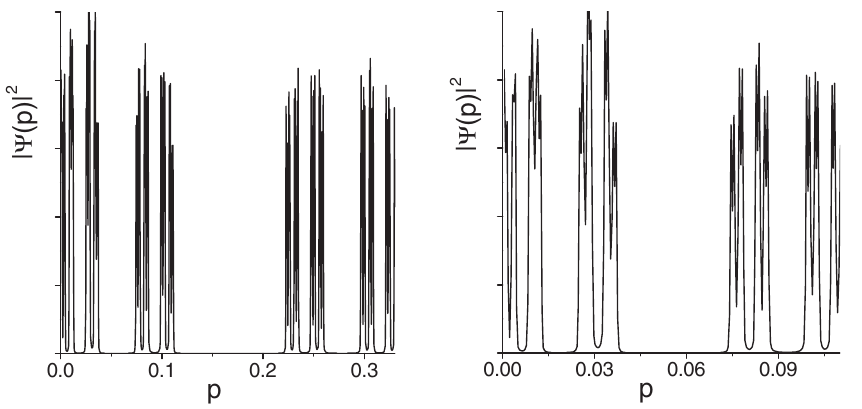}
\caption{Probability density of the leaky ternary baker map in momentum space from right eigenstates averaged over the longest-lived states. The magnification (right panel) 
illustrates the self-similar Cantor set character, from~\cite{KNPS:2006}.
}
\label{fig.quantum5}
\end{figure}

In another study, Pedrosa et al. considered the quantized version
of the binary baker map 
(\ref{eq:ld_baker1})
with leaks as stripes along the $p\equiv x$ axis, centered at $q_l\equiv y_l$ and of width $\Delta q$~\cite{Pedrosa:2009}. In the classical version, they found a
similar strong dependence of the escape rate on the leak position $q_l$ (for any fixed finite $\Delta q$), as discussed in
Sec.~\ref{ssec.dependence}. The influence of the shortest periodic orbit falling into the leak was also noticed. Looking for a quantum mechanical analogue of this, 
they determined the distribution of eigenvalues of the quantum evolution operator
in the complex plane, as shown in Fig.~\ref{fig.quantum6}. Since the map is open, moduli are less than 1, and the eigenvalues fall thus all inside the unit circle.  For a leak with smaller escape rate, the distribution of eigenvalues is found to be dense at the outer ring, while
the distribution is characterized by an increase of density near the origin for 
a case with faster escape.
The strong dependence of the classical escape rate on the position of the leak is
thus found to be reflected in the eigenvalue distribution of the quantum evolution operator. 

\begin{figure}[!h]
\includegraphics[width=\columnwidth]{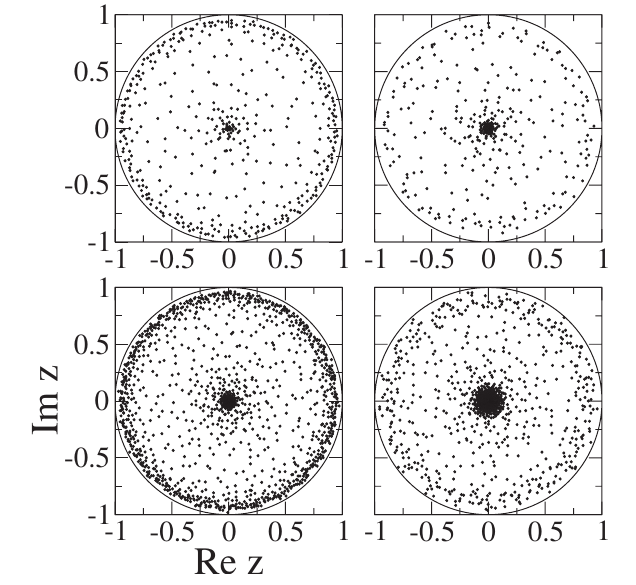}
\caption{
Eigenvalues of the leaky binary baker map in the complex plane. 
In the upper panels the dimension of the Hilbert space is $N=602$, 
while in the lower ones $N=2048$. 
In the left (right) column $q_l=0.3$ ($q_l=0.5$) with
classical escape rate $\gamma=0.090$ ($\gamma=0.165$). $\Delta q=0.1, \Delta p=0.5$ in all cases, from~\cite{Pedrosa:2009}.
}
\label{fig.quantum6}
\end{figure}

Weyl's law states that in closed systems the number $N(k)$ of energy levels with wave number smaller than $k$ grows as
\begin{equation}\label{eq.fractalweyls}
N(k) \sim k^{\beta_W},
\end{equation}  
where the exponent $\beta_W$ is an integer proportional to the dimension of the system (e.g., for $d$-dimensional symplectic maps $\beta_W=d/2$). 
In open chaotic maps the number $N(k)$ of resonances with wave numbers of real part smaller than $k$ scales with an
exponent~\cite{LSZ:2003}:  
\begin{equation}\label{eq.fracweyl}
{\beta_W}= \frac{D_0}{2}=D_0^{(1)},
\end{equation}  
where $D_0$ is the fractal dimension of the underlying classical chaotic 
saddle, and $D_0^{(1)}=D_0^{(2)}$ is the partial box-counting dimension
(see (\ref{DHam})). Equation~(\ref{eq.fracweyl}) shows that the original Weyl law is converted into a {\em
  fractal Weyl law}~\cite{Eberspracher2010,Spina2010, Nonnenmacher2005,LSZ:2003,Shep:2008,WM:2008,RPBF:2009,Kopp2010,Ermann:2010,Nonnenmacher:2011,Schomerus:2004,Pedrosa:2011,Novaes:2012,Koerber:new}. 
An investigation of this property was also carried out~\cite{Pedrosa:2009}.
As Fig. \ref{fig.quantum7} shows, they determined the fraction of
eigenvalues $N_{\nu}$ with modulus $\nu>0.3$ as a function of the Hilbert's space dimension, $N$, for different leaks.   
The prediction of the fractal Weyl law for this case is
$\log{N_{\nu}} \sim D_0^{(1)} \log{N}$, which is fulfilled for all graphs. An experimental investigation of fractal Weyl's law on open
microwave cavities was recently reported by~\cite{Potzuweit:2012}. It confirms a noninteger scaling of Eq.~(\ref{eq.fractalweyls}) and discusses the origin of
potential deviations from the classical prediction.

\begin{figure}[!h]
\includegraphics[width=0.8\columnwidth]{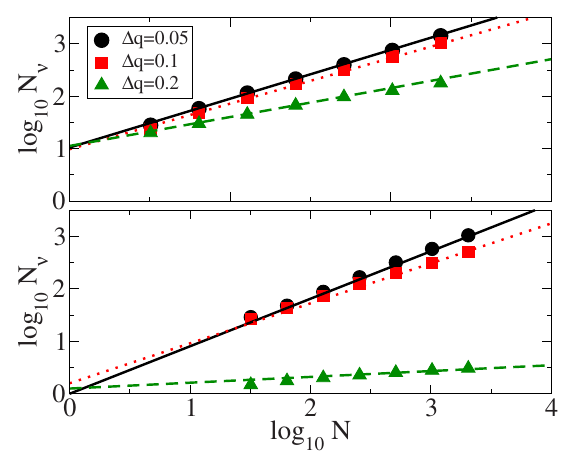}
\caption{ (Color online)
Logarithmic plot of the fraction of eigenstates $N_{\nu}$ for $\nu>0.3$
as a function of the dimension $N$ of the Hilbert space in the leaky binary baker map.
Lines correspond to the prediction of the fractal Weyl law.  
The different slopes reflect the sensitivity of the classical saddle to the position $(q_l,p_l)$ and size $(\Delta q,\Delta p)$ of
the leak.
In the upper (lower) panel $q_l=0.3$ ($q_l=0.5$), and 
circles correspond to $\Delta q=0.05$, squares to  $\Delta q=0.1$, 
and triangles to $\Delta q=0.2$ ($p_l=0,\Delta p=0.5$ in all cases).
From~\cite{Pedrosa:2009}.
}
\label{fig.quantum7}
\end{figure}
    
\subsubsection{Survival probability and quantum Poincar\'e recurrences}\label{sssec.survival}

In Sec.~\ref{ssec.poincare} we  related the escape of trajectories in systems with leaks to the problem of Poincar\'e
recurrences. In the same spirit, Casati, Maspero, and Shepelianski used the quantum survival probability $P_q(t)$  to
investigate what they define as Quantum Poincar\'e recurrences~\cite{Casati:Poincare}. This is performed introducing absorbing boundary
conditions (the leak) in quantized area-preserving maps in the weakly chaotic regime. 
The quantum decay~$P_q(t)$ is identical to the classical decay~$P_q(t)=P(t)$ up to a time $t_q \sim \sqrt{1/\hbar}$. This time is
  larger than the Ehrenfest time $t_E \sim \ln (1/\hbar)$~\cite{Casati:1997}, see also~\cite{Waltner:2008,Schomerus:2004} for important
  effects on the $t_E$ time scale. For times longer than the Heisenberg time $t>t_H\sim
1/\hbar$, one finds~$P_q(t)\sim 1/t$. This is valid up to a maximum time $t_{max}\sim\exp(1/\hbar)$ after which an exponential decay sets in,
whose rate is, however, different from the classical escape rate.  

The origin of the new quantum decay regimes can be understood following~\cite{Wimberger:2002} (see also
references therein) and writing 
\begin{equation}\label{eq.Pq}
P_q(t)=\sum_j \omega_j \exp(-\Gamma_j t),
\end{equation}
where $\Gamma_j>0$ is the decay rate of the eigenstates $|\psi_j\rangle$ and $\omega_j$ is the expansion coefficient $\omega_j=|\langle
\varphi_0 | \psi_j \rangle|^2$ for the initial state $|\varphi_0 \rangle$.  The asymptotic exponential decay corresponds to the smallest $\Gamma_j$ with $\omega_j\neq0$. In a mixed phase space this longest living state is localized
in the center of a KAM island and has $\omega_j\neq0$ due to chaotic tunneling. 
KAM islands (or sticky regions) are not the only source of power-law decays in $P_q(T)$, which can be originated through genuine quantum
effects such as the quantum mechanical (or dynamical) localization of states due to chaos~\cite{Stockmann:book}.
In particular, even classical systems showing exponential tails in~$P(t)$ 
have power-law tails in $P_q(t)$~\cite{Alt:1995,Alt1996,Fendrik:1997,Casati:Poincare}. 
The full description of $P_q(t)$ has to consider the effect of
the localization of many states, not only in the chaotic and regular components of the phase space but also at the hierarchical border of the KAM
islands~\cite{Ketzmerick:2000}.  
In general, $P_q$ depends both on the distribution of the decay rates
($\Gamma$'s in~(\ref{eq.Pq})) and on the initial condition
($\omega_j$), as seen from Eq.~(\ref{eq.Pq}) and emphasized by~\cite{Wimberger:2002}.   
The existence of regimes of power-law decay is generic, but exponents different from~$P_q(t) \sim 1/t$ are also typically found~\cite{Skipetrov:2006}. %
Altogether, the decay of~$P_q(t)$ and the previously mentioned Loschmidt echo (fidelity) decay show that the classical decay is still observed in quantum
systems, but quantum effects dominate the large and asymptotic behavior. 
These quantum effects are unfortunately not easy to detect in experiments [for more discussions on the time when $P_q(t)$ deviates from $P(t)$
see~\cite{Savin:1997,Puhlmann:2005}].

 \begin{figure}[!h]
 \centering
 \includegraphics[width=0.7\columnwidth]{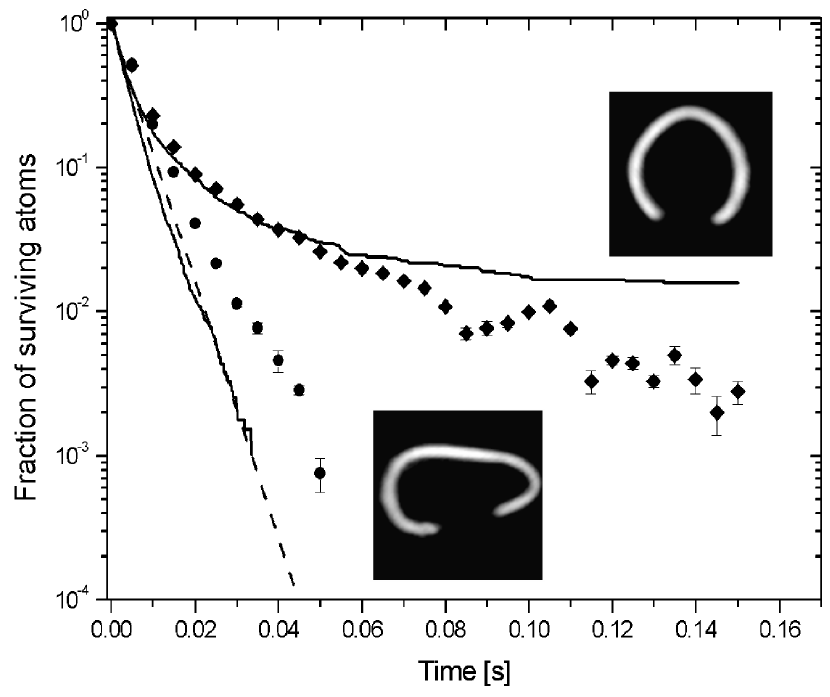}\\
 \includegraphics[width=1\columnwidth]{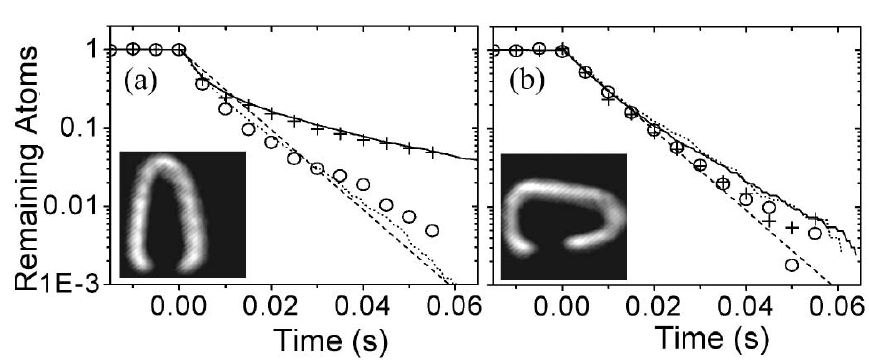}
 \caption{Survival probability of ultracold rubidium atoms confined by an optical potential. Upper panel: Diamonds (circles)
   correspond to the experimental results for a circular (stadium) billiard with a leak at the bottom,
as shown in the upper (lower) insets. Lower panels: Different potentials along the boundary of the stadium billiard, with two different
positions (a) and (b) of the leak. In both panels
results with harder ($\circ$) and softer ($+$) billiard walls are shown.
Full and doted lines correspond to numerical simulations and the dashed line indicates  Sabine
prediction~$\kappa=\mu(I)/\tcoll$, Eq.~(\ref{eq.mut}). From~\cite{Friedman:2001} and~\cite{Kaplan:2001}.} 
 \label{fig.quantumdecay} \label{fig.quantumdecay2}
 \end{figure}

The most detailed experiments of the survival probability are on atom-optics billiards as reported by~\cite{Friedman:2001,Kaplan:2001}.  In these experiments the survival probabilities of atoms inside cavities were measured for different
boundaries and leak positions. The connection to classical billiards becomes evident when comparing the
escape from integrable (circular) and strongly chaotic billiards (tilted stadium 
billiard\footnote{The usual parallel walls of the stadium billiard were tilted in 
  order to avoid the influence of the bouncing ball  orbits discussed in Sec.~\ref{ssec.weaksurvival}}). 
The measurements shown in the upper panel of Fig.~\ref{fig.quantumdecay} confirm that a much faster (exponential-like) decay is observed for the
chaotic case than for the integrable one.  A
subsequent study~\cite{Kaplan:2001} considered the effect of soft walls along the boundaries of the (tilted) stadium
billiard, which generically lead to the creation of KAM islands. The lower panels of Fig.~\ref{fig.quantumdecay2} show the experimental survival
probabilities. They confirm a slower decay for soft walls, which was shown to be related to the stickiness of trajectories around
the islands (see Sec.~\ref{ssec.weaksurvival}). A strong dependence on the leak position is also found. If the leak corresponds to a region
of the phase space that fully contains the island(s), no stickiness effect can be observed (see Fig~\ref{fig.quantumdecay2}, lower panel).

\section{Summary and Outlook}\label{sec.conclusions}

There are many different configurations in which a leak is introduced in a chaotic systems,  but in all cases the simplest theoretical approaches rely
on the same principle: the properties of the leaky system can be compared to the properties of the closed
system. Simple estimates 
based on this principle led to Sabine's law for acoustical reverberation at the end of the 19th century and, more 
than a century later, were used to explain the emission patterns of microlasers (see, e.g.,~\cite{Lebental:2007}) and the survival probability of cold atoms~\cite{Friedman:2001,Kaplan:2001}.
Here we presented a transient chaos based theory of the problem of leaky systems
that goes beyond the closed-system approximation (Sec.~\ref{ssec.closed}) and leads to a proper treatment of finite
leaks (Sec.~\ref{ssec.transient}, see also Tab.~\ref{tab.sum}), partial reflection (Secs.~\ref{ssec.partial} and~\ref{sec.3}), and weak chaos (Sec.~\ref{sec.5}).

Another aim of this review has been to illustrate the abundance of applications of leaky systems (Secs.~\ref{sec.1} and~\ref{sec.6}). 
In the regimes in which dynamical-system models apply, the results reviewed show that} all important quantities of the transient chaos
theory of leaky systems (Secs.~\ref{ssec.transient},~\ref{ssec.operators}, and~\ref{sec.fullchaos})
 have direct consequences to experiments and applications of contemporary
physics research: 

\begin{itemize}

\item The {\bf escape rate} (Sec.~\ref{ssec.escaperate}) dominates some regimes of the Loschmidt echo (Sec.~\ref{echo}) and was
directly measured in cold-atoms experiments (Sec.~\ref{sssec.survival});

\item  The {\bf chaotic saddle and its invariant manifolds} (Sec.~\ref{ssec.transient}) leave direct fingerprints on
astronomical problems (Sec.~\ref{ssec.astronomy}), hydrodynamical flows (Sec.~\ref{ssec.fluids}), magnetic field lines inside tokamaks
(Sec.~\ref{ssec.tokamaks}), optical microcavities (Sec.~\ref{ssec.microcavities}), and the distribution of eigenstates in quantum systems
(Sec.~\ref{sssec.eigenstates}). Many of these examples include weakly chaotic systems, where the observations often become easy to interpret 
in view of the {\bf division of the chaotic} saddle in hyperbolic and nonhyperbolic components (Sec.~\ref{ssec.saddles}).

\item The {\bf stickiness} due to the nonhyperbolic component  (Sec.~\ref{ssec.weaksurvival}) becomes observable 
in transport properties (Sec.~\ref{ssec.saddles}) and in quantum systems (Sec.~\ref{sssec.survival}). 

\item The dependence on {\bf leak position, size, and shape} (Sec.~\ref{ssec.dependence}) is clearly observable in hydrodynamical problems (Sec.~\ref{ssec.fluids}), in
optical microcavities (Sec.~\ref{ssec.microcavities}), and in quantum experiments (Sec.~\ref{sssec.survival}).

\item {\bf Basins of escape} (Sec.~\ref{ssec.basins}), shaped by an underlying chaotic saddle, prove to be concepts usefully applicable to
the three-body problem (Sec.~\ref{ssec.astronomy}), in asymmetric transport (Sec.~\ref{ssec.saddles}), and in the plasma problem
(Sec.~\ref{ssec.tokamaks}); 

\item Far-field {\bf emission} (Sec.~\ref{emission}) and {\bf partial reflection} (Sec.~\ref{ssec.partial}) play a crucial role in
experiments on optical microcavities (Sec.~\ref{ssec.microcavities}), and in
room acoustics (Sec.~\ref{ssec.sabine}).  

\end{itemize}

Our review also contains new results. 
They mainly arise within the framework of our operator description of true-time maps with partial leaks (Sec.~\ref{sec.operators}). The
general formula~(\ref{eq.kappacorrect}) relates averages taken with the c-measure to the energy escape rate $\langle e^{-\tilde{\kappa} t_{\text{coll}}}\rangle_{\tilde{c}} = \langle R \rangle_{\tilde{c}}$.  The idea of 
a partial reflection/absorption is not incorporated in traditional descriptions of dynamical systems, despite being physically
very natural, and also of easy implementation in ray simulations. Here we have shown that reflectivity can be naturally incorporated into the operator
formalism that avoids the artificial factorization of the total energy density~$\tilde{\rho}$ in ray density $\rho$ and intensity $J$. More
generally, the approach makes superfluous the attaching of {\em labels} (e.g., intensity or true time) to trajectories, that are
modified at each collision.
Also the numerical procedures (see Sec.~\ref{sec.ap-algorithm}) to estimate the different measures of leaky billiards should be
relevant not only for simulations but also in experiments for which data can be collected within a snapshot 
(as in the S set) 
or during some time interval (as in the Q set).

The significance of these results is not restricted to billiards or even to
Hamiltonian systems; they apply to {\em any} true-time map
obtained from a Poincar\'e surface of section. 
The collision times is then to be replaced by the return times to the Poincar\'e section.
It is a widespread practice to assume that results for 
maps extend immediately to flows or true-time maps. While this holds
for stroboscopic maps of periodically driven systems, autonomous problems, or more
generally, problems with Poincar\'e maps are different.  
Our results  further emphasize the message of~\cite{Kaufmann2001} that the connection between true-time and discrete-time maps in
open systems is different from the case of closed systems, and often involves surprising nontrivial results (e.g., Tab.~\ref{tab.sum}).  

Finally, 
we discuss unsolved problems and future research
directions. First we note that leaky systems are intimately related and provide further motivation to fundamental problems in
dynamical systems. In particular, in Hamiltonian systems we mention the understanding of the
mechanisms of stickiness  
in higher dimensions and its connection to Arnold diffusion~\cite{BunimovichNonlinearity}, and a formal treatment of the division of the 
chaotic saddle in hyperbolic and nonhyperbolic components (see e.g.~\cite{Altmann2008}). Among the questions directly related to leaky systems,
and in addition to \cite{Dettmann:OpenProblems}, the following general problems claim for further
investigations:

\begin{itemize}

\item 
Adapt and extend current results on the sensitivity on leak position to more generic situations. For instance, most rigorous mathematical 
results about the escape rate (see Sec.~\ref{ssec.dependence}) are demonstrated for one-dimensional systems with leaks in Markov
partitions. A similarly nongeneric feature is the standard explanation of directional emission in 2D-optical
microcavities (see Sec.~\ref{ssec.microcavities}), which relies on the existence of an unstable periodic orbit close to the leak. There is
an evident need for extending these results to generic leak positions and chaotic saddles.

\item Find the conditions for the existence of a single chaotic saddle (with a single physically relevant c-measure) after a leak is introduced~\cite{Collet:2000,Demers:2006}. The
hypothesis of the existence of a single saddle was used in our theory (Sec.~\ref{ssec.transient}). To explore the cases in which this is violated~\cite{Orestis2011,Gaspard:2001} remains a task for the future.

\item Consider novel configurations of leakage such as introducing a temporal dependence or a random choice on the properties of the leak~\cite{Nandori:2012,Bahsoun:2012,Orestis2012}.

\item Explore the implications of partial leak and true-time (Sec.~\ref{ssec.operators}) in operators approaches based on Ulam's method,
  which have been applied to the usual Perron-Frobenius operator~\cite{Tel:1992,Orestis2012,BunimovichWebb:2012,Cristadoro:2012} and
  received renewed interest in relation to the concept of almost invariant sets~\cite{Froyland:2009,Froyland:2010}.

\item Recent works suggest that certain nontraditional (e.g., location dependent) boundary conditions in problems of wave chaos can be
interpreted as generalization of leakage~\cite{BerryN2}. It remains to be explored how this can be related to the traditional leakage
discussed here.

\item Further investigate further the connection to transport properties~\cite{Orestis2011,Knight:2011,Kuhl2005} and the effect of stochastic perturbations~\cite{Faisst:2003,Altmann2010,DettmannNoise,Altmann2012,Bodai:new}.

\item Partial leaks: clarify the nature of the spectrum of dimensions of the invariant sets  (as mentioned in
Sec.~\ref{ssec.partial}), relate these dimensions to the escape rate and to the Lyapunov exponent. Extend the results to noninvertible systems, and develop efficient algorithms for the computation of chaos
characteristics (e.g., escape rate, dimensions, and Lyapunov exponents). For first results see~\cite{APT:2013}.

\item Quantum partial leaks: deepen the connection between quantum systems with absorption and classical dynamical systems with
  partial leaks. First results already revealed interesting phenomena, such as, e.g., a drastic modification of the fractal Weyl's
  law~\cite{WM:2008,Nonnenmacher:2008,Schenk:2009,Novaes:2012} and changes in the localization of eigenfunctions~\cite{Lippolis:2012}.
Additional interesting developments can be expected in view of the results presented in Sec.~\ref{ssec.partial} and~\cite{APT:2013,Arnoldi:new}.

\end{itemize}

Besides 
these general problems, a clear future research line is to adapt the models to make them more realistic to specific
applications. This involves
including new features in the dynamics, e.g., gain medium in lasing cavities, reactions of particles, or properties of the wave systems
(e.g., modified collision laws~\cite{Schomerus2006,AltmannEPL2008,Song:2010}).  The challenge here is to show how to connect experiments and observations to the theory proposed here.  Also important is
to show to what extent and to what level of detail the models match the experiments, e.g., to what extent nonhyperbolic properties of
chaotic dynamical systems play a role in the experimental results. 
In turn, these adaptations towards applications bring new problems and inspiration for theoreticians, a virtuous circle illustrated here by, e.g.,  the
case of partial leaks.
We are convinced that the collection of problems that can be modeled as leaky chaotic system will keep growing and pushing the research forward also in unforeseeable directions.

\section{Appendices}\label{sec.appendices}

\subsection{Projected measure and averages}\label{ssec.projected}

We turn now to the c-measure $\mu_{Pc}$ of the dynamics obtained by {\em projecting}
the flow on the boundary. This describes, in an ensemble of escaping trajectories, the probability of
the occurrence of collisions with  Birkhoff coordinates ${\bf x}$ {\em irrespective}
of the time of collision.  
Let the corresponding density be denoted as $\rho_{Pc}({\bf x})$, then
\begin{equation}
d \mu_{Pc} = \rho_{Pc} ({\bf x}) d{\bf x}.
\label{muP}
\end{equation}
The projected measure is obtained by integrating the flow measure $d \mu_{Fc}$ in Eq.~(\ref{mumuF2}) over coordinate time up to its maximum, $t_{\mathrm{coll}}({\bf x)}$ (see Fig.~\ref{fig.mapflow}).
This measure can thus be written as
\begin{equation}\label{mumuP}
\begin{array}{lll}
d \mu_{Pc} & = \int_0^{t_{coll}({\bf x})} d\mu_{Fc} = \int_0^{t_{coll}({\bf x})} A_c e^{-\kappa r} d r d\mu_c=\\ \\
 &=\dfrac{1-e^{-\kappa t_{\text{coll}}({\bf x})}}{\kappa} A_c d\mu_c.
\end{array}
\end{equation}
The relation between the densities is then
\begin{equation} 
\rho_{Pc}({\bf x}) = \rho_c({\bf x}) \frac{1-e^{-\kappa t_{\text{coll}}({\bf x})}}{\kappa} A_c
\rightarrow
\rho_{\mu}({\bf x}) \frac{t_{\mathrm{coll}}({\bf x})}{\tcoll_{\mu}},
\label{rhorhoP2}
\end{equation}
where the arrow indicates the limit of closed systems ($\kappa \rightarrow 0$).
Equivalently, 
\begin{equation} 
\rho_{Fc}({\bf x},r) = \rho_{Pc}({\bf x})\frac{\kappa e^{-\kappa r}}{1-e^{-\kappa t_{\mathrm{coll}}({\bf x})}}
\rightarrow 
\frac{\rho_{Pc}({\bf x})}{t_{\mathrm{coll}}({\bf x})}.
\label{rhorhoFP2}
\end{equation}

Using these relations, we can write average values in the projected and flow measures as a function of those in the true-time map
measure. Particularly interesting is the case of the average collision time. 
In the flow measure it is obtained as the average of the coordinate time $r$
\begin{equation} \label{tF2}
\begin{array}{lllllllll}
 \tcoll_{Fc} &= \int d{\bf x} \int _0^{t_{\mathrm{coll}}({\bf x})}dr \;r e^{-\kappa r} A_c \rho_c({\bf x}) \\ \\
&=\int d{\bf x} \dfrac{1-\et -\kappa t_{\mathrm{coll}}({\bf x})e^{-\kappa t_{\text{coll}}({\bf x})}}{\kappa^2} A_c\rho_c({\bf x}) \\ \\
&= \dfrac{1-\ec -\langle \kappa t_{\mathrm{coll}}e^{-\kappa t_{\text{coll}}}\rangle_c }{\kappa^2} A_c \rightarrow \dfrac{\langle t^2_{\mathrm{coll}}\rangle}{2 \tcoll}.
\end{array}
\end{equation}
In the projected measure the average collision time is obtained from (\ref{rhorhoP2}) as
\begin{equation} \label{tP2}
\begin{array}{lllllllll}
 \tcoll_{Pc} &=\int d{\bf x}\; t_{\mathrm{coll}}({\bf x}) \rho_{Pc}({\bf x})\\ \\
&=A_c \int d{\bf x}\; \dfrac{t_{\mathrm{coll}}({\bf x})(1-\et)}{\kappa} \rho_c({\bf x})\\ \\
&=\dfrac{\tcoll_c-\langle t_{\mathrm{coll}} e^{-\kappa t_{\text{coll}}}   \rangle_c}{\kappa} A_c \\ \\
&\rightarrow \dfrac{\langle t^2_{\mathrm{coll}}\rangle}{\tcoll} = 2\tcoll_F. 
\end{array}
\end{equation} 
The difference between the average collision time in the projection and in the true-time map is then 

\begin{equation}\label{difft2}
\begin{array}{ll}
\tcoll_{Pc} - \tcoll_c =  \\
\left( \tcoll_c \ec -\langle t_{\mathrm{coll}}\et\rangle_c\right) \dfrac{A_c}{\kappa}
\rightarrow \dfrac{\sigma^2_{t_{\mathrm{coll}}}}{\tcoll},
\end{array}
\end{equation}  
where $\sigma^2_{t_{\mathrm{coll}}}$ is the second cumulant of the distribution of the collisions times
in the closed map. The difference between the two averages is thus due to the 
inhomogeneity of the collision time distribution.

An even simpler relation can be obtained between the averages taken in the true-time map and the projected representation. 
Rearranging Eq.~(\ref{rhorhoP2}) and using the fact that~$\rho_c$ is normalized, we obtain 
\begin{equation} 
\int \rho_c({\bf x}) d{\bf x}= 1 = \int \rho_{Pc}({\bf x}) \frac{1}{A_c} \frac{\kappa}{1-\et} d{\bf x},
\label{Prhorho2}
\end{equation}  
from which it follows that
\begin{equation} 
\frac{1}{1-\ec}=\left\langle \frac{1}{1-e^{-\kappa t_{\mathrm{coll}}}}\right\rangle_{Pc}.
\label{1t2}
\end{equation}   
For closed systems this relation goes over into
\begin{equation} 
\frac{1}{\tcoll}=\left\langle\frac{1}{t_{\mathrm{coll}}}\right\rangle_{P}.
\label{1t}
\end{equation}   

\subsection{Algorithms for open billiards}\label{sec.ap-algorithm}

We describe the numerical simulations of open billiards.
 For the \limacon{} billiard considered map ${\bf f}$  was obtained using the standard geometric techniques~\cite{Robnik1983,Backer1997}: given a collision position ${\bf
  x}=(x,y)$ and velocity ${\bf v}$ (correspondingly, coordinates $(s,p)$), the next collision position is obtained as the first intersection
of the ray ${\bf x}+{\bf v}t$ with the billiard boundary. That is achieved by imposing that the point ${\bf x}+{\bf v}t$ satisfies
Eq.~(\ref{eq.limacon}) of the cardioid, which can be written for $S=1$ as $(x^2+y^2-\varepsilon x)^2 - (x^2+y^2) = 0$. The smallest solution
of the resulting fourth order polynomial in $t$ yields the collision time, $t_{\text{coll}}({\bf x'})$, and the next collision position,
${\bf x'}={\bf x}+{\bf v}t_{\text{coll}}({\bf x'})$.

\begin{figure}[!ht]
\centering
\includegraphics[width=\columnwidth]{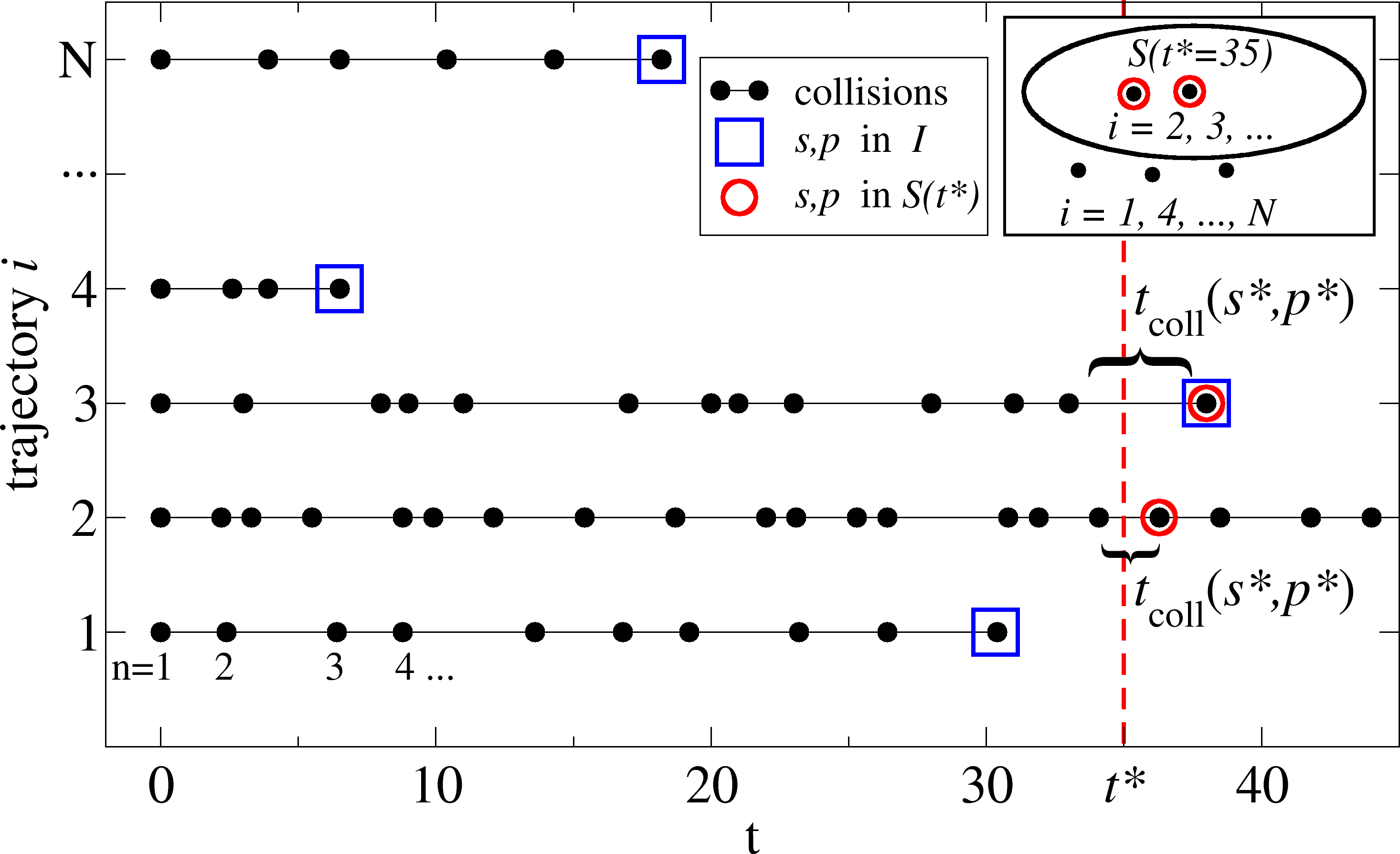}
\caption{(Color online) Schematic illustration of the numerical procedure to obtain the true-time
c-measure in open billiards. The time instants of collisions at the boundary are plotted, with black dots, for
  different trajectories. Trajectories $i=2$ and $i=3$ belong to $S(t^*=35)$ (inset) and have $n^*(i=2)=17$ and $n^*(i=3)=13$. $S(t^*)$
  denotes the set of trajectories that survive inside the billiard up to time~$t^*$ at least. Trajectories $i=1,4, N$ do not belong to $S(t^*=35)$.}  
\label{fig.ilust-numerics}
\end{figure}

We search for numerically efficient procedures to compute observable quantities in true-time open maps. In Sec.~\ref{ssec.transient} we learned that asymptotic quantities and invariant properties of the
escape process are given by the density~$\rho_c({\bf x})$ of the c-measure. Next we describe how this density
can be approximated in numerical simulations, related to the different representations of invariant measures. 

For the billiard, $\rho_c$ is approximated by taking all trajectories in an initial ensemble that survive up to a large
time~$t^*$. Mathematically, $t^*\rightarrow\infty$, but in practice, for $t^*> 2/\kappa$ a good approximation of $\rho_c$ is obtained.  For true-time maps, we have only the Birkhoff coordinates ${\bf x}_n=(s_n,p_n)$ and the time $t_n$ of collisions. It is natural to take for each trajectory
surviving up to time $t^*$ the values $s_{n^*}$ and $p_{n^*}$, where $n^*$ is the smallest~$n$ such that $t_{n^*}\ge
t^*$. Figure~\ref{fig.ilust-numerics} illustrates this procedure. Note that~$n^*$ is 
different for each trajectory and depends on the whole history of each trajectory [through
Eq.~(\ref{eq.tn})]. The points ${\bf x}^* \equiv(s_{n^*},p_{n^*})$ obtained 
through this procedure group together all points that at $t=t^*$ have coordinates ${\bf x}^*$ irrespective of their coordinate time $r$
after the last collision (small dots in Fig.~\ref{fig.mapflow})
Therefore, these points are then sampled according to~$\rho_{Pc}({\bf x})$, the projected density (which is insensitive to sliding orbits, see Eq.~(\ref{rhorhoP2})).
This means that the average~$\langle B \rangle_{Pc}$ of any observable~$B({\bf x})$ (e.g., $B({\bf x})=t_{\mathrm{coll}}({\bf x})$) can be numerically obtained
as 
$$\langle B \rangle_{Pc} \approx \frac{1}{N^*} \sum_{S(t^*)}B({\bf x}^*),$$
where $S(t^*)$ denotes the set of trajectories that survive inside the billiard up to time~$t^*$. Qualitatively speaking, the use of the ${\bf x}^*$ coordinates belonging to set $S(t^*)$ corresponds to an instantaneous observation of the system. In a simulation using $N$ initial
conditions, let the size of $S(t^*)$ be denoted by $N^*$. The survival probability 
can then be estimated as $P(t^*)=N^*/N$.
Furthermore, it is possible to use Eq.~(\ref{rhorhoP2}) and this numerical procedure to efficiently compute the true-time map average value
of any observable~$B({\bf x})$ as
\begin{equation}\label{avgB}
\begin{array}{lllll}
\langle B \rangle_c &\equiv \int d{\bf x}\; \rho_c({\bf x}) B({\bf x}) \\ \\
& = \int d{\bf x} \; \dfrac{1-\ec}{1-e^{-\kappa t_{\mathrm{coll}}({\bf x})}} \rho_{Pc}({\bf x}) B({\bf x})\\ \\
& \approx  \dfrac{1}{N^*}
\sum_{S(t^*)}  \dfrac{1-\ec}{1-e^{-\kappa t_{\mathrm{coll}}({\bf x^*})}} B({\bf x}^*),
\end{array}
\end{equation}
where the last relation is valid for $N^*,t^* \gg 1$.
In particular, 
with the choice $B({\bf x})\equiv1$, one recovers Eq.~(\ref{1t2}). Equation~(\ref{tP2})
 is recovered with the choice $B({\bf x})=t_{\mathrm{coll}}({\bf x}) (1-e^{-\kappa t_{\mathrm{coll}}({\bf x})})/\kappa$. 
Once the value of $\ec$ is known, Eq.~(\ref{avgB}) can be directly applied to an arbitrary observable.
 The measure of the leak $\mu_c(I)$ is computed, e.g., by choosing $B=1$ for ${\bf x} \in I$ and $B=0$ for
${\bf x} \notin I$.  

For partial leaks discussed in Sec.~\ref{ssec.partial}, Eq.~(\ref{avgB}) is generalized to
\begin{equation}
\langle B \rangle_{\tilde{c}} \approx \frac{1}{J^*}
\sum_{S(t^*)}  J_{t^*}  \frac{1-\langle e^{-\tilde{\kappa} t_{\text{coll}}} \rangle_c}{1-e^{-\tilde{\kappa} t_{\mathrm{coll}}({\bf x^*})}} B({\bf x}^*),
\label{avgBJ}
\end{equation}
where $J_{t^*}$ is the intensity of the trajectory over which summation runs, and $J^*=\sum_i^{N^*}\!\!J_{t^*}$ is the total intensity of the rays in
$S(t^*)$ (i.e., of all trajectories in the system at
time~$t^*$), which are all trajectories if $R({\bf x})\neq 0$. The advantage of this procedure, and of any application using the $S(t^*)$ set, or of projected densities, is that we can have
good control on the time~$t=t^*$ at which the measures are approximated numerically.
This is particularly important for
systems with mixed phase space discussed in Sec.~\ref{sec.weak}, where a crossover from exponential to power-law decay is observed in $P(t)$.

An alternative procedure for generating averages in the true-time map is based on
sampling a set 
of collision points 
according to the density~$\rho_c$ of the map. 
This is done by collecting in a set
$Q(t^*,\Delta t^*)$ all collisions in a time interval
$\Delta t^*=[t^*,t^*+\Delta t^*]$ for $\Delta t^* > \max\{t_{\text{coll}}({\bf x})\}$ and over all $N$ trajectories.
The use of the $Q$ set assumes an observation of the system over an extended time period.

To see how this procedure generates the $c$-measure, first consider $q$ different time instants $t_j^* \in \Delta t^*$, $j=1,...,q$, and let
 $S^q(t^*)=\{S(t^*_j)\}_{j=1}^{q}$. Since $\rho_{Pc}$ is independent of time,  points in all $S(t^*_j)$'s and in $S^q(t^*)$ are distributed according to $\rho_{Pc}$. However, the coordinates ${\bf x}^*$ of a single
  trajectory~$i$ will typically appear multiple times in $S^q(t^*)$ because different time instants $t^*\in\Delta t^*$ may have the same $n^*$. 
This can be seen by considering two closeby values of~$t^*$ in Fig.~(\ref{fig.ilust-numerics}). 
(For $t^*_j-t^*_{j+1}<\tcoll_c$, most trajectories will lead to repeated values
of ${\bf x}^*$.) For $q\rightarrow\infty$, the number of times a given ${\bf x}^*$  appears in $S^q(t^*)$ is proportional to
$\int_0^{t_{\text{coll}}({\bf x}^*)} e^{-\kappa t} dt$. 

A point ${\bf x}^*$ of a single trajectory is included only once in $Q(t^*,\Delta t^*)$ and possibly many times in $S^q(t^*)$. 
Therefore, noting that the $S$ set is sampled according to $\rho_{Pc}$, we find that points in $Q(t^*,\Delta t^*)$ are sampled
according to
$$\rho_Q({\bf x})= \frac{\rho_{Pc}({\bf x})}{A_c \int_0^{t_{\text{coll}}({\bf x})} e^{-\kappa t} dt}= \frac{\kappa}{A_c(1-e^{-\kappa t_{\text{coll}}({\bf x})})} \rho_{Pc}({\bf x})$$
where $A_c$ is the normalization constant (\ref{eq.Ac2}). 
Comparing to Eq.~(\ref{rhorhoP2}), we
obtain  $\rho_Q({\bf x}) \equiv \rho_c({\bf x})$.  
[Numerically it is efficient to consider the case $Q(t^*)\equiv Q(t^*,\Delta
t^*\rightarrow\infty)$, i.e, to consider all collisions until escaping the system.]
 Using this sampling procedure, Eq.~(\ref{avgB}) is rewritten as
\begin{equation}
\langle B \rangle_c \approx \frac{1}{N_Q}
\sum_{Q(t^*)}   B({\bf x}^*),
\label{avgBQ}  
\end{equation}
where $N_Q\gg1$ is the number of points in the set $Q(t^*)$. The case of partial leak~(\ref{avgBJ}) is in this case simply
\begin{equation}
\langle B \rangle_{\tilde{c}} \approx \frac{1}{J_{Q,t^*}} \sum_{Q(t^*)} J_{t^*}  B({\bf x}^*),
\label{avgBJQ}
\end{equation}
where $J_{Q,t^*}$ is the sum of the intensities of all collisions in the set $Q$.

\begin{figure}[!ht]
\includegraphics[width=1\columnwidth]{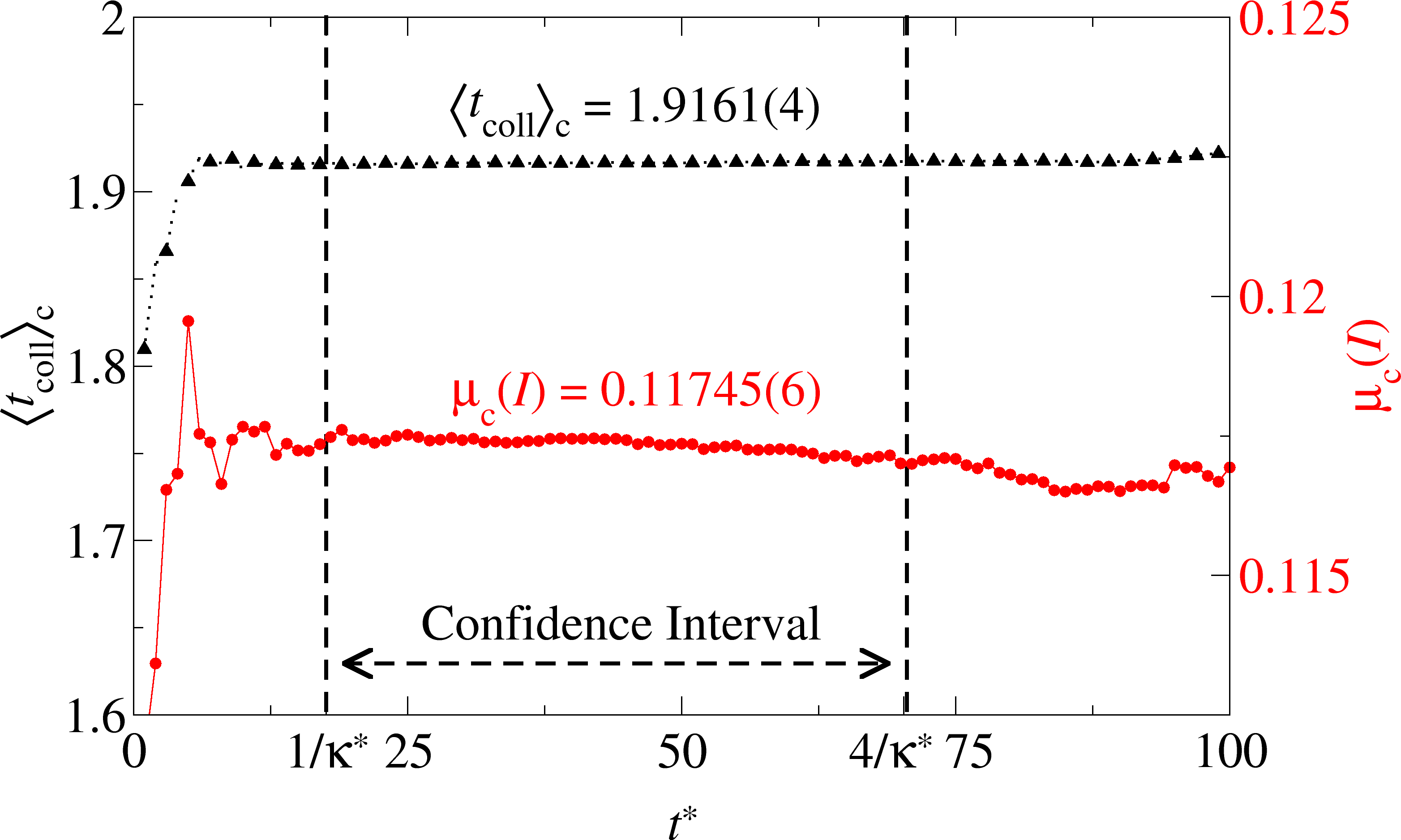}
\caption{(Color online) Convergence of the numerical estimation of $\mu_c(I)$ and $\tcoll_c$.  These results were obtained using Eq.~(\ref{avgBQ})  for the
  Cardioid billiard [$\varepsilon=1$ in Eq.~(\ref{eq.limacon})] with the leak as in Fig.~\ref{fig:open} 
using  $2.5\;\cdot 10^7$   initial conditions uniformly distributed in the full phase space $[-1,1] \times [-1,1]$. Short time fluctuations are present during the convergence 
to the c-measure. The results
$\mu_c(I)=0.11745 \pm 0.00006$ and  $\tcoll_c=1.9161 \pm 0.0004$ follow as
the mean and the standard deviation of the values obtained for
  integers $t^*$ in the interval $t^* \in [1/\kappa^*,4/\kappa^*]$ with the naive estimate $\kappa^*=0.05693$ obtained
  with Eq.~(\ref{eq.kappastar}) using~$\mu(I)=0.10$ and $\tcoll=3 \pi^2/16=1.85055\ldots$. The $Q$ sets at each time $t^*$ include all collisions for
  $t>t^* (\Delta t^*=\infty)$.
The actual decay rate $\kappa$ was obtained through a direct fit
  of~$P(t)$ (not shown) and resulted in  
  $\kappa=0.06559 \pm 0.00001$.
  The procedures described here have been used with finite $\Delta t^*$-s in all computations throughout.} 
\label{fig.convergence}
\end{figure}

In our simulations we  used the set $Q$ for calculating $\mu_c(I)$ and $\tcoll_c$,  as illustrated in Fig.~\ref{fig.convergence}. While the
set $Q$ provides better statistics and should typically be used, the results for the set $S$ are conceptually interesting and may find applications in different
simulations or experimental applications.

\subsection{Computation of invariant manifolds and densities}\label{ssec.appendixnew}

Numerical approximations of the invariant sets of transiently chaotic maps can be obtained using the sprinkler
method~\cite{TelGruiz-book,LaiTel-book}. The idea is to use initial conditions distributed uniformly in the phase space and keep
track of the trajectories that never escape up to a time~$t^*$. Coherent with the notation of Sec.~\ref{sec.ap-algorithm}, $t^*$ corresponds to the
transient time needed for the convergence to $\rho_c({\bf x})$. For $t^* \gg
1/\kappa$, the surviving trajectories necessarily start close to the stable manifold of the saddle (at $t=0$), approach the saddle (at
$t\approx t^*/2$), and most of them will be about to leave the system through the unstable manifold of the saddle (at $t=t^*$). The
invariant sets are approximated by the position of these surviving trajectories at the times $t_s=\{0,t^*/2,t^*\}$. More precisely, the stable
manifold, the chaotic saddle, and the unstable manifolds build the support of densities $\rho({\bf x},t_s)$ obtained from the position of
the trajectories ${\bf x}(t_s)$ at the different times $t_s=\{0,t^*/2,t^*\}$, respectively. For $t_s=t^*$ we recover the c-density
$\rho_c({\bf x})$. 

Modifications of the methods above are needed to address true-time maps with partial leaks. In order to address the true-time aspect of
the map, the modifications discussed in the previous section are needed. 
In the simulations shown here we considered an approach based on the $Q$-set of the
previous section, with a fixed $t^*\gg1/\kappa$ and a finite $\Delta t^* > \max\{t_{\text{coll}}({\bf x})\}$.
From trajectories surviving up to $t^*$, the positions~${\bf x}$ of all collisions (until escape) in the time interval $t \in  [t_s,t_s+\Delta t^*]$ 
 with $t_s=\{0,t^*/2,t^*\}$ generate the densities $\rho({\bf x},t_s)$.
 This procedure ensures that the requirement of surviving up to time $t^*$ is satisfied by all considered points.
(Note that when initial distributions reach the saddle
relatively fast within the chosen interval $\Delta t^*$, the initial positions are a better approximation to the stable manifold than the positions in the time interval $t \in  [0,\Delta t^*]$.)
For the case of partial leaks the same procedure is employed, and the position ${\bf x}$ of each collision 
with $t\in[t_s,t_s+\Delta t^*]$ is counted with a weight given by the intensity $J(t^*+t-t_s)$
 (at a given time $t>t^*$, the same intensity is used for the computation of the densities along the saddle, stable, and unstable manifolds). Note that for the case of full leak ($R=0$) and fixed
collision time ($t_{\text{coll}}=1$) the usual definitions are practically recovered.  
This general procedure was employed to compute 
Figs.~\ref{fig.saddle2} and \ref{fig.partial}.

\subsection{Difference between Poincar\'e and true-time maps}\label{ssec.diff}

The distinction between the usual Poincar\'e map and the true-time map is crucial in all open    
systems because the c-measure in both maps are different~\cite{Kaufmann2001}. This difference becomes
dramatic in the case of billiards with concave borders such as the \limacon{} billiard considered here, because of the existence of
trajectories with $t_{\text{coll}}({\bf x})=0$ sliding along the boundaries (whispering gallery) characterized by $p \approx \pm 1$. 
In the Poincar\'e map, these orbits build one-parameter families 
(the boundary) of nonhyperbolic trajectories. However, these orbits have little influence on the flow, where time is counted by~$t$
and not by the number $n$ of collisions.

\begin{figure}[!ht]
\centering
\includegraphics[width=0.9\columnwidth]{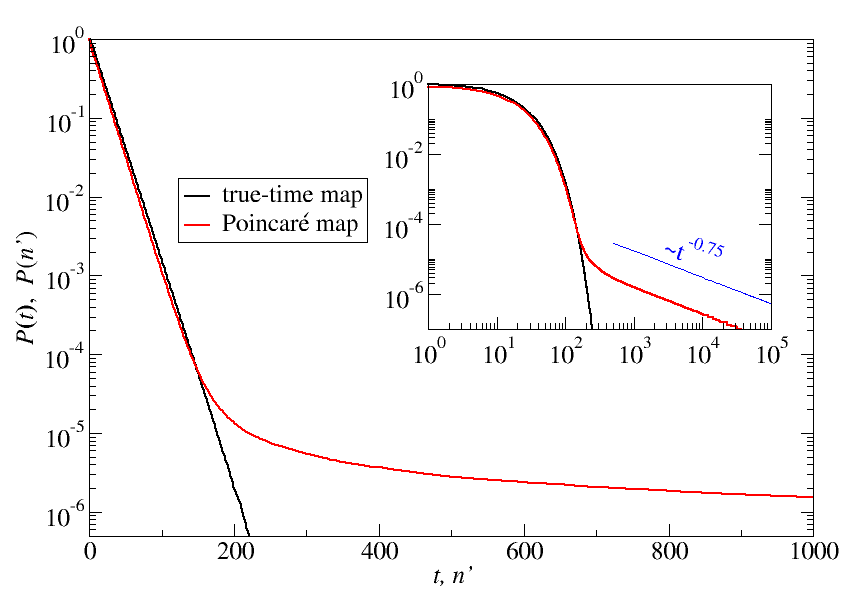}
\caption{(Color online) Difference between the true-time and the Poincar\'{e} map
of the cardioid billiard. Survival probability $P(t)$ as a function of the physical time $t$ (black) and
the discrete-time survival probability $P(n)$ given in terms of iteration number $n$ of the Poincar\'e map (gray/red), multiplied by the average
collision time $\tcoll_c$ for comparison, $n'=n\tcoll_c$. The leak is as in Fig.~\ref{fig:open} and Tab.~\ref{tab.ic.num}. 
 $10^8$  initial conditions have been taken, uniformly distributed in $[-0.9,0.9] \times [-0.9,0.9]$, to avoid sliding initial conditions. The log-log inset highlights the power-law tail of the Poincar\'{e} map.}
\label{fig:app}
\end{figure}

\begin{figure}[!ht]
\centering\includegraphics[width=0.65\columnwidth]{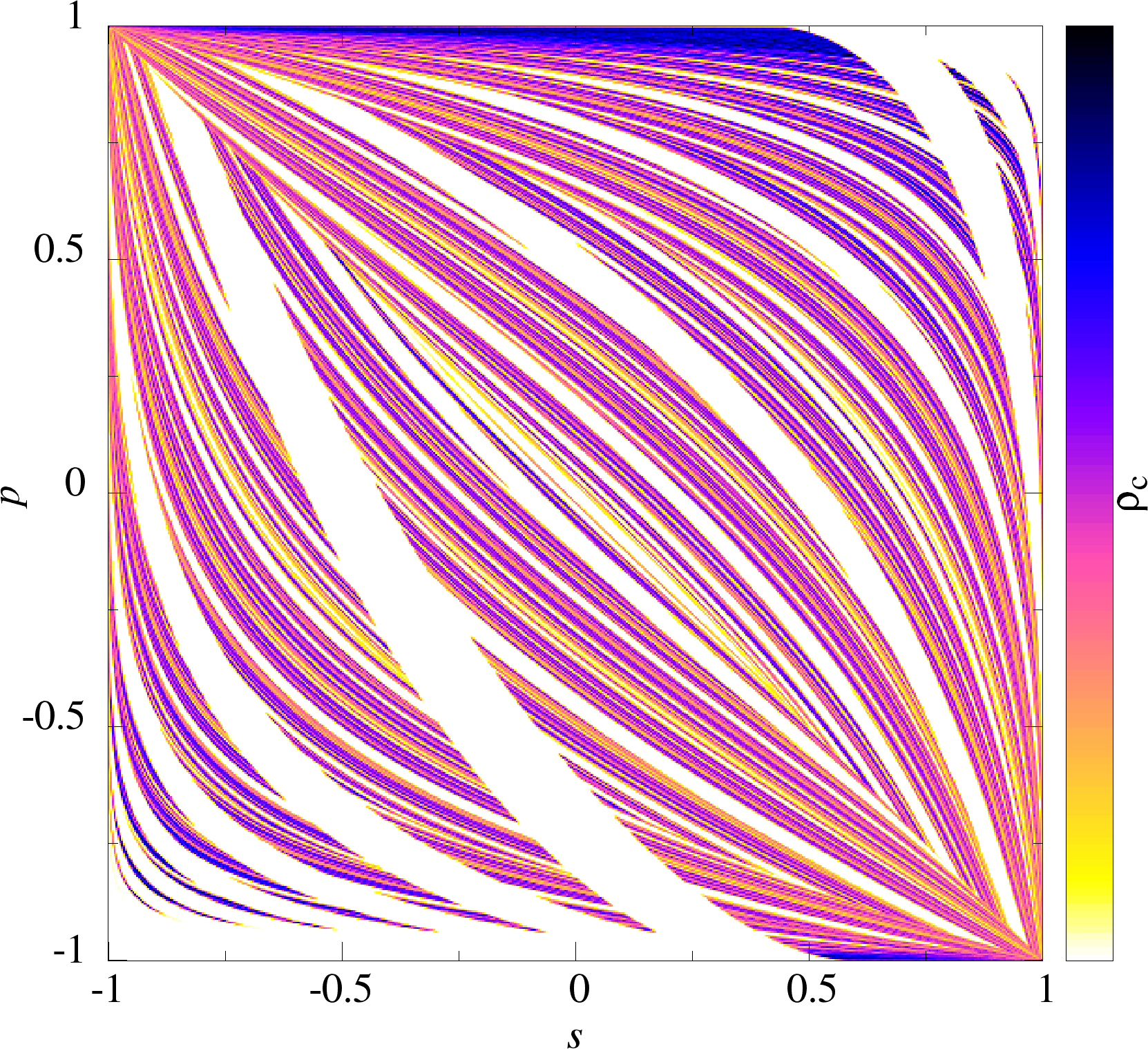}
\caption{(Color online) Conditionally invariant measure for the Poincar\'{e} map 
of the cardioid billiard obtained for initial conditions uniformly distributed in $[-0.9,0.9] \times [-0.9,0.9]$. This measure was obtained
by recording, during an interval $\Delta n^*=30$, the coordinates of the orbits that survived at least $n^*=10$ iterations.
 In comparison to the corresponding true-time map measure, Fig.~\ref{fig.saddle2}, obtained with $t^*=\Delta t^*=80$, points in the sliding-orbit region ($|p|\approx1$) have a bigger weight, although in the range $10<n<40$, escape is essentially exponential
(cf. Fig.~\ref{fig:app}).} 
\label{fig:rhoMap}
\end{figure}

The flow of the closed
cardioid billiard is proved to be ergodic and strongly mixing~\cite{Robnik1983,Wojtkowski1986}, and we expect 
a well-defined c-measure in the leaky case both in the flow and in the true-time map. Figure~\ref{fig:app} shows that the
usual  Poincar\'e map is not fully (uniformly) hyperbolic due to the sliding trajectories.
They convert the long time decay to non-exponential in the leaky case. The c-measure of
the Poincar\'e map is thus ill-defined. From this example it is clear that the simple correspondence $n \langle t_{\text{coll}} \rangle_c
\mapsto t$
between the physical time $t$ and the number of iterations $n$ of the map is not able to
explain the qualitative difference between the (physically meaningful) survival
probability~$P(t)$ and the \emph{discrete-time} survival probability~$P(n)$, defined as the fraction of initial conditions that survive
inside the system up to~$n$ collisions at least. From Fig.~\ref{fig:app} one can have the impression that the differences appear only for long times.
In Fig.~\ref{fig:rhoMap} we present, however, $\rho_c$
of the Poincar\'e map numerically obtained as usual, which should be compared to 
the true-time map result shown in Fig. \ref{fig.saddle2}. 
There is a disagreement which can be clearly seen in (but is not restricted to) the region of the sliding orbits ($|p|\approx 1$).

The c-measure of the Poincar\'e map and of the true-time map are generically different (even in the absence of sliding orbits)
$\kappa \neq \gamma_{\text{Pmap}}/\langle t_{\mathrm{coll}}  \rangle_c$, and there is no simple relationship between~$\kappa$ of the flow
and $\gamma_{\text{Pmap}}$. For the cardioid billiard $\gamma_{\text{Pmap}}$ is not even defined, as $P(n)$ decays asymptotically as a
power law. If, nevertheless, we consider 
an effective $\gamma_{\text{Pmap}}$ extracted from the intermediate time behavior in
 Fig.~\ref{fig:app}, we obtain that $\gamma_{\text{Pmap}}$ and $\kappa\tcoll{}_c$ differ by about 2\%, as can be seen from
 Tab.~\ref{tab.ic.num}, and the data reported in the caption.

\section*{Acknowledgments}
The authors thank A. B\"acker, M. V. Berry, C. Dettmann, G. Dr\'otos, O. Georgiou, A. Goussev, Z. Kauffmann, M. Novaes, S. Shinohara, S. Szanyi, S. Wimberger, and
D. A. Wisniacki for their valuable contributions, and all the authors who kindly allowed reproduction of their results in Sec.~\ref{sec.applications}. This work was supported by 
OTKA Grant No. NK100296, the von Humboldt Foundation, and the Fraunhofer Society.


%
\end{document}